\documentclass[preprint,10pt]{elsarticle}
\usepackage{amsfonts}
\usepackage{epsfig,amsmath,latexsym,amssymb}
\usepackage{amscd}
\usepackage{multirow}
\usepackage{graphicx}
\usepackage{geometry}
\usepackage{lscape}
\usepackage{amsmath}
\usepackage{amssymb}
\usepackage{bm}
\usepackage{subfigure}

\providecommand{\U}[1]{\protect\rule{.1in}{.1in}}
\newcommand{\Remm}[1]{}

\newtheorem{model ass}[theo]{Model Assumptions}

\numberwithin{equation}{section}

\begin{document}

\begin{frontmatter}

\title{Impact of Insurance for Operational Risk: Is it worthwhile to insure or be insured for severe losses?}
\author{Gareth W.~Peters$^{1,2}$ \quad Aaron D.~Byrnes$^{1}$ \quad Pavel V.~Shevchenko$^{2}$} 
\date{{\footnotesize {Working paper, version from \today }}}
\maketitle

\begin{abstract}
\noindent Under the Basel II standards, the Operational Risk (OpRisk) advanced measurement approach allows a provision for reduction of capital as a result of insurance mitigation of up to 20\%. This paper studies different insurance policies in the context of capital reduction for a range of extreme loss models and insurance policy scenarios in a multi-period, multiple risk settings. A Loss Distributional Approach (LDA) for modelling of the annual loss process, involving homogeneous compound Poisson processes for the annual losses, with heavy-tailed severity models comprised of $\alpha$-stable severities is considered. There has been little analysis of such models to date and it is believed, insurance models will play more of a role in OpRisk mitigation and capital reduction in future. The first question of interest is when would it be equitable for a bank or financial institution to purchase insurance for heavy-tailed OpRisk losses under different insurance policy scenarios?  The second question pertains to Solvency II and addresses quantification of insurer capital for such operational risk scenarios. Considering fundamental insurance policies available, in several two risk scenarios, we can provide both analytic results and extensive simulation studies of insurance mitigation for important basic policies. The intention being to address questions related to VaR reduction under Basel II, SCR under Solvency II and fair insurance premiums in OpRisk for different extreme loss scenarios. In the process we provide closed-form solutions for the distribution of loss process and claims process in an LDA structure as well as closed-form analytic solutions for the Expected Shortfall, SCR and MCR under Basel II and Solvency II. We also provide closed-form analytic solutions for the annual loss distribution of multiple risks including insurance mitigation. 
\end{abstract}

\begin{keyword}
Operational Risk, Loss Distributional Approach, Insurance Mitigation, Capital Reduction, $\alpha$-Stable, Basel II, Solvency II.
\end{keyword}

\begin{center}
{\footnotesize {\ \textit{$^{1}$ UNSW Mathematics and Statistics
Department, Sydney, 2052, Australia; \\[0pt]
email: peterga@maths.unsw.edu.au \\[0pt]
(Corresponding Author) \\[0pt]
$^{2}$ CSIRO  Mathematics, Informatics and Statistics, Locked Bag
17, North Ryde, NSW, 1670, Australia \\[0pt] } } }
\end{center}

\end{frontmatter}

\pagebreak

\section{Motivation}
Modelling the impact of insurance mitigation for different risk cells and business units is an important challenge in the setting of operational risk (OpRisk) management yet to be fully understood and therefore adopted in practice. The slow uptake of insurance policies in OpRisk for capital mitigation can be partially attributed to the limited understanding of their impact in complex multi-risk, multi-period scenarios, under heavy-tailed losses and the fair premium to charge for such policies, as well as a relatively conservative Basel II regulatory cap of 20\%. Therefore, although OpRisk models are maturing, OpRisk insurance mitigation is still in its infancy (\cite{brandts2004operational}, \cite{bazzarello2006modeling}).

The Basel II OpRisk regulatory requirements for the Advanced Measurement Approach, BIS (2006) p.148, states \textquotedblleft \textit{Under the AMA, a bank will be allowed to recognise the risk mitigating impact of insurance in the measures of operational risk used for regulatory minimum capital requirements. The recognition of insurance mitigation will be limited to 20\% of the total operational risk capital charge calculated under the AMA}.
\textquotedblright\ Therefore from the perspective of a financial intstitution, such as a bank, there is a strong incentive to understand the effect of insurance mitigation on the OpRisk capital. 

From the insurers perspective a quantitative understanding of the impact of insurance in OpRisk extreme loss scenarios will allow for accurate pricing of insurance premiums. In addition by studying the risk transfer from bank to insurer, this will aid in modelling of the required capital for an insurer under Solvency II. As discussed in the initiatives developed by the International Association of Insurance Supervisors (\cite{kawai2005iais} and \cite{linder2004solvency}), the Solvency II framework was developed as a similar system to the Basel II three pillar system. It specifies the financial resources that a company must hold to be considered solvent. In \cite{sandstrom2006solvency} the IAIS guidances under Principal 8 discuss minimum capital where by
\textquotedblleft \textit{A minimum level of capital has to be specified}
\textquotedblright\
, this is then a quantitative challenge. In the second phase of the EU project Solvency II, the commision introduced two distinct levels of solvency: these are measured according to an upper level, the Solvency Capital Requirement (SCR) and lower level, the Minimum Capital Requirement (MCR), see \cite{sandstrom2006solvency}. 

The framework we develop in this paper for the Basel II LDA capital reduction analysis, will also naturally extend to allow for calculation of required insurer capital under Solvency II. That is we can utilise the LDA claim process models from the Basel II OpRisk studies to estimate the MCR and SCR of the insurer. The MCR and SCR measures for an insurance underwriter differ to the Value at Risk (VaR) that a banking institution must hold. In particular, unlike the capital measure obtained by a 99.95\% VaR for banking institutions under Basel II, the SCR or (target capital requirement) is the target level of corrections in a going concern. The SCR as proposed under Solvency II is treated as a `soft' level since  
\textquotedblleft \textit{there are no intervention measures restricting management of the business, ..., there might be measures taken by the authority to let the company, for examle, submit a plan about restoring the capital level.}
\textquotedblright\
(\cite{sandstrom2006solvency}, p.186).
Contrary to the SCR, the MCR measures the normal target level of capital that enables an institution to absorb significant unforseen losses. As such MCR is considered to be a `hard' solvency margin which defines the level at which management of a company is taken over by the supervisory authority. 

In this paper, we make explicit the estimation of the MCR which we define to be the Solveny II analog of the capital requirement in Basel II. That is, given our LDA OpRisk model under the Basel II framework, with any particular insurance policy, we can define the MCR as the 95\% VaR or Expected Shortfall (ES) of the claims process. This is obtained as a by-product of the simulation of the annual loss distribution in OpRisk under the LDA framework when estimating the insurance mitigated bank capital. In the OpRisk context, measuring the MCR according to the percentile of the claims process, simulated from the OpRisk LDA model claims proces allows for diversification in the form of the four levels proposed in (\cite{sandstrom2006solvency}, p.188). These relate to accounting for risk exposures, subportfolios, main risk catergories and subrisk classes and business units. In some insurance settings we will even obtain closed-form analytic solutions for evaluation of MCR. 
 
Therefore, in this paper we will study from the perspective of both the banking sector and the insurer, the basic transfer of risk and therefore capital requirements from financial institution to insurer for several important examples of OpRisk insurance policies. In particular we will focus on the interplay between the capital reduction in OpRisk for different insurance polices as measured by VaR of the annual loss in an LDA OpRisk model and the MCR in Solvency II as measured by the VaR of the claims process\footnotetext[1]{However we ignore the fact that insurers can transfer the risk to a `collective' by diversifying through selling many policies. This is somewhat justified for some risks where policies are client specific as would be the case in extreme events in OpRisk.}. An understanding of such processes for fundamental policies will also provide instructive analysis for regulatory bodies who will be better able to understand the extent that capital redcution can be offset by such insurance policies, allowing for an informed conservatism to be applied. 

In particular we focus our analysis on the scenarios involving heavy-tailed severity models in the rare-event extreme consequence context. Thereby providing analysis of the loss process most likely to have significant consequences on a financial institution, those which may lead to ruin. This involves introducing to OpRisk modelling an important family of severity models, utilised in insurance claims reserving in \cite{adler1998practical}, given by the $\alpha$-stable severity model. This family of severity model are flexible enough to incorporate light-tailed Gaussian loss models throught to infinitie mean, infinite variance severity loss models such as the Cauchy model. 

The three questions of particular interest to this paper are posed as:
\begin{itemize}
\item When would it be equitable for a bank or financial institution to purchase insurance for heavy-tailed OpRisk losses under different insurance policy scenarios?
\item How does the SCR and MCR capital measures for insurers under Solvency II behave relative to the Basel II VaR captial mitigation due to OpRisk insurance under different insurance policy structures? 
\item From an insurers perspective, what is the fair premium to charge as a percentage above the expected annual claim for basic building blocks of different insurance policies?
\end{itemize}

To address these questions we consider the standard LDA Basel II structures, involving an annual loss in a risk cell (business line/event type) modelled as a compound random variable,%
\begin{equation}
Z_{t}^{\left( j\right) }=\sum\limits_{s=1}^{N_{t}^{\left( j\right)
}}X_{s}^{\left( j\right) }\left( t\right).  \label{Model1}
\end{equation}%
Here $t=1,2,\ldots,T,T+1$ in our framework is discrete time (in annual units) with $T+1$ corresponding to the next year. The upper script $j$ is used to identify the risk cell. The annual number of events $N_{t}^{(j)}$ is a random variable distributed according to a frequency counting distribution $P^{(j)}(\cdot) $, typically Poisson. The severities in year $t$ are represented by
random variables $X_{s}^{(j)}(t)$, $s\ge1$, distributed according to a severity distribution $F^{(j)}(\cdot)$. The total bank's loss in year $t$ is calculated as
\begin{equation}
Z_{t}=\sum\limits_{j=1}^{J}Z_{t}^{\left( j\right) },
\label{totalLoss}
\end{equation}%
where formally for OpRisk under the Basel II requirements $J=56$
(seven event types times eight business lines). However, this may
differ depending on the financial institution and type of problem.

\section{Insurance Models}
This section presents different versions of insurance model speficied by Top Cover Limits (TCL) under multiple risk modelling scenarios. These are selected to be fundamental insurance models that provide information about the building blocks for more advanced policy structures. Building on these we also consider two advanced insurance policy structures, the first involving a Basel II haircut with a linearly increasing TCL over the duration of the year \cite{bazzarello2006modeling}. The second policy involves a stochastic banding loss structure for the TCL, which can be considered an extension of the model proposed in \cite{bazzarello2006modeling}, to a stochastic insurance structure. 

We examine the competing trade-off between bank capital mitigation and the MCR attributed to the insurer due to the OpRisk loss process. To accomplish this, we will investigate the behaviour of the risk mitigated loss over one year for the $j$-th risk process and $i$-th insurance policy, $Z^{(j,i)}_{t}$, as well as that of the value of claims over the same period, denoted $C^{(j,i)}_{t}$.

\subsection{Basic Insurance Models}
\label{BasicIns}
We begin with individual loss processes, followed by a policy applicable to combined risk processes, which may be used in practice to exploit knowledge of dependence properties of particular risk processes in Basel II. For studies of the impact of depedence models in OpRisk; see \cite{peters2009dynamic}. In addition, we assume a simple setting in which deductible excess is zero.

\subsubsection{Policy 1 - Individual Loss Policy (ILP)}
\label{ILPSect}
\begin{figure}[ht]
\includegraphics[width = 0.8\textwidth, height = 5cm]{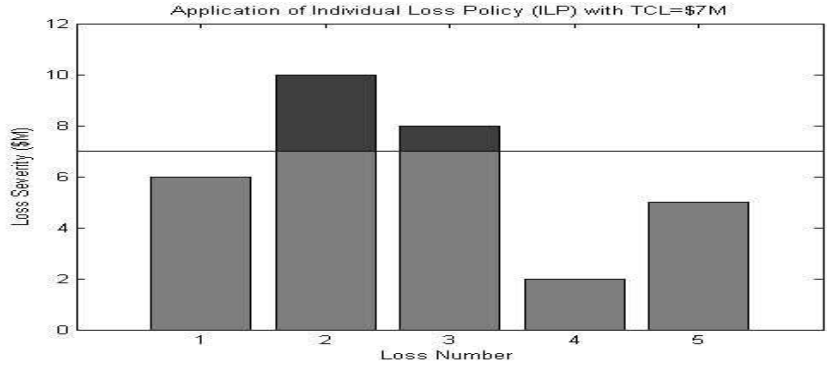}
\caption{Individual Loss Policy and the application of the Top Cover Limit.}
\label{Fig1}
\end{figure}

The ILP provides a specified maximum compensation, a TCL on a per event basis. This concept can be illustrated via the following example in which we consider a year consisting of five OpRisk losses of $\{\$6M,\$10M,\$8M,\$2M,\$5M\}$. For each loss, the insurer will provide compensation on the loss up to the value TCL, as illustrated in Figure \ref{Fig1}.

As can be seen in losses 2 and 3, the value of the loss exceeds the TCL of $\$7$M, hence the insurer provides compensation of $\$7$M (highlighted in gray) and the bank still incurs the loss above this value (highlighted in black). Therefore one can write the ILP risk mitigated loss LDA model from the perspective of the banking institution according to the loss process
\begin{equation}
Z_{t}^{\left(j,ILP\right) }=\sum\limits_{s=1}^{N^{\left( j\right)}(t)}\max{(X_{s}^{\left( j\right) }\left( t\right)-TCL,0)}.  
\label{ILPLoss}
\end{equation}
From the perspective of the insurer, the claims process arising from this ILP risk mitigated loss LDA model can be presented according to
\begin{equation}
C_{t}^{\left( j, ILP\right) }=\sum\limits_{s=1}^{N^{\left( j\right)
}(t)}\min{(X_{s}^{\left( j\right) }\left( t\right),TCL)}.  
\label{ILPClaim}
\end{equation}

\subsubsection{Policy 2 - Accumulated Loss Policy (ALP)}
\begin{figure}[ht]
\includegraphics[width = 0.8\textwidth, height = 5cm]{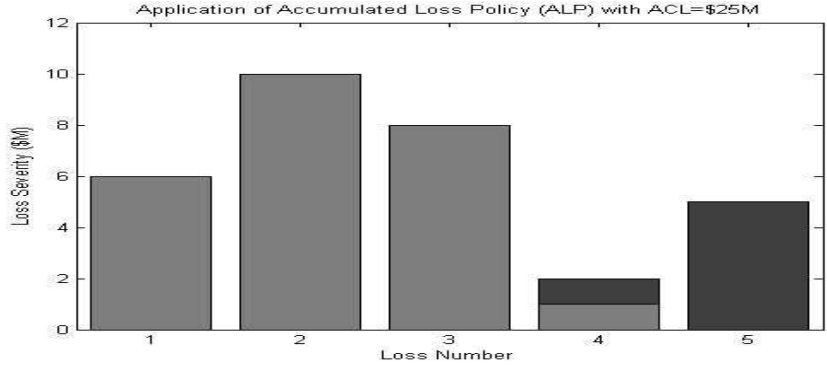}
\caption{Accumulated Loss Policy and the application of the Accumulated Cover Limit.}
\label{FigALP}
\end{figure}

The ALP provides a specified maximum compensation, which is a transformation of the TCL to account for frequency, on loss experienced over a one year insurance period. Again, we illustrate the application of such a policy with the annual loss example presented in the ILP Section \ref{ILPSect} and ACL = $\$25$M. In this case the insurer will provide complete compensation of all losses over the year until the value of compensation reaches the limit ACL, as depicted in Figure \ref{FigALP}. 

In this setting the insurer compensates the bank for losses 1 to 3. However, the $4^{\text{th}}$ loss brings the total claim value to $\$26$M which will exceed the ACL cap of $\$25$M, hence the insurer only compensates the bank for the first $\$1$M of the $4^{\text{th}}$ loss and the bank is exposed to the entirety of the $5^{\text{th}}$ loss.

The risk mitigated loss of the bank can be expressed according to Equation \ref{ALPLoss}. This formulation can be simplified, though this representation is particularly useful when application of the policy on an event by event basis in the simulation studies an also if ACL is time dependent. 
{\small{
\begin{equation}
\begin{split}
Z_{t}^{\left( j, ALP\right) } = \sum\limits_{s=1}^{N^{\left( j\right)
}(t)}& X_{s}^{(j)}\left( t\right) \times \mathbb{I}\left(\sum\limits_{k=1}^{s-1}X_{k}^{(j)}\left( t\right)\geq ACL\right) \\
&+ \left(\left(\sum\limits_{k=1}^{s}X_{k}^{(j)}\left( t\right)\right)-ACL\right) \times \mathbb{I}\left(0 < ACL-\sum\limits_{k=1}^{s-1}X_{k}^{(j)}\left( t\right)<X_{s}^{(j)}\left( t\right)\right).  
\end{split}
\label{ALPLoss}
\end{equation}
}}
with $\mathbb{I}\left(\cdot\right)$ denoting the indicator function. The claims process for this period can also be expressed according to
{\small{
\begin{equation}
\begin{split}
C_{t}^{\left( j,ALP\right) } = \sum\limits_{s=1}^{N^{\left( j\right)
}(t)} & \left[X_{s}^{(j)}\left( t\right)\times \mathbb{I}\left(\sum\limits_{k=1}^{s}X_{k}^{(j)}\left( t\right)\leq ACL\right) \right.\\
& \left. + \left(ACL-\left(\sum\limits_{k=1}^{s}X_{k}^{(j)}\left( t\right)\right)\right)\times \mathbb{I}\left(ACL-\sum\limits_{k=1}^{s-1}X_{k}^{(j)}\left( t\right)<X_{s}^{(j)}\left( t\right)\right)\right].  
\end{split}
\label{ALPClaim}
\end{equation}
}}
\subsubsection{Policy 3 - Combined Loss Policy (CLP)}
\begin{figure}[ht]
\includegraphics[width = 0.8\textwidth, height = 5cm]{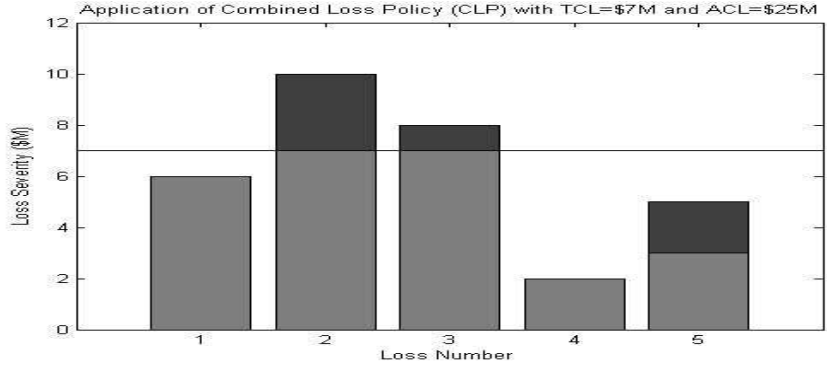}
\caption{Combined Loss Policy and the application of the ILP and ACL.}
\label{CLPFig}
\end{figure}

A CLP insurance contract provides a specified maximum compensation, TCL, on a per event basis up to a maximum per year loss, ACL. Considering again the example in Subsection \ref{ILPSect}, we illustrate the application of such a policy in Figure \ref{CLPFig}. Under the CLP, the insurer will provide compensation on the loss up to the value TCL. However, once the total value of claims exceeds the aggregate limit ACL the insurer will not provide any further compensation.

As can be seen, the insurer only compensates the bank for the first $\$7$M of losses 2 and 3. In addition, the total value of claims is exceeded by the $5^{\text{th}}$ loss. Hence, the insurer will only compensate the first $\$3$M of the $5^{\text{th}}$ loss and the remaining exposure is incurred fully by the bank. The risk mitigated loss process can be expressed as
{\small{
\begin{align}
Z_{t}^{\left( j,CLP\right) } & = \sum\limits_{s=1}^{N^{\left( j\right)
}(t)}\left[X_{s}\left( t\right) \times \mathbb{I}\left(\sum\limits_{k=1}^{s-1}\min{\left(X_{k}^{(j)}\left( t\right),TCL\right)}\geq ACL\right)\right. \\ \nonumber
&  + \max{\left(X_{s}^{(j)}\left( t\right)-TCL,0\right)} \times \mathbb{I}\left(\sum\limits_{k=1}^{s}\min{\left(X_{k}^{(j)}\left( t\right),TCL\right)}\leq ACL\right) \\
& +\left(X_{s}^{(j)}\left( t\right)-\left(ACL-\sum\limits_{k=1}^{s-1}\min{\left(X_{k}^{(j)}\left( t\right),TCL\right)}\right)\right) \\
& \left. \times \mathbb{I}\left(ACL-\sum\limits_{k=1}^{s-1}\min{\left(X_{k}^{(j)}\left( t\right),TCL\right)}<\min{\left(X_{s}^{(j)}\left( t\right),TCL\right)}\right)\right]. \nonumber 
\label{CLPLoss}
\end{align}
}}
Conversely, the claims process experienced by the insurer in such a model, for this period can also be expressed according to
\begin{align}
C_{t}^{\left( j,CLP\right) } & = \sum\limits_{s=1}^{N^{\left( j\right)
}}\left[\min{\left(X_{s}\left( t\right),TCL\right)} \times \mathbb{I}\left(\sum\limits_{j=1}^{s}\min{\left(X_{j}\left( t\right),TCL\right)}\leq ACL\right)\right] \\
& +\left[\left(ACL-\sum\limits_{j=1}^{s-1}\min{\left(X_{j}\left( t\right),TCL\right)}\right) \times \mathbb{I}\left(ACL-\sum\limits_{j=1}^{s-1}\min{\left(X_{j}\left( t\right),TCL\right)}<\min{\left(X_{s}\left( t\right),TCL\right)}\right)\right]. \nonumber 
\label{CLPClaim}
\end{align}

\subsubsection{Policy 4 - Accumulated Loss Policy - Two Risk Exposures (ALP2)}
Here we extend the CLP to the multi-risk setting. We illustrate this concept for a bi-variate risk process, where a cap on compensation for accumulated losses across two loss processes over the year is imposed, denoted ACL. Under this approach, the insurer will provide compensation to the bank in the same method as in Policy 2, however, the ACL limit is placed over claims on both risk exposures. As applied to the basic LDA model, a simplified risk mitigated loss can be expressed according to 
\begin{equation}
Z_{t}^{\left( j,ALP2\right) } = \max{\left[\left(\sum\limits_{s=1}^{N^{\left(1\right)}(t)}X^{(1)}_{s}\left( t\right)\right) + \left(\sum\limits_{s=1}^{N^{\left(2\right)}(t)}X^{(2)}_{s}\left( t\right)\right) - ACL,0\right]} 
\label{ALP2Loss}
\end{equation}
The resulting combined claims process generated by the multi-risk setting for this period can also be expressed according to Equation \ref{ALP2Claim}.
\begin{equation}
C_{t}^{\left( j,ALP2\right) } = \min{\left[\left(\sum\limits_{s=1}^{N^{\left(1\right)}(t)}X^{(1)}_{s}\left( t\right)\right) + \left(\sum\limits_{s=1}^{N^{\left(2\right)}(t)}X^{(2)}_{s}\left( t\right)\right), ACL\right]}  
\label{ALP2Claim}
\end{equation}

\subsection{Advanced Insurance Models}
Basel II requires other insurance modelling conditions as outlined in (\cite{BIS}(2006); p.148), two of these being residual term of a policy and payment uncertainty. To address these we consider two advanced insurance models, the first following guidelines proposed in Basel II relating to the insurance premium haircut and the second based on a stochastic banding strucutre. In particular, the second stochastic model extends the model of \cite{bazzarello2006modeling}, allowing one to capture the notion of payment uncertainty. This is a critical aspect of both Basel II and Solvency II modelling, see (\cite{BIS}(2006); p.148).

\subsubsection{Policy 5 - Haircut Loss Policy (HLP)}
Under the Basel II framework, it is clearly specified that ``for policies with a residual term of less than one year, the bank must make appropriate haircuts'' \cite{BIS} Section 678, p.148. Modelling the haircut complicates the LDA model since now one requires explicit knowledge of the arrival time process. In this paper we model the inter-arrival time of losses in a year as exponentially distributed. Under this model, we consider the simplest scenario in which a basic haircut is applied to the single risk LDA model (\ref{Model1}).

Under the HLP, insurance is applied to the loss process however, each compensation amount is specified by a proportion of the TCL. In particular the insurance mitigation follows a discounted time sensitive factor, or haircut factor. For simplicity, we consider a linear function increasing from 0\% insurance mitigation at the beginning of the year up to 100\% of the TCL at the end of the year.

We can therefore write the risk mitigated loss process according to 
\begin{equation}
Z_{t}^{\left( j,HLP\right) }=\sum\limits_{s=1}^{N^{\left( j\right)
}(t)}\max{(X_{s}^{\left( j\right) }\left( t\right)-\alpha\left( t\right)TCL,0)}.  
\label{HLPLoss}
\end{equation}

Additionally, the claims process for this period can also be expressed as 
\begin{equation}
C_{t}^{\left( j,HLP\right) }=\sum\limits_{s=1}^{N^{\left( j\right)
}(t)}\min{(X_{s}^{\left( j\right) }\left( t\right),\alpha\left( t\right)TCL)},  
\label{HLPClaim}
\end{equation}
where the function $\alpha(t) \in [0,1], \; \forall t>0$ describes the haircut of the policy limit TCL as a percentage.

\subsubsection{Policy 6 - Banded Loss Policy (BLP)}
To model insurer payment uncertainty for OpRisk, we consider a banding model. Payment uncertainty as discussed in \cite{brandts2004operational} generally arises as a result of disagreements between a bank or financial institution and its insurer as to the true value of loss that will be realized. As such, when modelling the resulting processes for both annual loss from the banking perspective and claims from the insurers perspective, it is important to account for such uncertainty to ensure appropriate capitilization and solvency.

As proposed in \cite{bazzarello2006modeling}, a banded structure for payment uncertainty allows for accounting for the fact that severe losses will typically attract more disagreement from insurers and are more likely to be affected by payment delays on such claims on these losses. Therefore, severe losses may be more realistically modelled as being discounted by larger values due to their heightened likelihood of payment uncertainty arising from counter party disputes on larger claims. Previously such models were deterministic, we extend these models to treat payment uncertainty as a stochastic process. To achieve this, we consider a stochastic banding structure across different levels of severity in which we can reflect higher probabilities of reductions in total coverage of losses as severity of such losses increases. In addition, we demonstrate the influence this has on the insurers solvency as measured by MCR. As applied to the basic loss model, Equation (\ref{Model1}), the risk mitigated loss process can be expressed according to Equation (\ref{BLPLoss}).
\begin{equation}
Z_{t}^{\left( j,BLP\right) }=\sum\limits_{s=1}^{N^{\left( j\right)
}(t)}\max{\left(X_{s}^{\left( j\right) }\left( t\right)-CL\left( X_{s}^{\left( j\right) }\left( t\right)\right),0\right)}.  
\label{BLPLoss}
\end{equation}

The resulting claims process for this period can also be expressed according to 
\begin{equation}
C_{t}^{\left( j,BLP\right) }=\sum\limits_{s=1}^{N^{\left( j\right)
}(t)}\min{\left(X_{s}^{\left( j\right) }\left( t\right),CL\left( X_{s}^{\left( j\right) }\left( t\right)\right)\right)},  
\label{BLPClaim}
\end{equation}
where function $CL(X)$ is defined below. We will segment the top level of cover for the policy as quantified by the $TCL$ into $D$ bands of equal length $L$ (possibly unequal length depending on the application). Under this segmentation, we can define a function that identifies in which band $X$ is located, denoted by the indicator function $d(X)$ for a given band and defined according to 
\begin{equation}
d(X)= \min\left(D,\lfloor\left(\frac{X}{L}\right)\rfloor+1\right)
\label{BLP_BASIND}
\end{equation}
According to the definition of $d(X)$ we can now define $CL(X)$ as
\begin{equation}
CL(X) = (d(X)-1)L+\delta_{d(X)}\min{\left(L,X-(d(X)-1)L\right)}
\label{BLPTCL}			
\end{equation}
where $\delta_X \sim Be(\alpha(X),\beta(X))$ is a random variable from Beta distribution with
\begin{equation*}
\begin{split}
\alpha(X)&=\mathbb{I}\left(d(X)\geq\lceil\left(\frac{D+1}{2}\right)\rceil\right)+\left(\lceil\left(\frac{D+1}{2}\right)\rceil-d(X)\right)\left[\frac{2}{\left(D-\lceil\left(\frac{D+1}{2}\right)\rceil\right)}\right]\times \mathbb{I}\left(d(X)<\lceil\left(\frac{D+1}{2}\right)\rceil\right),\\		\beta(X)&=\mathbb{I}\left(d(X)\leq\lfloor\left(\frac{D+1}{2}\right)\rfloor\right)+\left(d(X)-\lfloor\left(\frac{D+1}{2}\right)\rfloor\right)\left[\frac{2}{\left(D-\lfloor\left(\frac{D+1}{2}\right)\rfloor\right)}\right]\times \mathbb{I}\left(d(X)>\lfloor\left(\frac{D+1}{2}\right)\rfloor\right).	
\end{split}
\label{BLPAlpha}
\end{equation*}

\begin{figure}[ht]
\includegraphics[width = 0.9\textwidth, height = 5cm]{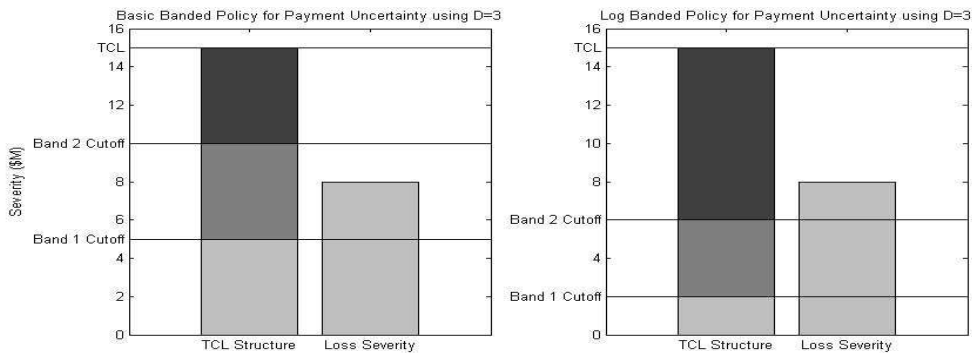}
\caption{Linear and logarithmic stochastic banding policies.}
\label{BLPFig}
\end{figure}

However, in actual application of payment uncertainty to an observed claims processes, it is unlikely that the bands of cover limit would have equal length. Therefore to account for this we will convert the $i$-th band on the $[0,1]$ basic scale with length $l=\frac{L}{TCL}$ to the $i$-th band on the $[0,1]$ log-scale with length $B_i$ via the transformation
\begin{equation}
B_i=\frac{\exp{\left(il\right)}-\exp{\left((i-1)l\right)}}{\exp{\left(1\right)}-1}, \; \text{for } i=1,\ldots,D.
\label{BLP_LOGTRAN}
\end{equation}
Hence, the band identifying function (\ref{BLP_BASIND}), $d(X)$, can be redefined in the log-scale case to $b(X)$ which identifies the log-band in which $X$ is located
\begin{equation}
b(X)= \mathbb{I}\left(X\leq B_1 TCL\right)+\sum^{D-1}_{i=2}{\left[i \; \mathbb{I}\left(\sum^{i-1}_{j=1}{B_i}<\frac{X}{TCL}\leq \sum^{i}_{j=1}{B_i}\right)\right]}+D \; \mathbb{I}\left(X> B_{D-1}TCL\right).
\label{BLP_LOGIND}
\end{equation}
From here, the calculation of $CL(X)$ is the same where $d(X)$ is replaced by $b(X)$. To illustrate the calculation of $CL(X)$ under the two different banding structures, refer to Figure \ref{BLPFig}.
As can be seen, this figure illustrates the application of the 3-Banded insurance structure to the same loss of value $\$8$M. Under the Basic Banded Policy, $L=5$ and hence the loss is categorized into the $2^{\text{nd}}$ band. This means the insurer will provide complete compensation of the first band $\$5$M, plus a proportion of the remaining loss $\$3\text{M}=\$8\text{M} - \$5\text{M}$ as determined by $\delta_2 \sim Be(1,1)$.

However, under the Log Banded Policy (note: for simplicity the bandwidths have been selected as integer values), the loss is categorized into the $3^{\text{rd}}$ band. As such, the insurer will provide complete compensation for the first 2 bands $\$6\text{M}=\$2\text{M} - \$4\text{M}$, plus a proportion of the remaining loss $\$2\text{M}=\$8\text{M} - \$6\text{M}$ as determined by $\delta_3 \sim Be(1,3)$.

\section{Loss Distributional Approach Model Specifications}
\label{LDA_Models}
In this section we develop the statistical model that will be utilised in analysis of the loss and claims processes. OpRisk LDA models are discussed widely in the literature; see e.g. \cite{Cruz}, \cite{Chavez-Demoulin} \cite{Frachot}, \cite{Shevchenko09}. Under the LDA Basel II requirements, banks should quantify distributions for frequency and severity of OpRisk for each business line and event type over a one-year time horizon. 

To reflect both the nature of OpRisk data with extreme but rare events, the severity models selected for this analysis are chosen to exhibit extreme heavy-tails, with particular interest in distributions with infinite mean/variance. To accomplish this we consider the family of $\alpha$-Stable severity distributions. These particular models have been proposed as suitable models for insurance claims modelling and finance in previous papers, such as \cite{embrechts1994modelling}, \cite{mcneil2000estimation}, \cite{fama1968some}, \cite{peters2010bayesian} and \cite{peters2009likelihood}.

In particular we develop novel results relating to closed-form analytic expressions for the annual loss distribution of the LDA model given by the Poisson-Stable compound process. We use the properties of the $\alpha$-stable severity model to derive a closed-form expression for the distribution for the ES. Consequently, we are able to derive an analytic expression for the ES and MCR for the insurer under Solvency II. 

\subsection{$\alpha$-Stable Severity}
Considered as generalizations of the Gaussian distribution, $\alpha$-Stable models are defined as the class of location-scale distributions which are closed under convolutions. We restrict to the class of truncated $\alpha$-stable models to ensure we only work with a non-negative loss processes. In an OpRisk context, $\alpha$-stable distributions possess several useful properties, including infinite mean and infinite variance, skewness and heavy tails \cite{zolotarev1986one} and \cite{samorodnitsky1994stable}. 

We assume the $i$-th loss of the $j$-th risk process in year $t$ is a random variable with $\alpha$-stable distribution, denoted by $X^{(j)}_i(t) \sim S_{\alpha}\left(x; \beta, \gamma, \delta, 0\right)$. Where, $S_{\alpha}\left(x; \beta, \gamma, \delta, 0\right)$ denotes the univariate four parameter stable distribution family under parameterization $S(0)$, see \cite{peters2009likelihood} for details.  

The univariate $\alpha$-stable distribution we consider is specified by four parameters: $\alpha \in (0, 2]$ determining the rate of tail decay; $\beta \in [-1, 1]$ determining the degree and sign of asymmetry (skewness); $\gamma > 0$ the scale (under some parameterizations); and $\delta \in \mathbb{R}$ the location. The parameter $\alpha$ is termed the characteristic exponent, with small and large $\alpha$ implying heavy and light tails respectively. Gaussian $(\alpha = 2, \beta = 0)$, Cauchy $(\alpha = 1, \beta = 0)$ and Levy $(\alpha = 0.5, \beta = 1)$ distributions provide the only analytically tractable sub-members of this family. Except these special cases, in general $\alpha$-stable models admit no closed-form expression for the density which can be evaluated point-wise, inference typically proceeds via the characteristic function, see discussions in \cite{peters2009likelihood}. However, intractable to evaluate point-wise, importantly for OpRisk applications, simulation of random variates is very efficient, see \cite{chambers1976method}. From the perspective of application of such models in OpRisk LDA settings, there are efficient estimation routines available for parameter estimation given OpRisk data; see a review in \cite{peters2009likelihood}. 

From \cite{Nolan}, a random variable $X$ is said to have a stable distribution, $S(\alpha,\beta,\gamma,\delta;0)$, if its characteristic function has the following form:
\[ E[\text{exp}(i\theta X)] = 	\left\{              
			\begin{array}{ll}
				\text{exp}\{ -\gamma^\alpha|\theta|^\alpha(1+i\beta(\text{sign}(\theta))\tan({\pi \alpha \over 2})(|\gamma \theta|^{1-\alpha}-1))+i\delta \theta\} & \text{if   } \alpha \neq 1\\ 
				\text{exp}\{ -\gamma|\theta|(1+i\beta({2 \over \pi})(\text{sign}(\theta))\text{ln}(\gamma|\theta|))+i\delta \theta\} & \text{if   } \alpha = 1.
				
			\end{array}
			\right.
	\]

The following Lemmas will provide review of relevant properties that $\alpha$-stable random variables posses. These will be used to construct an analytic exact Poisson mixture representation of the annual loss process for a bank under $\alpha$-stable severity models with the required positive support. This will be achieved by considering a special sub-family of $\alpha$-stable models. Then this analytic expression for the annual loss distribution will be extended to an exact expression for the ILP insurance model in Section \ref{ILPSect}. In addition we will provide an analytic expression for the tail distribution of these models and the properties of the ES in special cases of this LDA model.
		
\textbf{Lemma 1} \textit{If $Y \sim S(\alpha,\beta,\gamma,\delta;0)$, then for any $a \neq 0, b \in \Re$, the transformation $Z=aY+b$ is a scaled version of the $\alpha$-stable distribution. That is $Z \sim S(\alpha,(\text{sign}(a)\beta,|a|\gamma,a\delta+b;0)$. In addition, the characteristic functions, densities and distribution functions are jointly continous in all four parameters $(\alpha,\beta,\gamma,\delta)$ and in $x$. These results follow from \cite{samorodnitsky1994stable} and \cite{Nolan} Proposition 1.16.}

\textbf{Lemma 2} \textit{If for all $i \in \left\{1,\ldots,N\right\}$ one has random variables $X_i \sim S(\alpha,\beta_i,\gamma_i,\delta_i;0)$ then the distribution of the linear combination, given N, is 
\begin{equation}
\begin{split}
Z &= \sum_{i=1}^N X_i \sim S(\alpha,\tilde{\beta},\tilde{\gamma},\tilde{\delta};0)\\
\tilde{\gamma}^{\alpha} &= \sum_{i=1}^N |\gamma_i|^{\alpha}, \; \; \; \; \tilde{\beta} =  \frac{\sum_{i=1}^N \beta_i|\gamma_i|^{\alpha}}{\sum_{i=1}^N |\gamma_i|^{\alpha}} \; \; \; \;
\tilde{\delta} = 	\left\{              
			\begin{array}{ll}
			\sum_{i=1}^N \delta_i + \tan \frac{\pi \alpha}{2}\left(\tilde{\beta}\tilde{\gamma} - \sum_{i=1}^N \beta_j\gamma_j\right) & \text{if   } \alpha \neq 1\\  
			\sum_{i=1}^N \delta_i + \frac{2}{\pi}\left(\tilde{\beta}\tilde{\gamma}\log\tilde{\gamma} - \sum_{i=1}^N \beta_j\gamma_j\log|\gamma_i|\right) & \text{if   } \alpha = 1
			\end{array}
			\right. 
\end{split}
\end{equation}
This result follows from (\cite{samorodnitsky1994stable} Section 1.2, Property 1.2.1) and (\cite{Nolan}, Proposition 1.17).}

\begin{figure}[!ht]
\includegraphics[width = 0.4\textwidth, height = 5cm]{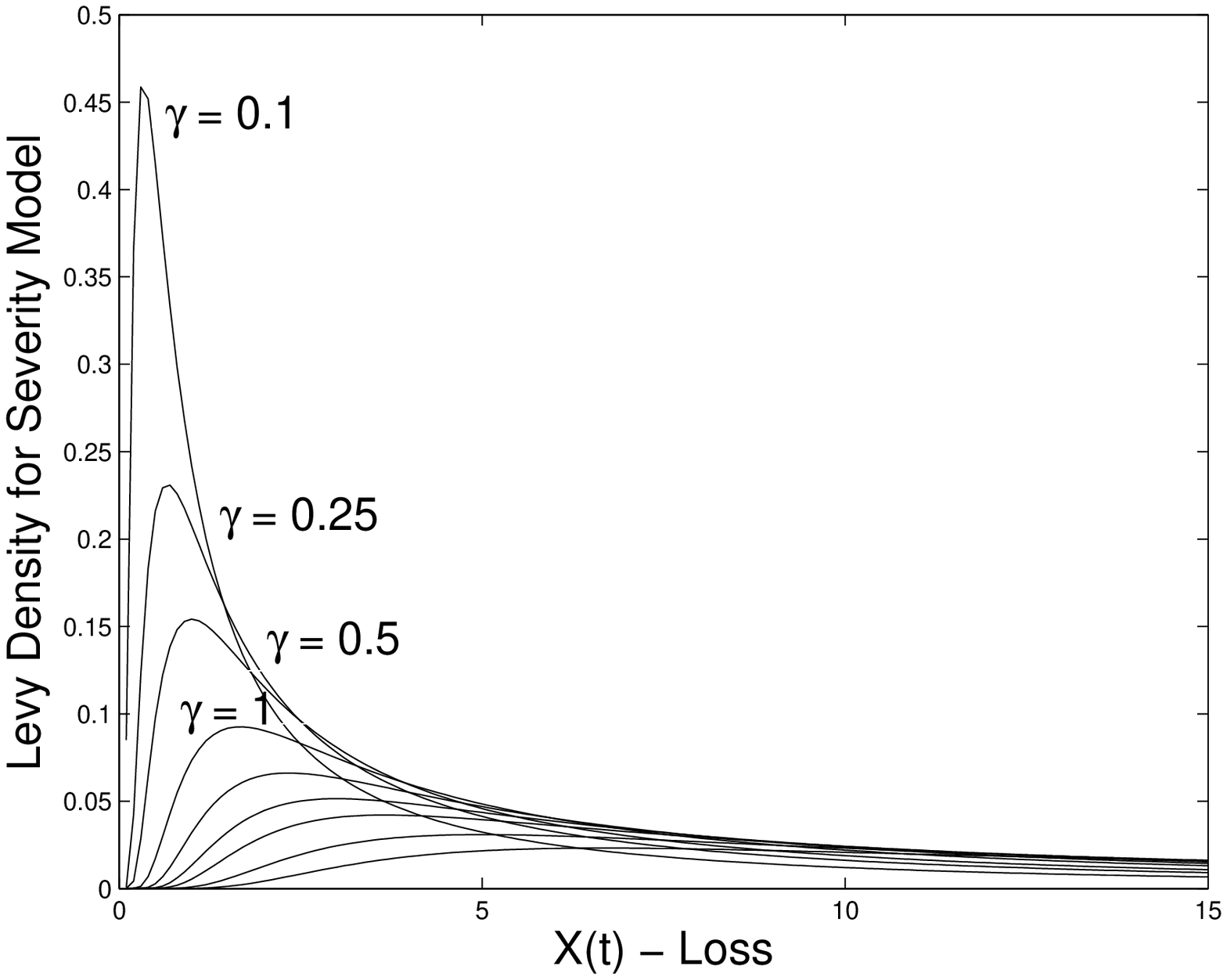}
\includegraphics[width = 0.7\textwidth, height = 5cm]{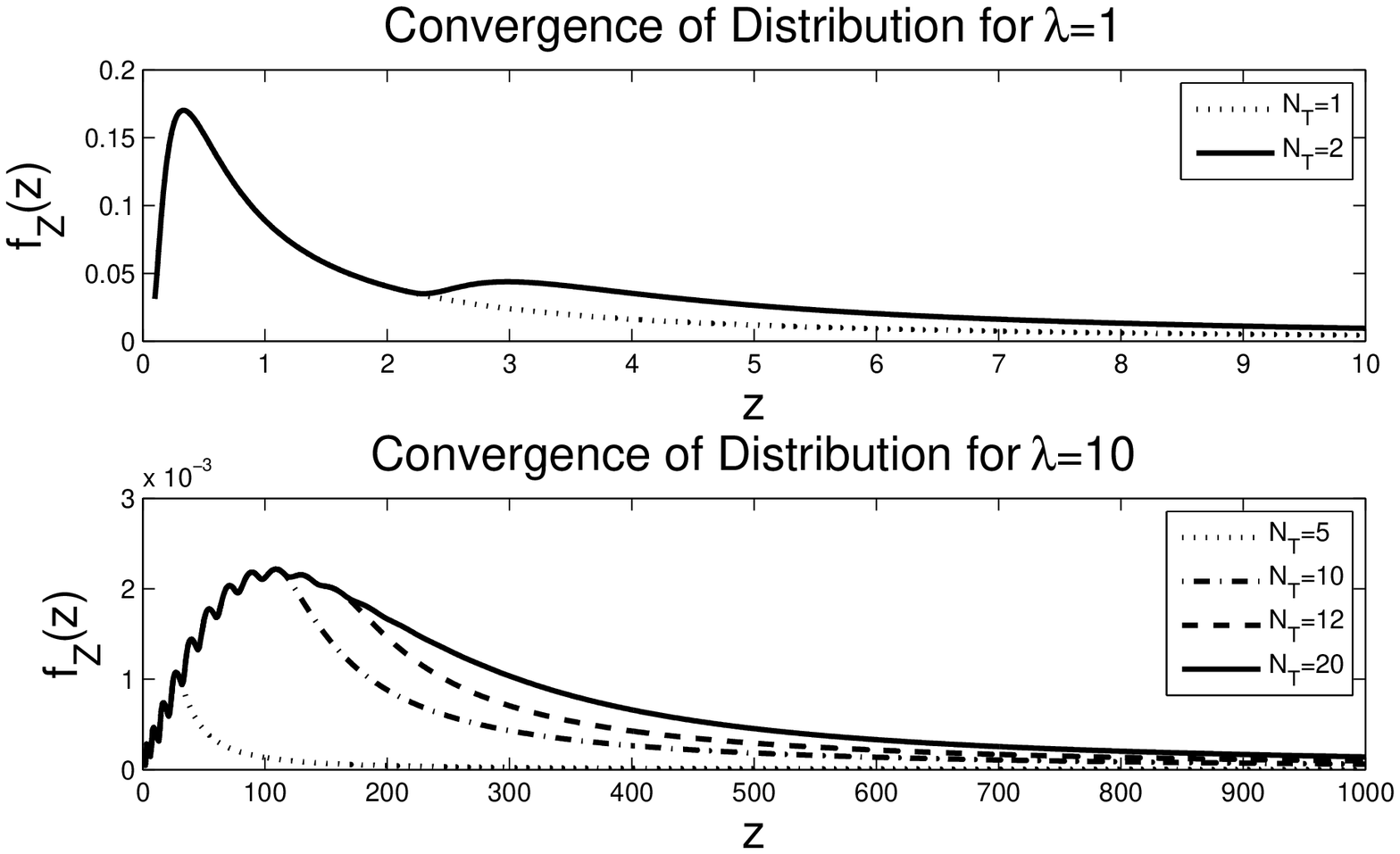}
\caption{\textbf{Left} - Levy Severity Model as a function of scale parameter, with location $\delta = 0$. \textbf{Right} - Truncated sum annual loss distribution approximations.}
\label{Fig_Levy}
\end{figure}

The property in Lemma 2, of closure under convolution for random variables with identical $\alpha$ parameter is lost under truncation to positive support. The implications of this are that in OpRisk the convolution property will in general be lost for general members of the truncated severity models ($X > 0$). This is with one notable exception given by the sub-family of Levy distributions. Examples of loss distributions for this sub-family are illustrated in Figure \ref{Fig_Levy} as a function of the scale parameters $\gamma$. 
 
\textbf{Lemma 3} \textit{If $X \sim S(0.5,1,\gamma,\delta;0)$ this model specifies the sub-family of $\alpha$-stable models with postive real support $x \in [\delta,\infty)$. The density and distribution functions are analytic and given resepectively, for $\delta < x < \infty$, by
\begin{equation*}
f_{X}(x) = \sqrt{\frac{\gamma}{2 \pi}}\frac{1}{\left(x-\delta\right)^{3/2}}\exp\left(-\frac{\gamma}{2\left(x-\delta\right)}\right), \; \; F_{X}(x) = \text{erfc}\left(\sqrt{\frac{\gamma}{2\left(x-\delta\right)}}\right).
\end{equation*}
The median is given by $\tilde{\mu} = \frac{\tilde{\gamma}_n}{2}\left(\text{erfc}^{-1}\left(0.5\right)\right)^2$ and the mode $M = \frac{\tilde{\gamma}_n}{3}$,
where $\text{erfc}(x) = 1-\text{erf}(x) = 1-\frac{2}{\sqrt{\pi}}\int_{0}^xe^{-t^2}dt$. This result follows from \cite{Nolan} (Chapter 1. p.5).}

The combination of Lemma 1,2 and 3 allows us to state Theorem 1.

\textbf{Theorem 1} \textit{The distribution of the annual loss process $Z^{(j)}$ represented by a compound process model with LDA structure in which the frequency is $N^{(j)}(t) \sim Po(\lambda^{(j)})$ and the severity model \\$X_i^{(j)}(t) \sim S(0.5,1,\gamma^{(j)},\delta^{(j)};0)$, then the exact density of the annual loss process can be expressed analytically as a mixture density comprised of $\alpha$-stable components with Poisson mixing weights for $N_t^{(j)} > 0$,
\begin{equation}
f_Z(z) = \sum_{n=1}^{\infty} \exp(-\lambda)\frac{\lambda^n}{n!} \left[ \sqrt{\frac{\tilde{\gamma}_n}{2\pi}}\frac{1}{\left(z-\tilde{\delta}_n\right)^{3/2}}\exp\left(-\frac{\tilde{\gamma}_n}{2\left(z-\tilde{\delta}_n\right)}\right) \right] \times \mathbb{I}\left[\tilde{\delta}_n<z<\infty\right]
\label{PoissMix}
\end{equation}
with
\begin{equation*}
\begin{split}
\tilde{\gamma_n}^{0.5} = \sum_{i=1}^n |\gamma_i|^{0.5} = n|\gamma|^{0.5}, \; \; \; \; \tilde{\beta_n} = 1 \; \; \; \;
\tilde{\delta_n} = \sum_{i=1}^n \delta_i + \tan \frac{\pi}{4}\left(\tilde{\gamma}_n - \sum_{j=1}^n \gamma_j\right) = n\delta + \tan \frac{\pi}{4}\left(n^2|\gamma| - n\gamma\right),  
\end{split}
\end{equation*}
and $f_Z(0) = \text{Pr}(N_t^{(j)}=0)=\exp(-\lambda)$ for $N=0$. The exact form of the annual loss cummulative distribution function is also expressable in closed-form,
\begin{equation}
\Pr\left(Z^{(j)} < z\right) = F_Z(z) = \sum_{n=1}^{\infty} \exp(-\lambda)\frac{\lambda^n}{n!} \text{erfc}\left( \sqrt{\frac{\tilde{\gamma}_n}{2\left(z-\tilde{\delta}_n\right)}}\right)\times \mathbb{I}\left[\tilde{\delta}_n<z<\infty\right] + \exp(-\lambda)\times \mathbb{I}\left[z=0\right]. 
\label{PoissMix}
\end{equation}
This result follows directly from application of Lemma 1, Lemma 2 and Lemma 3.}

Theorem 1 can then be applied to the case of the ILP insurance model to obtain Corollary 1 and can be extended to other forms of compound process model as mentioned in Corollary 2.

\textbf{Corollary 1} \textit{Furthermore, under the ILP insurance model, with $\delta \geq TCL$, one can also obtain a closed-form solution for the annual loss distribution as a mixture distribution comprised of $\alpha$-stable components with Poisson mixing weights as given in Equation (\ref{PoissMix}).}

\textbf{Corollary 2} \textit{The results in Theorem 1 are trivially extended to Binomial and Negative Binomial compound process LDA models, with appropriate adjustments to the mixing weights.}

\textbf{Theorem 2} \textit{The distribution of the annual loss process represented by multiple risks, eg. $j \in \left\{1,2\right\}$, in a LDA compound process structure in which the frequency is $N^{(j)}(t) \sim Po(\lambda^{(j)})$ and the severity model $X_i^{(j)}(t) \sim S(0.5,1,\gamma^{(j)},\delta^{(j)};0)$, can be expressed analytically as a mixture distribution comprised of $\alpha$-stable components with Poisson mixing weights. Define the total annual loss random variable as,
\begin{equation}
Z_{t} = Z^{(1)}_{t} + Z^{(2)}_{t} = \sum_{i=1}^{N^{(1)}(t)}X_{i}^{(1)}(t) + \sum_{j=1}^{N^{(2)}(t)}X_{j}^{(2)}(t).
\label{combined}
\end{equation}
Furthermore, if each of these loss processes has an insurance mitigation applied under an ILP policy with top cover limits, $\delta^{(1)} \geq TCL^{(1)}$ and $\delta^{(2)} \geq TCL^{(2)}$, then the closed-form analytic expression for the annual loss density for $N_t^{(1)} + N_t^{(2)} > 0$ is given by
\begin{equation}
\begin{split}
f_{Z_T}(z) &= \sum_{m=0}^{\infty}\sum_{n=0}^{\infty} \exp(-\lambda^{(1)}-\lambda^{(2)})\frac{\left(\lambda^{(1)}\right)^n\left(\lambda^{(2)}\right)^m}{n!m!} \left[ \sqrt{\frac{\tilde{\gamma}_{nm}}{2\pi}}\frac{1}{\left(z-\tilde{\delta}_{nm}\right)^{3/2}}\exp\left(-\frac{\tilde{\gamma}_{nm}}{2\left(z-\tilde{\delta}_{nm}\right)}\right) \right] \times \mathbb{I}\left[z>\tilde{\delta}_{nm}\right]\\
\tilde{\gamma}_{nm}^{0.5} &= \sum_{i=1}^n |\gamma^{(1)}_i|^{0.5} + \sum_{j=1}^m |\gamma^{(2)}_j|^{0.5} = n|\gamma^{(1)}|^{0.5} + m|\gamma^{(2)}|^{0.5}, \; \; \; \; \tilde{\beta}_{nm} = 1 \\
\tilde{\delta}_{nm} &= \sum_{i=1}^n \delta^{(1)}_i + \sum_{j=1}^m \delta^{(2)}_j + \tan \frac{\pi}{4}\left(\tilde{\gamma}_{nm} - \sum_{j=1}^n \gamma_j^{(1)} - \sum_{k=1}^m \gamma_k^{(2)}\right) \\
&= n\delta^{(1)} + m\delta^{(2)} + \tan \frac{\pi}{4}\left(\left( n|\gamma^{(1)}|^{0.5} + m|\gamma^{(2)}|^{0.5}\right)^2 - n \gamma^{(1)} - m \gamma^{(2)}\right), 
\end{split}
\label{PoissMix2}
\end{equation}
and $f_{Z_T}(0) = \text{Pr}(N_t^{(1)} + N_t^{(2)}=0)=\exp(-\lambda_1-\lambda_2)$. The exact form of the insurance mitigated annual loss cummulative distribution function is also expressable in closed-form,
\begin{equation}
\begin{split}
\Pr\left(Z_T < z\right) &= F_{Z_T}(z) = \sum_{m=0}^{\infty}\sum_{n=0}^{\infty} \exp(-\lambda^{(1)}-\lambda^{(2)})\frac{\left(\lambda^{(1)}\right)^n\left(\lambda^{(2)}\right)^m}{n!m!} \text{erfc}\left( \sqrt{\frac{\tilde{\gamma}_{nm}}{2\left(z-\tilde{\delta}_{nm}\right)}}\right)\times \mathbb{I}\left[z>0\right] \\
&+ \exp(-\lambda_1 -\lambda_2)\times \mathbb{I}\left[z=0\right].
\end{split}
\label{PoissMix}
\end{equation}
This result follows directly from Equation (\ref{totalLoss}), Theorem 1 and Corrollary 1.}

\textbf{Corollary 3} \textit{The result in Theorem 2 can be generalized to any number of risk processes believed to have severity models in the same class of $\alpha$-stable model. This allows one to obtain a closed-from analytic expression for the annual loss pdf and cdf of multiple risk processes in a bank, each with their own insurance mitigation under ILP policies.}

\textbf{Corollary 4} \textit{The result in Theorem 2 when each independent compound process random variable in a given year $Z^{(j)}(t)$ is combined into a compound process vector $\left(Z^{(1)}(t) ,\ldots,Z^{(J)}(t) \right)$ can have dependence introduced in the form of a Levy copula producing a closed from analytic expression for the annual loss process of the insurance mitigated, dependent aggregated loss processes. An example of an appropriate copula model in this family with closed-form density is the Clayton-Levy copula proposed in \cite{Bocker} for OpRisk.}

In Section \ref{RiskMeasures} we will consider the estimation of risk measures that will form estimators that will quantify capital requirements for both Basel II OpRisk and the MCR, SCR and premium calculations. These estimators of VaR and ES will be based on tail properties of the severity distribution, given in Lemma 4. This shows, that although one could estimate these quantities for any model, it may not always be a meaningful exercise to report such estimates in practice.

\textbf{Lemma 4} \textit{If $X \sim S(\alpha,\beta,\gamma,\delta;0)$ then the expected loss given by, 
\begin{equation}
\mathbb{E}\left[|X|^p\right] < \infty \; \; \text{ if } 0<p<\alpha,
\end{equation}
is finite. In addition, for $1<\alpha\leq2$ the mean loss is given by,
\begin{equation}
\mathbb{E}\left[X\right] = \delta - \beta\gamma\tan\left(\frac{\pi \alpha}{2}\right).
\end{equation}
This result follows from \cite{Nolan} Chapter 1. Proposition 1.13.}

\textbf{Corollary 5} \textit{A consequence of Lemma 4 is that the expected loss is infinite for models in which the estimated or expert elicited value of the tail index $\alpha$ is found to be less than 1. This also results in the ES being infinite for any such model, where as the VaR will in such cases always be finite.}

Furthermore, we can also state the asymptotic properties of the general stable distribution model, which will be useful in analysis of the risk measures of VaR and ES in Section \ref{RiskMeasures}.

\textbf{Lemma 5} \textit{Given a random variable for the severity of a loss $X \sim S(\alpha,\beta,\gamma,\delta;0)$ then as $x \rightarrow \infty$ one can write the limiting tail distribution
\begin{equation}
\begin{split}
P(X>x) &\sim \gamma^{\alpha}c_{\alpha}(1+\beta)x^{-\alpha}, \; \text{as x $\rightarrow \infty$} \\
f_{X}(x|\alpha,\beta,\gamma,\delta;0) &\sim \alpha\gamma^{\alpha}c_{\alpha}(1+\beta)x^{-(\alpha+1)}, \; \text{as x $\rightarrow \infty$.}
\end{split}
\end{equation}
where $c_{\alpha} = \sin\left(\frac{\pi \alpha}{2}\right)\frac{\Gamma(\alpha)}{\pi}$. This result follows from (\cite{Nolan},Theorem 1.12)}

\textbf{Theorem 3} \textit{Given the LDA structure in which the frequency is $N^{(j)}(t) \sim Po(\lambda^{(j)})$ and the severity model $X_i^{(j)}(t) \sim S(0.5,1,\gamma^{(j)},\delta^{(j)};0)$, the exact expressions for the median and tail asymptotic of the resulting infinite Poisson mixture are given by,
\begin{equation*}
\begin{split}
\tilde{\mu}\left(Z^{(1)}\right) &= \left(\frac{1}{2}\exp(-\lambda)\left(\text{erfc}^{-1}(0.5)\right)^2|\gamma|\right)\sum_{n=1}^{\infty}\frac{\lambda^n}{n!}n^2 = C_{\tilde{\mu}} \sum_{n=1}^{\infty}W_n\\
\text{Pr}\left(Z^{(1)}>z_q\right) &\sim x^{-1.5}\exp(-\lambda)c_{0.5} \sum_{n=1}^{\infty}\frac{\lambda^n}{n!} \tilde{\gamma}_n^{0.5} = C_{\text{Tail}}\sum_{n=1}^{\infty} T_n 
\end{split}
\end{equation*}
where $z_q$ is an upper tail quantile of the annual loss distribution. These results allows us to show that a unique maximum of $\frac{d}{dn}\log(W_n) = 0$ corresponding to the term $n$ with the maximum contribution to the compound processes median is the term that is the solution to $n\left[\log(\lambda) - \log(n)\right] + \frac{3}{2} = 0$. In the case in which $\lambda \geq 1$ this will correspond to a maximum contribution at index $n \approx \lambda + 1$ and if $\lambda < 1$ the maximum term will correspond to $n \approx 1$.
}

\textbf{Proof} Given the LDA structure of Theorem 1 and Lemma 2 in which the frequency is $N^{(j)}(t) \sim Po(\lambda^{(j)})$ and the severity model $X_i^{(j)}(t) \sim S(0.5,1,\gamma^{(j)},\delta^{(j)};0)$, the exact expression for the mode and median of the resulting infinite Poisson mixture can be written as,
\begin{equation*}
\begin{split}
\tilde{\mu}\left(Z^{(1)}\right) &= \sum_{n=1}^{\infty} \exp(-\lambda)\frac{\lambda^n}{n!} \left[\frac{\tilde{\gamma}_n}{2}\left(\text{erfc}^{-1}(0.5)\right)^2\right] =\left(\frac{1}{2}\exp(-\lambda)\left(\text{erfc}^{-1}(0.5)\right)^2|\gamma|\right)\sum_{n=1}^{\infty}\frac{\lambda^n}{n!}n^2 = C \sum_{n=1}^{\infty}W_n\\
\end{split}
\end{equation*}
Considering the terms $W_n$ and apply Stirling's approximation to $n! = \Gamma(n+1) \approx \sqrt{2\pi}n^{n+0.5}e^{-n}$ to give 
\begin{equation*}
\begin{split}
\log W_n &= n\log(\lambda) + 2\log(n) -\frac{1}{2}\log(2\pi) -\left(n + 0.5\right)\log(n) + n\\
\Rightarrow \; \; \; \frac{d}{dn}\left[\log W_n\right] &=  \log(\lambda) + \frac{2}{n} - \left(\log(n) + 1\right) - \frac{1}{2n} + 1\\
\Rightarrow \; \; \; \frac{d^2}{dn^2}\left[\log W_n\right] &=  -\frac{2}{n^2} - \frac{1}{n} + \frac{1}{2n^2}\\
\Rightarrow \; \; \text{$n$ large} \; \; \frac{d}{dn}\left[\log W_n\right] &\approx  \log(\lambda) - \log(n).\\
\end{split}
\end{equation*}
This tells us that if $\lambda \geq 1$, then the maximum occurs at approx $n = \lambda$. If $\lambda < 1$, then max occurs at approx $n=1$. \qed

\textbf{Remark 1} \textit{The sequence $W_n$ is unimodal in $n$ and the aim is to find $W_L < W_0 < W_U$ such that the truncated sum approximation is sufficiently accurate for use in the expression of the annual loss distribution of the compound process for $Z^{(1)}$.}

\textbf{Corrollary 2} \textit{Using the result of Theorem 1, Lemma 3, Lemma 5 and Theorem 3, we can bound the number of terms in the Poisson mixture expression in order to control the rate of change of the mode of the annual loss distribution as truncated for some value when the contributions to the expressions for the median do not change by a precision amount $\epsilon$. This is expressed analytically as a mixture density comprised of $\alpha$-stable components with Poisson mixing weights,
\begin{equation}
f_Z(z) = \sum_{n=N_L}^{N_U} \exp(-\lambda)\frac{\lambda^n}{n!} \left[ \sqrt{\frac{\tilde{\gamma}_n}{2\pi}}\frac{1}{\left(x-\tilde{\delta}_n\right)^{3/2}}\exp\left(-\frac{\tilde{\gamma}_n}{2\left(x-\tilde{\delta}_n\right)}\right) \right]
\label{PoissMix}
\end{equation}
with $N_L$ and $N_U$ are determined from $\max(W_L \leq e^{-37}W_{0},1)$ and $W_U \leq e^{-37}W_{0}$ and shown in Figure \ref{Fig_Levy}.}

The result of Theorem 1, Lemma 4 and Lemma 5 allows us to state Theorem 4, relating to ES calculations in the $\alpha$-stable setting in the next section. For all other cases one can simulate the annual loss process via a Monte Carlo based procedure. To achieve this we will require an efficient simulation algorithm of losses from an $\alpha$-stable severity model, that can be used in generation of the annual loss process in an LDA model, such an algorithm is standard; see \cite{chambers1976method}.

\section{Risk measures - Basel II capital measures and Solvency II SCR and MCR}
\label{RiskMeasures}
By working with the most fundamental insurance policies in Section \ref{BasicIns} in several two risk scenarios, we can provide an extensive simulation study of performance for important basic policies under consideration by financial institutions. To measure the insurance mitigation in a meaningful fashion for the annual loss process we must consider quantification of capital. Under Basel II regulation, the capital of a bank is defined as the 0.999 VaR which is the quantile of the distribution for the next year annual loss $Z_{T + 1}$:
$$\mathrm{VaR}_q[Z_{T + 1}] = \inf
\{z\in\mathbb{R}:\Pr [Z_{T + 1}
> z] \le 1 - q\}$$ at the level $q = 0.999$. Here, index $T+1$ refers to the next year. Then, the expected shortfall  is
$$\mathrm{ES}_\alpha[Z_{T + 1}]=\mathrm{E}[Z_{T + 1}|Z_{T + 1} \geq \mathrm{VaR}_\alpha[Z_{T + 1}]],$$
which is the conditional expected loss given that the loss exceeds
$\mathrm{VaR}_\alpha [Z_{T + 1}]$. Note that formally the above
definition is for the case when there are no discontinuities in
distribution function at level $q$.

Under Solvency II guidelines, we must also consider quantification of the SCR and MCR. The MCR will be defined to be either the VaR or the ES at a 95\% quantile of the annual claims process generated by the OpRisk loss process, for a given insurance policy. As discussed in \cite{sandstrom2006solvency} we will define the SCR as the expected insurer liability plus three standard deviations. In our context, this will correspond to the claims process generated by the insurance policy on the OpRisk loss process.

Having defined these risk measures, we can demonstrate that under the LDA Poisson-Stable model, we can derive closed-form exact analytic expressions for the expected claim (for premium calculation purposes) and the ES for potential MCR measure and the SCR based on the first and second moments of the ILP claims process under Solvency II. These are stated in Theorem 4.

\textbf{Theorem 4} \textit{In an LDA model for risk process $j$, with annual loss $Z^{(j)}_t = \sum_{i=1}^{N^{(j)}(t)}X_{i}^{(j)}(t)$, in which the severity model has losses distributed as $X \sim S(0.5,1,\gamma,\delta;0)$ according to the sub-family of $\alpha$-stable models with postive real support $x \in [\delta,\infty)$. In this case the closed-form analytic solution for the annual loss distribution given in Theorem 1, one can derive an analytic expression for the expected claim, the SCR and the MCR as measured by the ES of the claims process, via the distribution of the loss process, under the ILP policy. \\
The expected claim for the $j$-th risk process is given by
\begin{equation}
\begin{split}
\mathbb{E}\left[C^{(j)}\right] &=\sum_{n=1}^{\infty}\exp(-\lambda)\frac{\lambda^n}{n!}  \sqrt{\frac{\tilde{\gamma}_n}{2\pi}} \left[\frac{e^{-\frac{\tilde{\gamma}_n}{2y}}\left(\sqrt{2\pi}(\tilde{\delta}_n+\tilde{\gamma}_n)e^{\frac{\tilde{\gamma}_n}{2y}}\text{erf}\left(\frac{\sqrt{\tilde{\gamma}_n}}{\sqrt{2y}}\right) + 2 \sqrt{\tilde{\gamma}_ny}\right)}{\sqrt{\tilde{\gamma}_n}}\right]_{0}^{TCL^{(j)}-\tilde{\delta}_n}.
\end{split}
\end{equation}
The SCR under ILP insurance, quantified according to $\mathbb{E}(C^{(j)}) + 3\sqrt{\mathbb{V}ar\left[C^{(j)}\right]}$ is then analytic given the variance of the claims process,
{\small{
\begin{equation*}
\begin{split}
\mathbb{V}ar(C^{(j)}) &= \sum_{n=1}^{\infty}w_n \frac{1}{3\sqrt{\tilde{\gamma}_n}}e^{-\frac{\tilde{\gamma}_n}{2y}}\left( 
2\sqrt{\tilde{\gamma}_n y}\left(-6\tilde{\delta}_n -\tilde{\gamma}_n +y\right)
-\sqrt{2\pi}\left(3\tilde{\delta}^2_n + 6 \tilde{\delta}_n\tilde{\gamma}_n + \tilde{\gamma}^2_n\right)e^{\frac{\tilde{\gamma}_n}{2y}}\text{erf}\left(\frac{\sqrt{\tilde{\gamma}_n}}{\sqrt{2y}}\right) \right) \\
&- w_n \left(\left[\frac{e^{-\frac{\tilde{\gamma}_n}{2y}}\left(\sqrt{2\pi}(\tilde{\delta}_n+\tilde{\gamma}_n)e^{\frac{\tilde{\gamma}_n}{2y}}\text{erf}\left(\frac{\sqrt{\tilde{\gamma}_n}}{\sqrt{2y}}\right) + 2 \sqrt{\tilde{\gamma}_ny}\right)}{\sqrt{\tilde{\gamma}_n}}\right]_{0}^{TCL^{(j)}-\tilde{\delta}_n}\right)^2\\
\end{split}
\end{equation*} }}
with $w_n = \exp(-\lambda)\frac{\lambda^n}{n!}  \sqrt{\frac{\tilde{\gamma}_n}{2\pi}}$.
Finally, we can also state the MCR as measured by the ES at quantile $q$. Consider the estimated quantile value of the annual loss distribution $q$ which is a point estimate of the VaR of $m\%$, typically $m \in \left\{95, 99.5, 99.9, 99.95\right\}$ depending on whether the risk measure is used for example as regulatory capital or economic capital. In this case the ES for the bank is infinite. However, under the ILP insurance policy, we can obtain a closed-form expression for the Solvency II MCR as, 
\begin{equation}
\begin{split}
ES_{q}\left[C^{(j)}\right] &\approx \sum_{n=1}^{\infty} \exp(-\lambda)\frac{\lambda^n}{n!} \left(1-F_Z(VaR_q)\right)^{-1} \sqrt{\tilde{\gamma}_n} 2c_{0.5}\left[\frac{2}{3}\left(TCL^{(j)}\right)^{3/2} - \frac{2}{3}\left(VaR_q\right)^{3/2}\right]\\
\end{split}
\end{equation}
where $F_Z(VaR_q) = \sum_{n=1}^{\infty} \exp(-\lambda)\frac{\lambda^n}{n!} \text{erfc}\left(\sqrt{\frac{\tilde{\gamma}_n}{2(VaR_q-\tilde{\gamma}_n)}}\right)$ and
$c_{0.5} = sin\left(\frac{\pi}{4}\right)\Gamma(0.5)/\pi$.
}

\textbf{Proof:} The proof of these results follows from Lemmas 1,2 and 3 and Theorem 1. Define $p_n = \exp(-\lambda)\frac{\lambda^n}{n!}$. The expected claim for the $j$-th risk process is given by
\begin{equation}
\begin{split}
\mathbb{E}\left[C^{(j)}\right] &=\sum_{n=1}^{\infty}p_n \sqrt{\frac{\tilde{\gamma}_n}{2\pi}} \int_{0}^{TCL^{(j)}} x \frac{1}{\left(x-\tilde{\delta}_n\right)^{3/2}}\exp\left(-\frac{\tilde{\gamma}_n}{2\left(x-\tilde{\delta}_n\right)}\right) dx\\
&=\sum_{n=1}^{\infty}p_n  \sqrt{\frac{\tilde{\gamma}_n}{2\pi}} \int_{0}^{TCL^{(j)}-\tilde{\delta}_n}  \frac{y-\tilde{\delta}_n}{\left(y\right)^{3/2}}\exp\left(-\frac{\tilde{\gamma}_n}{2\left(y\right)}\right) dy\\
&=\sum_{n=1}^{\infty}p_n \sqrt{\frac{\tilde{\gamma}_n}{2\pi}} \left[\frac{e^{-\frac{\tilde{\gamma}_n}{2y}}\left(\sqrt{2\pi}(\tilde{\delta}_n+\tilde{\gamma}_n)e^{\frac{\tilde{\gamma}_n}{2y}}\text{erf}\left(\frac{\sqrt{\tilde{\gamma}_n}}{\sqrt{2y}}\right) + 2 \sqrt{\tilde{\gamma}_ny}\right)}{\sqrt{\tilde{\gamma}_n}}\right]_{0}^{TCL^{(j)}-\tilde{\delta}_n}\\
\end{split}
\end{equation}
The second moment for the resulting ILP claims process is derived as,
{\small{
\begin{equation}
\begin{split}
\mathbb{E}&\left[(C^{(j)})^2\right] = \sum_{n=1}^{\infty}p_n  \sqrt{\frac{\tilde{\gamma}_n}{2\pi}} \int_{\tilde{\delta}_n}^{TCL^{(j)}} x^2 \frac{1}{\left(x-\tilde{\delta}_n\right)^{3/2}}\exp\left(-\frac{\tilde{\gamma}_n}{2\left(x-\tilde{\delta}_n\right)}\right) dx \\
&= \sum_{n=1}^{\infty}p_n  \sqrt{\frac{\tilde{\gamma}_n}{2\pi}} \left[\frac{1}{3\sqrt{\tilde{\gamma}_n}}e^{-\frac{\tilde{\gamma}_n}{2y}}\left( 
2\sqrt{\tilde{\gamma}_n y}\left(-6\tilde{\delta}_n -\tilde{\gamma}_n +y\right) 
-\sqrt{2\pi}\left(3\tilde{\delta}^2_n + 6 \tilde{\delta}_n\tilde{\gamma}_n + \tilde{\gamma}^2_n\right)e^{\frac{\tilde{\gamma}_n}{2y}}\text{erf}\left(\frac{\sqrt{\tilde{\gamma}_n}}{\sqrt{2y}}\right) \right)\right]_{0}^{TCL^{(j)}-\tilde{\delta}_n}
\end{split}
\end{equation}
}}
The MCR as measured by the ES is derived by utilizing the expression for the tail of the $\alpha$-stable model presented in Lemma 5, combined with Theorem 1 to give
\begin{equation}
\begin{split}
ES_{q}\left[C^{(j)}\right] &= \mathbb{E}\left[Z^{(j)}\mid VaR_q \leq Z^{(j)} \leq TCL^{(j)}\right]\\
&= \int_{VaR_q}^{TCL^{(j)}} \frac{x f_{Z^{(j)}}\left(x\right)}{1-F_{Z^{(j)}}(VaR_q)}dx\\
&\approx \sum_{n=1}^{\infty} p_n \left(1-F_Z(VaR_q)\right)^{-1} \int_{VaR_q}^{TCL^{(j)}} \sqrt{\tilde{\gamma}_n} 2c_{0.5}\sqrt{x} dx\\
&= \sum_{n=1}^{\infty} p_n \left(1-F_Z(VaR_q)\right)^{-1} \sqrt{\tilde{\gamma}_n} 2c_{0.5}\left[\frac{2}{3}\left(TCL^{(j)}\right)^{3/2} - \frac{2}{3}\left(VaR_q\right)^{3/2}\right]\\
\end{split}
\end{equation}
where $F_Z(VaR_q) = \sum_{n=1}^{\infty} \exp(-\lambda)\frac{\lambda^n}{n!} \text{erfc}\left(\sqrt{\frac{\tilde{\gamma}_n}{2(VaR_q-\tilde{\gamma}_n)}}\right)$ and
$c_{0.5} = sin\left(\frac{\pi}{4}\right)\Gamma(0.5)/\pi$. \qed

For all other severity model settings and insurance policies we perform numerical analysis which are provided in detail in Section \ref{NumericalAnalysis}.

\section{Analysis of Insurance Mitigation and the interplay between Basel II OpRisk Capital and Solvency II Insurer Capital.}
\label{NumericalAnalysis}
In this section we are interested in studying the interplay between insuring OpRisk losses, the capital reduction and the resulting insurers exposure for such insurance policies when applied to these OpRisk scenarios. We systematically analyse the impact on capital mitigation on Basel II OpRisk capital under each insurance policy proposed in Section \ref{BasicIns}. We note that we measure the capital in terms of the VaR. We recognise that the Basel II framework requires a much stricter 99.95 \% quantile of the annual loss distribution. However, we make the point, that although such a quantile is now routinely reported as a loss measure to regulators, it is seldom considered in the context of the fact that it represents a point estimator and should be therefore reported with appropriate confidence intervals for the numerical uncertainty associated with this capital measure. The key reason for this is typically that the number of simulated annual loss years required to reduce the uncertainty in this capital measure to a reasonable level is far too great. In this paper the estimated uncertainty in such an extreme tail measure as a 99.95 \% VaR is very large, especially in the case of the heavy-tailed models we consider in this paper in which the resulting uncertainty in the estimates is sufficiently large to be restrictive. Therefore we report the 95\% which is much more accurately estimated in our models when estimated from one million simulated annual years of losses.
  
\subsection{Simulation Study Design}
\label{Design}
The experiments consider both univariate and bivariate Poisson-Stable LDA risk models for the annual losses. We consider both high and low frequency with $\lambda=1$ and $\lambda = 10$ scenarios for the Poisson intensity model, in the context of rare, yet extreme severity loss modelling. Then we varied systematically the degree of severity in the simulated loss models, as parameterized by the tail index of the $\alpha$-stable family models as well as the measure of total insurance applied, typically captured by the TCL. The parameters of the $\alpha$-Stable severity model were set as $\beta=0.8$, $\gamma=10000$ and $\delta=0$ and the tail index ranged over $\alpha=\{2.0,1.9,1.8,\ldots,1.0,0.75,0.5,0.25\}$.

Therefore under each of these settings, through the application of each of the insurance policies to simulated OpRisk losses we can re-evaluate the capital requirements for an insured OpRisk exposure and compare these requirements to those of an uninsured exposure. Studying this application, it is possible to evaluate the capital relief generated by each of the above policies as a result of a variety of selected levels of TCL. 
 
For the purposes of this study, the value of ACL was set as the value of $TCL$ for the ILP and CLP multiplied by the $70^{th}$ percentile of the frequency distribution in order to make the investigations undertaken purely a function of the individual $TCL$.

For the basic insurance policy studies the $TCL$ was segmented into 51 strata ranging from $0$, indicating an insurance free loss, to a top limit which provided a complete mitigiation of the specified risk, that is the $VaR$ of the loss when the maximum $TCL$ was applied was $0$. These studies were performed on a grid computer with 42 SGI Altix XE320 nodes each with $42\times2$ Intel Xeon X5472 Quad Core 3.0GHz each with 16GB of RAM. Each simulated risk process was run with $1,000,000$ simulated loss years which provided a total of $8,568$ simulation studies exceeding of 200hrs of compute time.

\subsection{Risk Transfer: OpRisk Capital Mitigation (Basel II) versus Insurer MCR (Solvency II)}
First we present the complete investigations of capital reduction under each insurance policy, as measured by VaR under Basel II and insurer risk transfer as measured by MCR under Solvency II. A detailed analysis of the Banks ability to reduce capital under insurance mitigation as the magnitude of the losses realized increases, via increasing $\alpha$-stable tail index is performed thus allowing us to consider the question of viability of different insurance policies. We then study in detail a selection of three severity model settings for the tail index of the $\alpha$-stable models, corresponding to $\alpha \in \left\{2, 1.3, 0.5\right\}$ with light (Gaussian), heavy-tailed with finite mean and infinite mean loss cases. In this context we also consider combining the basic insurance policies with advanced policy structures of the form involving haircuts and stochastic banding. This provided an addtional $1,404$ simulation studies.

To ensure we can make general statements about performance accross each policy type and modelling scenario, we first consider a comparative measure of VaR (that is the risk exposure of the $i$-th insured loss process $Z^{(j,i)}$ to the uninsured loss process $Z^{(j,0)}$) denoted by $VaR_{0.95}\left[Z^{(j,i)}\right]$. The comparative measure of capital mitigation provided by the $i$-th policy is given by the \% reduction in VaR for the bank in question, denoted by $VaR_{0.95,Mit}\left[Z^{(j,i)}\right]$. Analagously, the MCR on the claims made on the insurer are quantified via the 95-th percentile of the claims process under each policy type, and have been adjusted to provide a comparative measure, denoted by $MCR_{0.95}\left[C^{(j,i)}\right]$ and presented in Equation \ref{MCRTran}.
\begin{equation}
\label{MCRTran}
\begin{split}
\%VaR_{0.95}\left[Z^{(j,i)}\right]=\frac{VaR_{0.95}\left[Z^{(j,i)}\right]}{VaR_{0.95}\left[Z^{(j,0)}\right]}, \; &\%VaR_{0.95,Mit}\left[Z^{(j,i)}\right]=1-\frac{VaR_{0.95}\left[Z^{(j,i)}\right]}{VaR_{0.95}\left[Z^{(j,0)}\right]} \\
\%MCR_{0.95}\left[C^{(j,i)}\right]&= \frac{VaR_{0.95}\left[C^{(j,i)}\right]}{VaR_{0.95}\left[Z^{(j,0)}\right]} 
\end{split}
\end{equation}

\subsubsection{Basic Insurance Policies}
\begin{figure}[!ht]
\includegraphics[width=0.5\textwidth, height = 6cm]{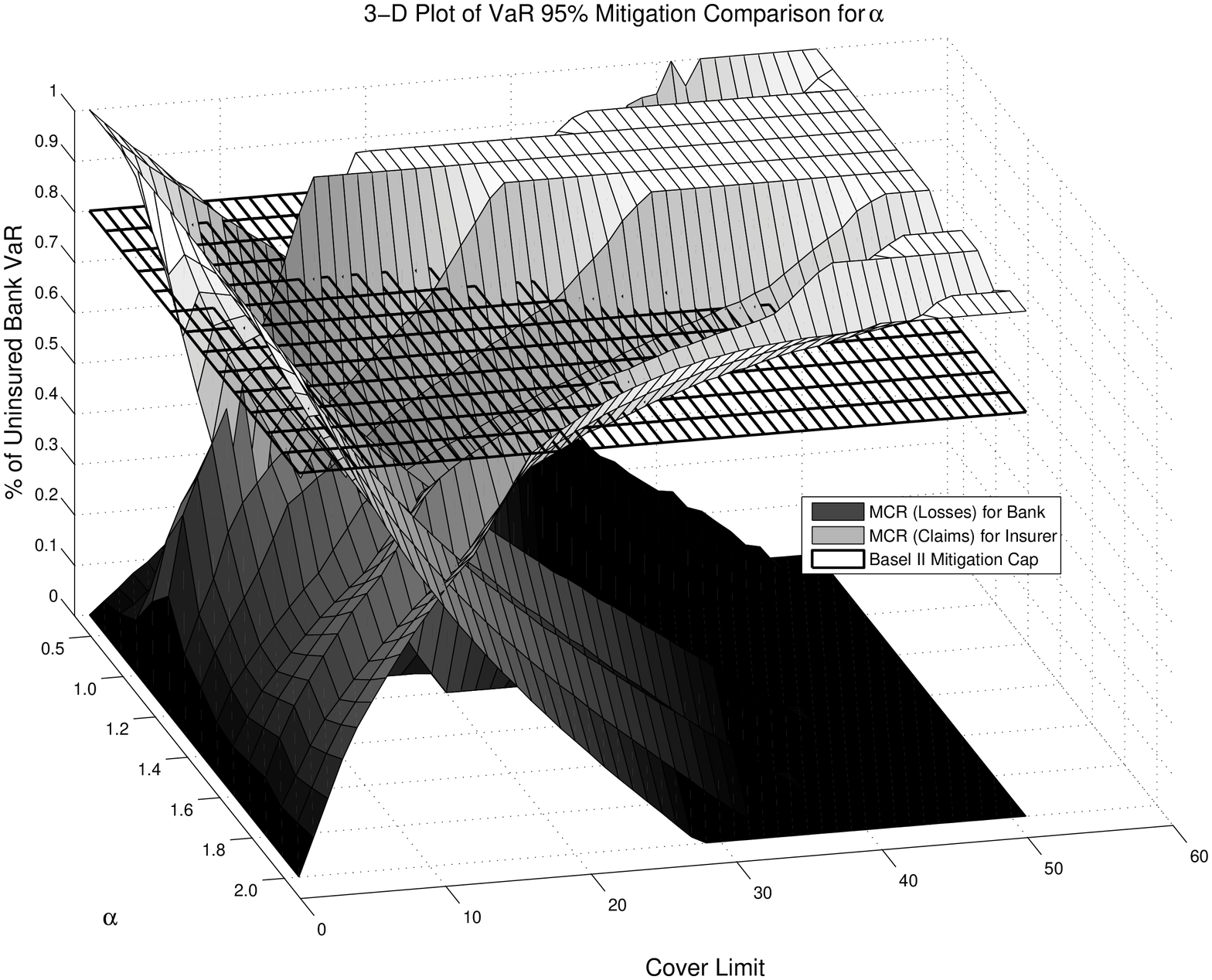}
\includegraphics[width=0.5\textwidth, height = 6cm]{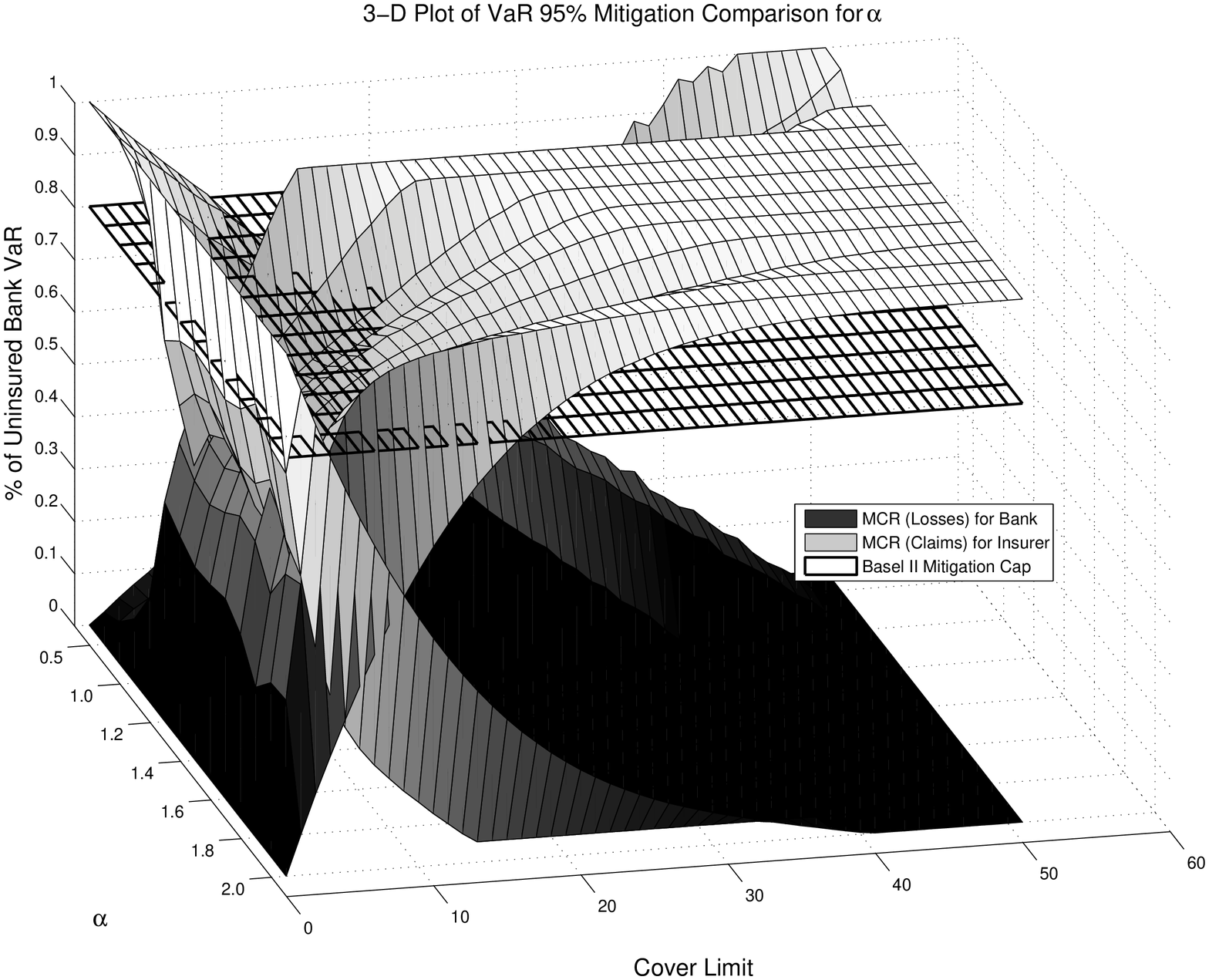}
\caption{ Risk Mitigated Bank VaR vs. MCR ($MCR^{(j)}=VaR_{0.95}\left[C^{(j)}\right]$) for $\alpha$-Stable Model : Individual Loss Policy $\lambda = 1$ (Left) and $\lambda = 10$ (Right).}
\label{Fig3d_LowAlphaStabVaR_ILP}
\end{figure}

\begin{figure}[!ht]
\includegraphics[width=0.5\textwidth, height = 6cm]{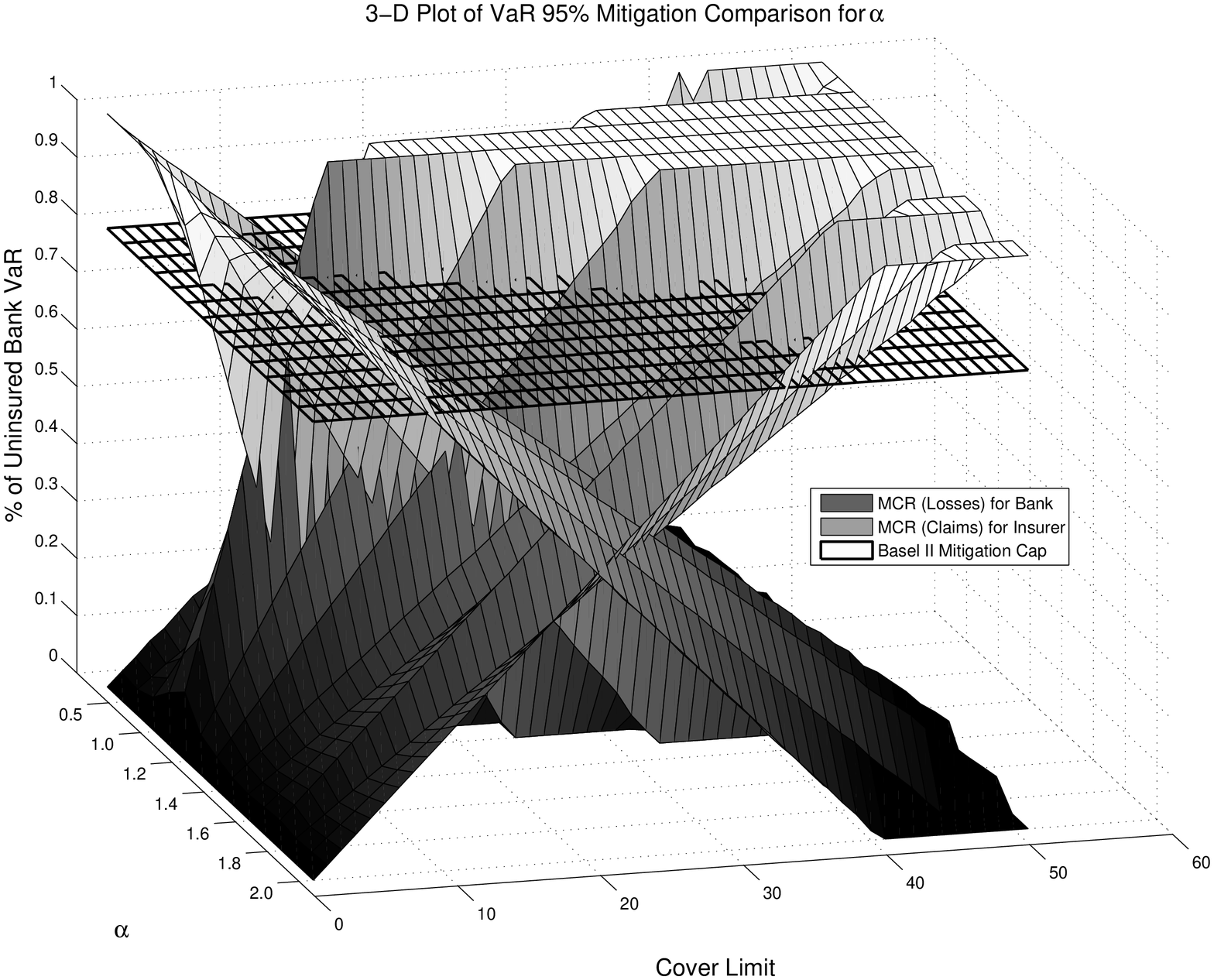}
\includegraphics[width=0.5\textwidth, height = 6cm]{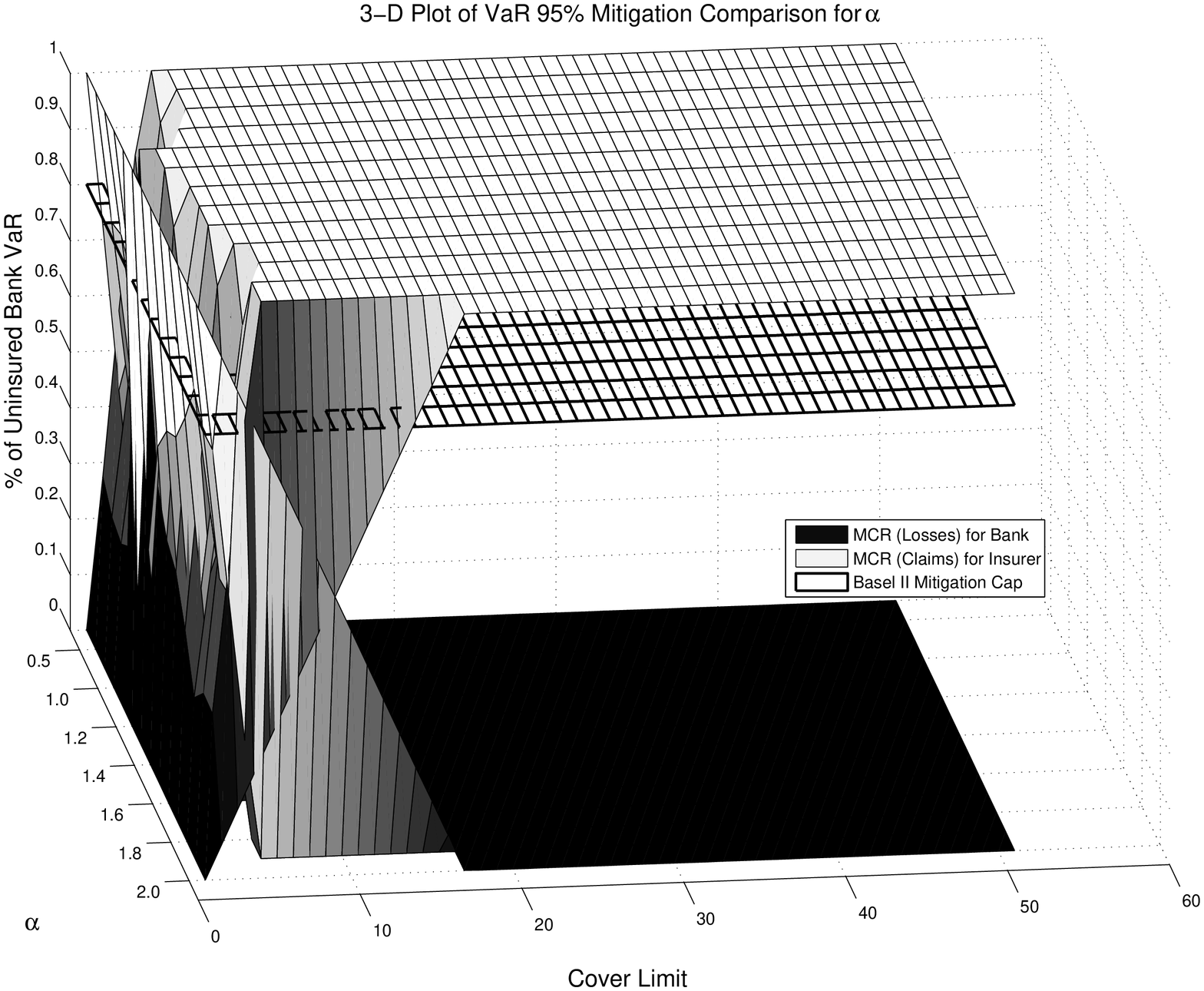}
\caption{ Risk Mitigated Bank VaR vs. MCR ($MCR^{(j)}=VaR_{0.95}\left[C^{(j)}\right]$)  for $\alpha$-Stable Model : Accumulated Loss Policy $\lambda = 1$ (Left) and $\lambda = 10$ (Right).}
\label{Fig3d_LowAlphaStabVaR_ALP}
\end{figure}

\begin{figure}[!ht]
\includegraphics[width=0.5\textwidth, height = 6cm]{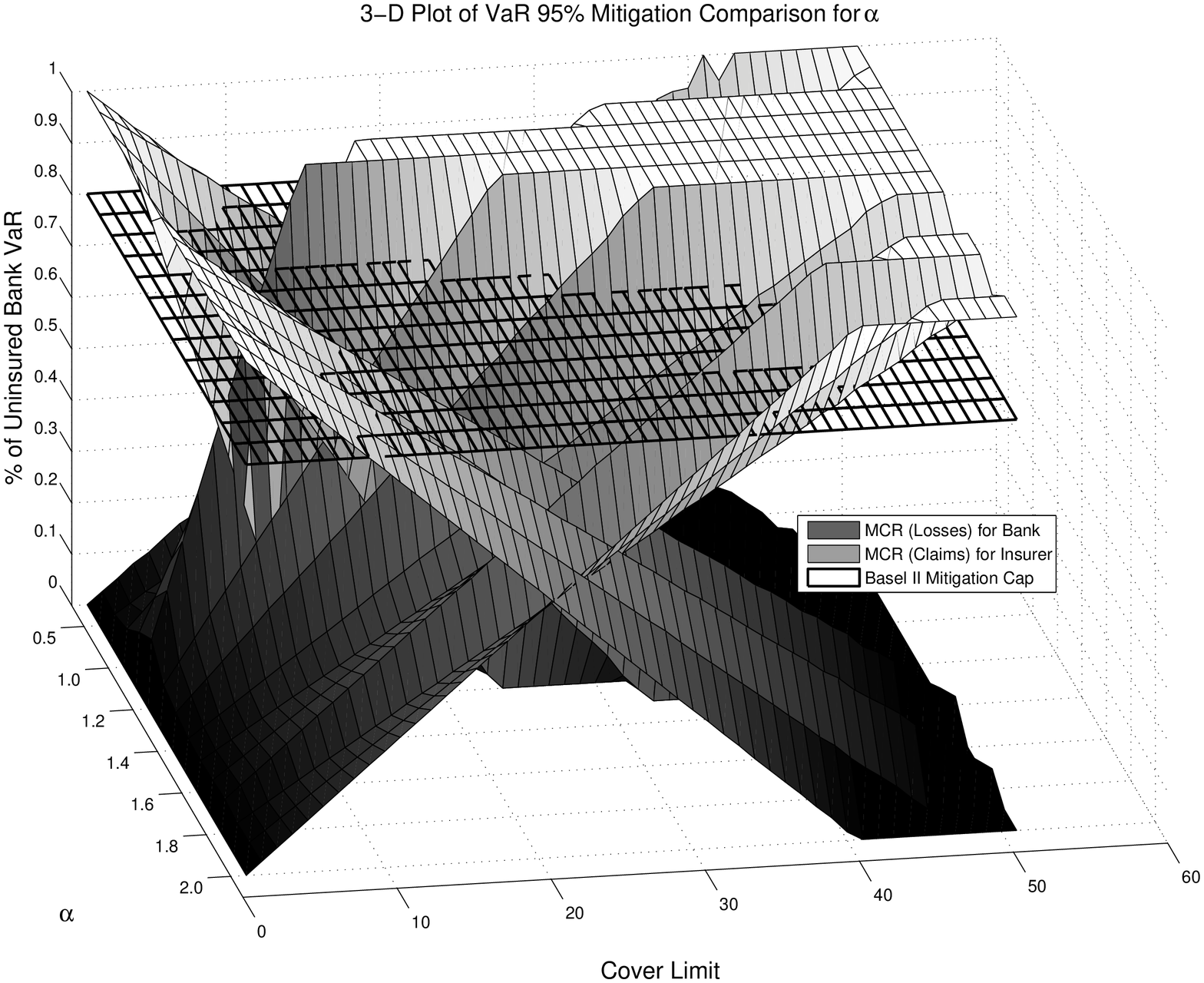}
\includegraphics[width=0.5\textwidth, height = 6cm]{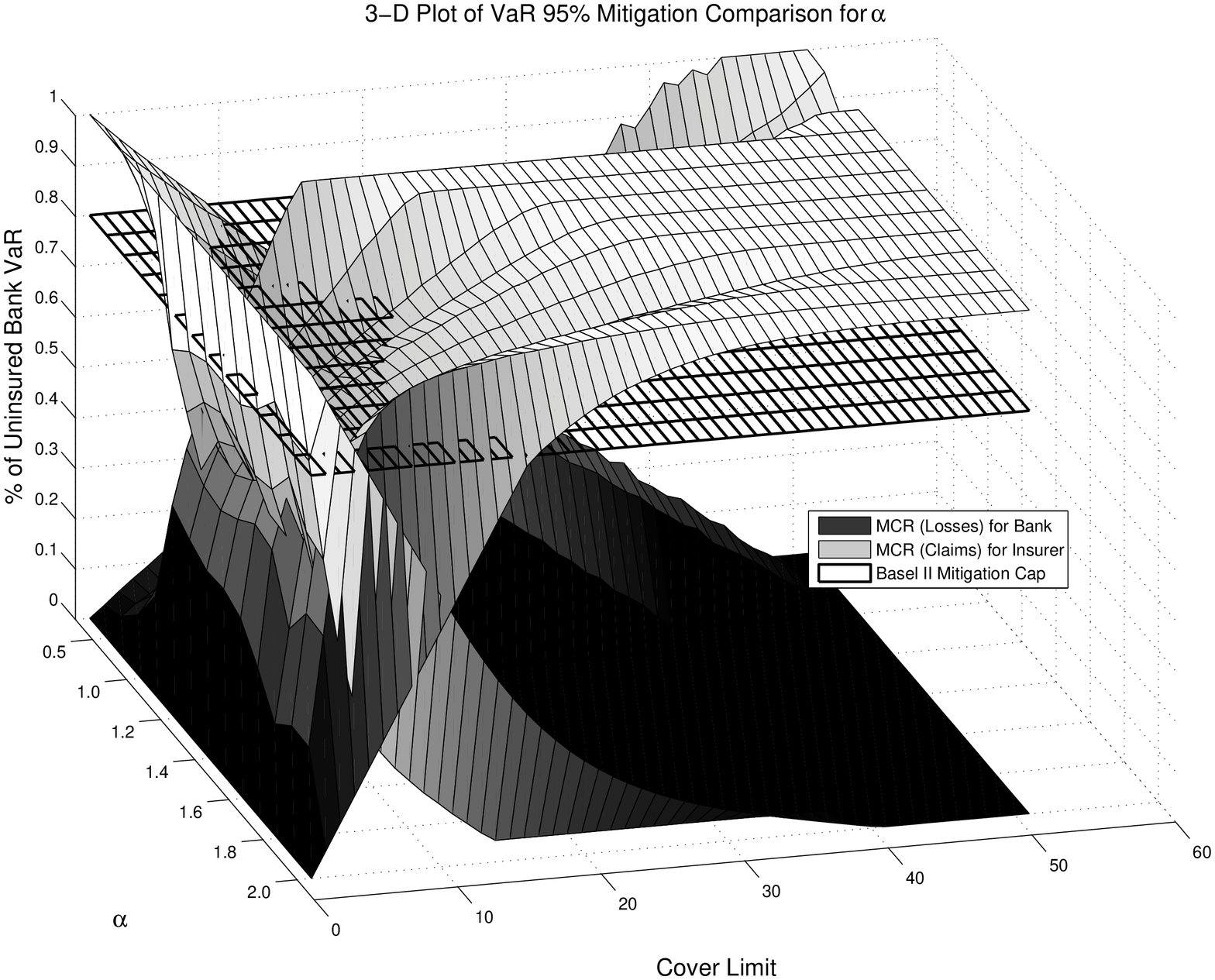}
\caption{ Risk Mitigated Bank VaR vs. MCR ($MCR^{(j)}=VaR_{0.95}\left[C^{(j)}\right]$)  for $\alpha$-Stable Model : Combined Loss Policy $\lambda = 1$ (Left) and $\lambda = 10$ (Right).}
\label{Fig3d_LowAlphaStabVaR_CLP}
\end{figure}

\begin{figure}[!ht]
\includegraphics[width=0.5\textwidth, height = 6cm]{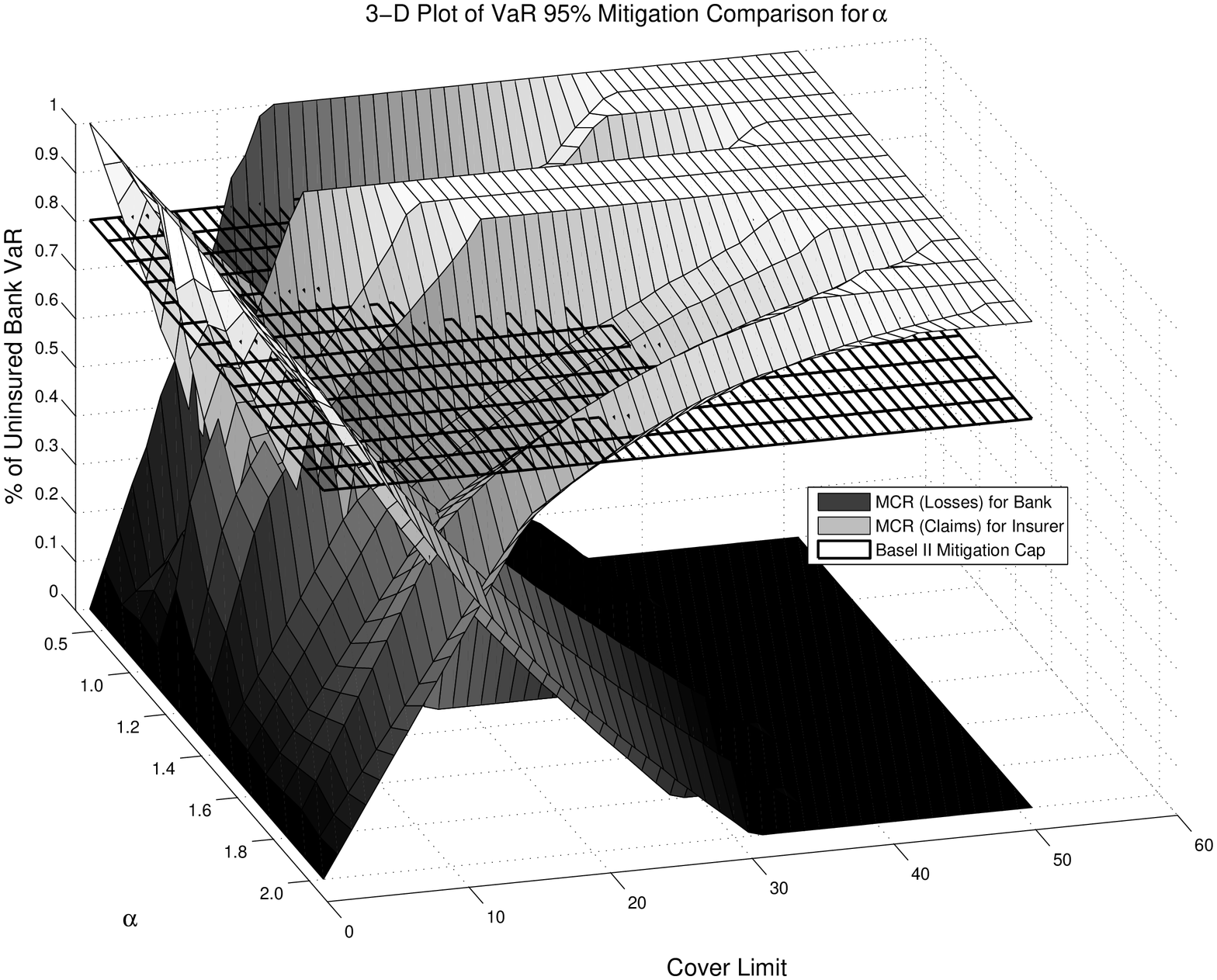}
\includegraphics[width=0.5\textwidth, height = 6cm]{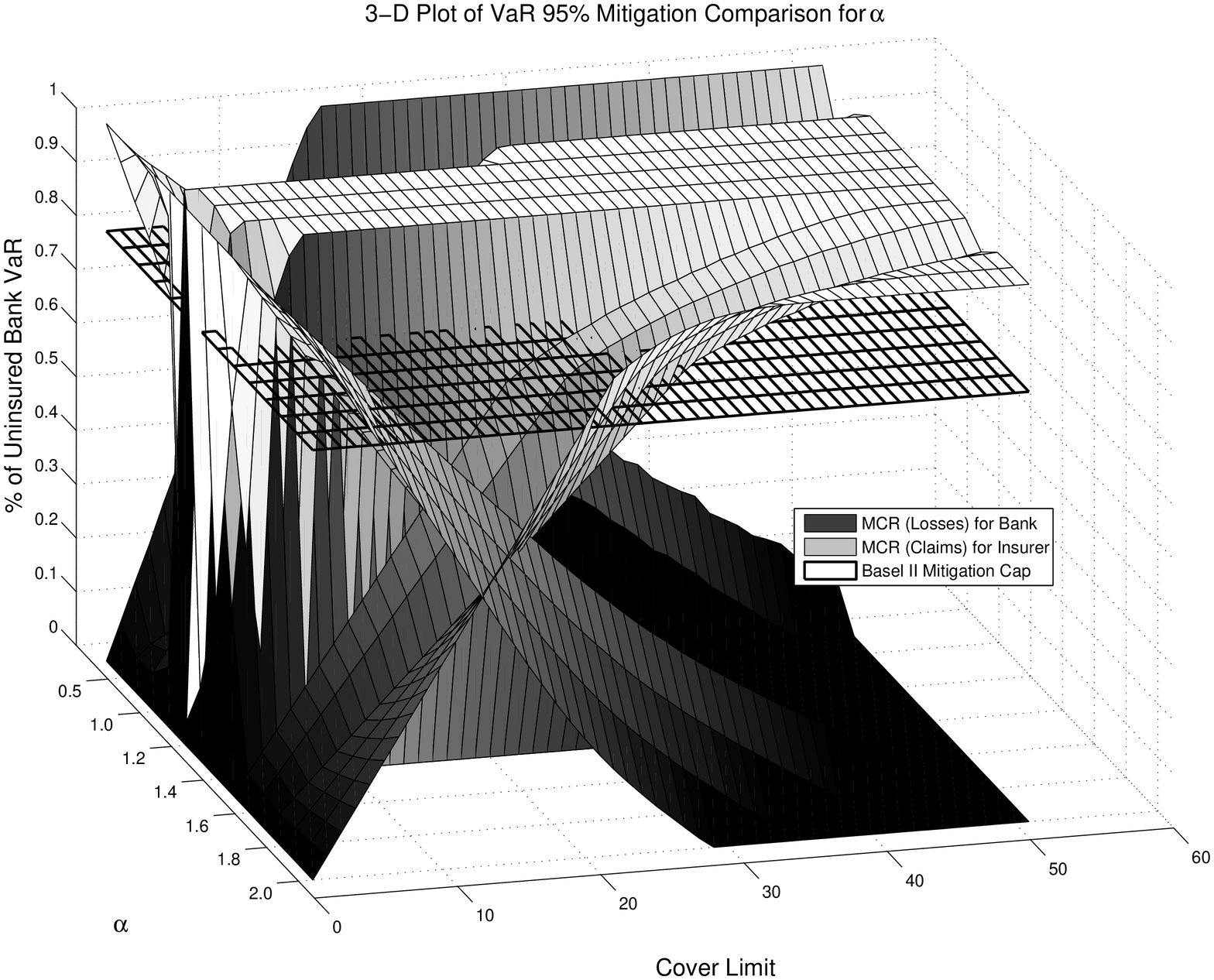}
\caption{ Risk Mitigated Bank VaR vs. MCR ($MCR^{(j)}=VaR_{0.95}\left[C^{(j)}\right]$)  for $\alpha$-Stable Model : Accumulated Loss Policy (Bivariate) $\lambda = 1$ (Left) and $\lambda = 10$ (Right).}
\label{Fig3d_LowAlphaStabVaR_ALP2}
\end{figure}

The bank's capital reduction, as measured by \%VaR reduction under the insurance mitigation for the loss process are compared to the surface corresponding to the insurer's \%MCR measure under Solvency II, as determined by the resulting claims process. The results of this analysis are presented for the ILP, ALP, CLP for single risk and ALP for bi-variate risk models under low frequency and high frequency LDA models in Figures \ref{Fig3d_LowAlphaStabVaR_ILP}, \ref{Fig3d_LowAlphaStabVaR_ALP}, \ref{Fig3d_LowAlphaStabVaR_CLP} and \ref{Fig3d_LowAlphaStabVaR_ALP2}. These results summarise the simulation studies for each basic insurance policy. We have identified the risk mitigation cap as specified by Basel regulation, which restricts the reduction in the measured risk exposure of the bank to be at a minimum $80\%$ of the uninsured risk (this is identified by the mesh grid).

\begin{figure}[!ht]
\includegraphics[width=0.5\textwidth, height = 4cm]{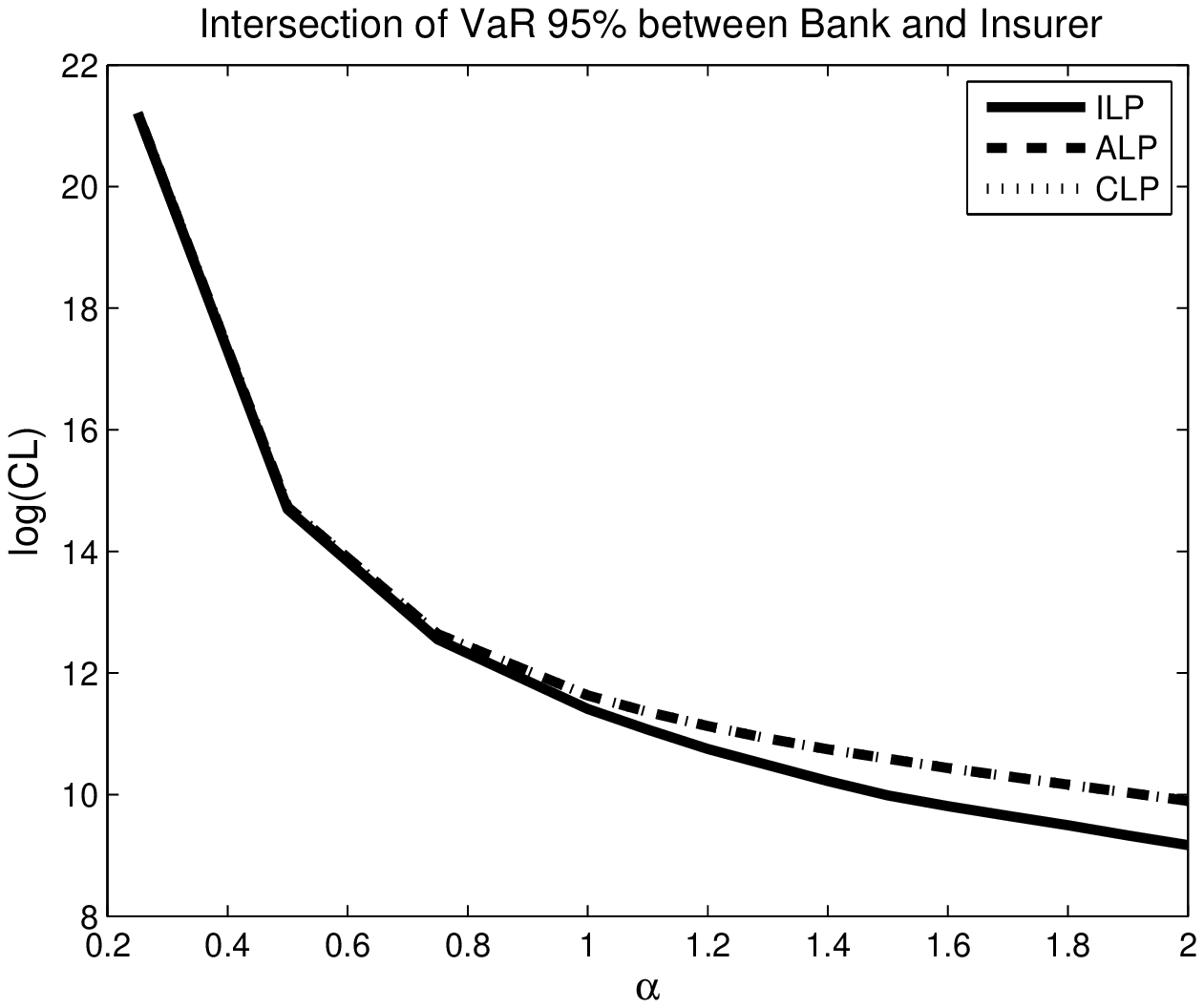}
\includegraphics[width=0.5\textwidth, height = 4cm]{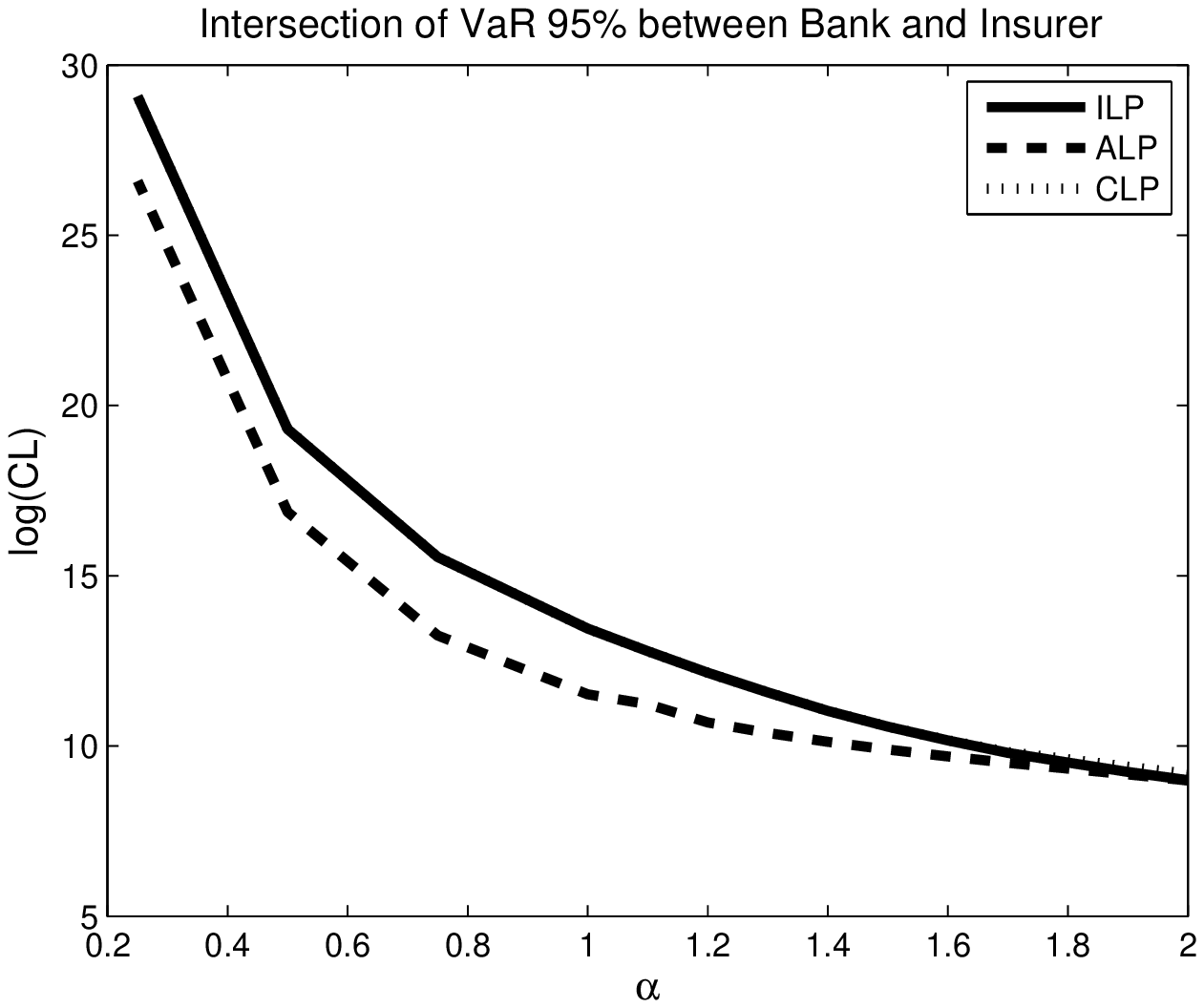}
\includegraphics[width=0.5\textwidth, height = 4cm]{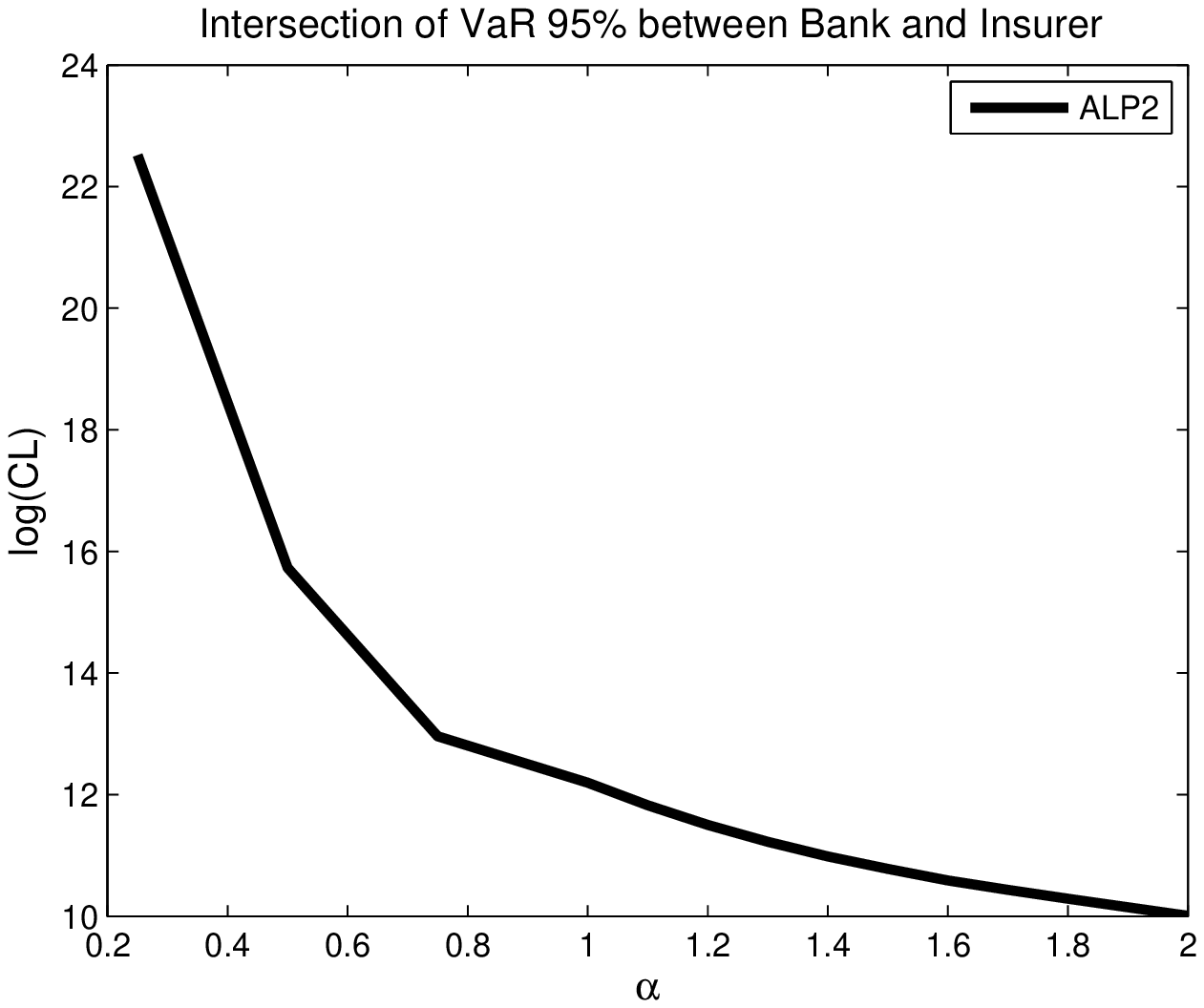}
\includegraphics[width=0.5\textwidth, height = 4cm]{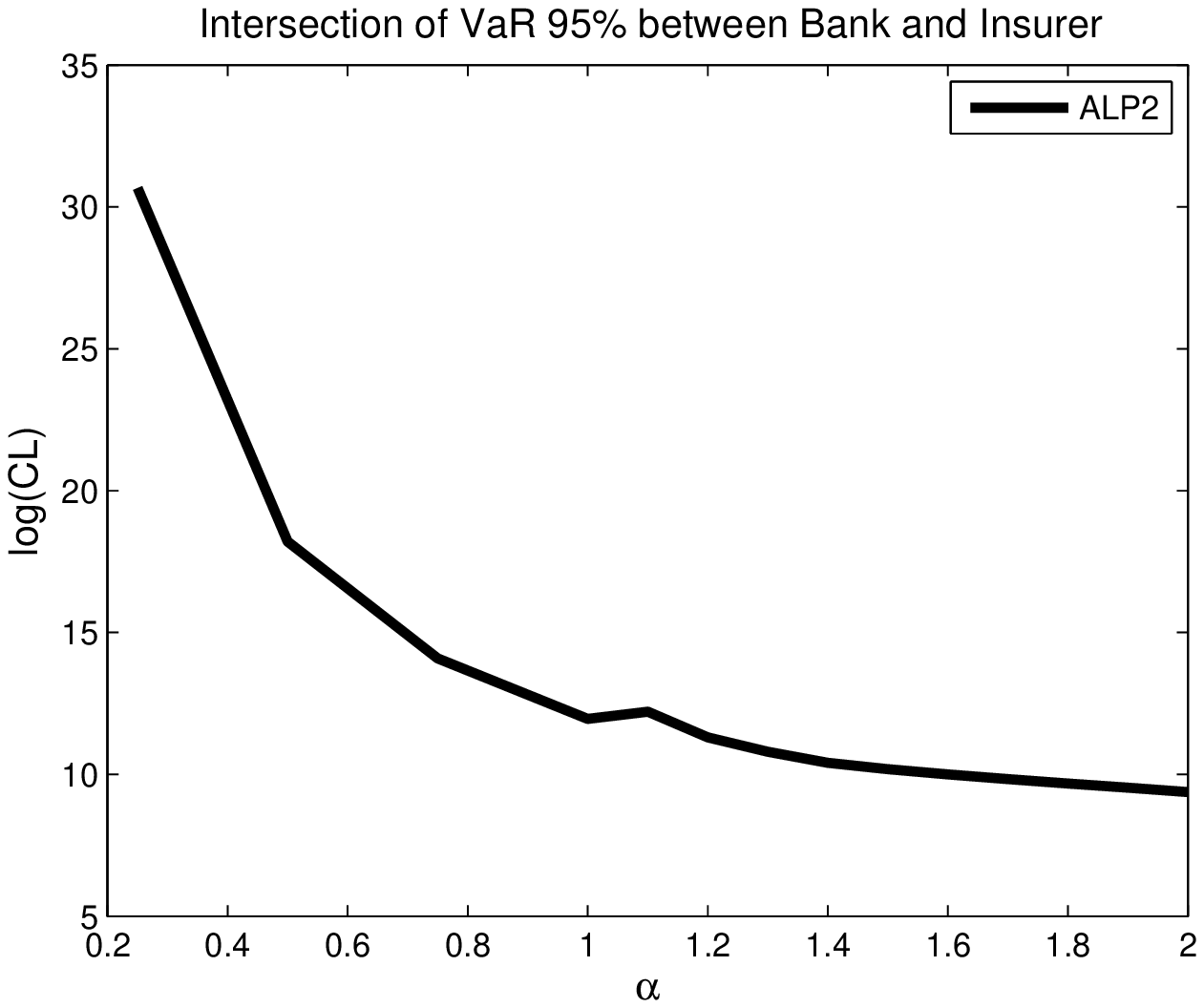}
\caption{Risk Equality Point between Bank and Insurer Risk $\lambda = 1$ (Left) and $\lambda = 10$ (Right).}
\label{FigINT_LowAlphaStabVaR}
\end{figure}

From each of these surfaces we extract information corresponding to the \textquotedblleft break-even point\textquotedblright between Bank VaR and Insurer MCR. Figure \ref{FigINT_LowAlphaStabVaR} corresponds to this \textquotedblleft break-even point\textquotedblright of Bank VaR and Insurer MCR plotted with $\alpha$ versus TCL for each of the basic policies. This quantity provides important information about the relationship between tail severity ($\alpha$) and insurance coverage (TCL), as it demonstrates the point at which the insurer bears an equal risk exposure to that of the bank being insured, as measured by the capital required under regulation. As such, this quantity will hitherto be referred to as the \textquotedblleft risk equality point\textquotedblright.

In the low frequency setting ($\lambda=1$), in Figure \ref{FigINT_LowAlphaStabVaR} a coincidence of ALP and CLP can be observed. This can be attributed to the dominance of ACL over TCL for the CLP case. Under the low frequency setting ($\lambda=1$), as our evaluation of ACL was set as the product of TCL and the $70^{th}$ percentile of the frequency distribution, we will achieve $ACL<TCL$ for all levels of TCL. As such, the risk mitigation offered under the CLP will coincide with the risk offered under the ALP, as the ACL will be completely exhausted for all exposures prior to the TCL being exhausted. Subsequently, a distinct difference between the application of ILP (with insurance coverage capped on a per event basis, TCL) and ALP/CLP (with insurance coverage capped on an annual basis, ACL) can be observed. However, we see a convergence in the \textquotedblleft risk equality point\textquotedblright between the ALP and ILP as we increase the heaviness of the tails (that is reduce $\alpha$). This can be explained by the fact that in extreme severity cases, the difference between the TCL and ACL becomes less significant and consequently, the value of risk mitigation begins to coincide.

An important caveat for this comparison is to note the dependence of ACL on the level of TCL. Alternatively, if ACL were to increase for fixed TCL we would observe a divergence between the ALP and CLP and, for sufficiently large ACL, a convergence of the ILP and CLP results.

In the high frequency setting ($\lambda=10$), from Figures \ref{Fig3d_LowAlphaStabVaR_ILP}, \ref{Fig3d_LowAlphaStabVaR_ALP}, \ref{Fig3d_LowAlphaStabVaR_CLP} we observe a significant difference in the risk mitigation provided by the ALP and ILP/CLP. This discrepancy is also apparent in Figure \ref{FigINT_LowAlphaStabVaR} and can be attributed to a so called \textquotedblleft risk equality point\textquotedblright  achieved in the application of the ALP. This benefit is best understood by considering the insurance mitigation provided on a single extreme event. Due to the high frequency setting, our derivation yields $ACL>TCL$. Under the ILP (and subsequently the CLP), the insurer will provide risk mitigation up to the specified TCL, while under the ALP, the insurer will provide risk mitigation up to the specified ACL. As such, the ALP proves a more efficient method of mitigating risk. However, the greatest divergence between the policies occurs for $\alpha \in (0.6,1)$ and we can observe a convergence in the \textquotedblleft risk equality point\textquotedblright for all three policies as we reduce the heaviness of the tails and more subtlely as we increase the heaviness of the tails. The justification for the convergence as $\alpha$ decreases is similar to that given in the low frequency setting. As $\alpha$ increases, we see a distinct (although small) difference in the \textquotedblleft risk equality point\textquotedblright of all three policies. This can be attributed to the fact that as $\alpha$ decreases the probability of experiencing extreme losses decreases. As discussed above, the comparative advantage of ALP is its ability to mitigate extreme losses (in other it has the capability of mitigating a larger portion of extreme losses), hence as the probability of extreme losses decreases we see a convergence in the risk mitigation provided under ALP and ILP. Additionally, we can see a divergence between the ILP and CLP, this is due to the fact that as we decrease the probability of extreme losses, the TCL no longer dominates the mitigation in the CLP and it becomes an interplay of the per event basis cover (TCL) and the annual cover (ACL). Consequently, we see a divergence in the mitigation between the two.

Finally, under the bi-variate risk setting, by comparign Figures \ref{Fig3d_LowAlphaStabVaR_ILP} and \ref{Fig3d_LowAlphaStabVaR_ALP} we can observe a similar behaviour across low and high frequency settings for the ALP. In other words, the application of an aggregate policy cap (annually) can be seen to have similar impact as that of applying an annual limit on insurance coverage for a univariate risk exposure.

\begin{figure}[!ht]
\includegraphics[scale=0.35]{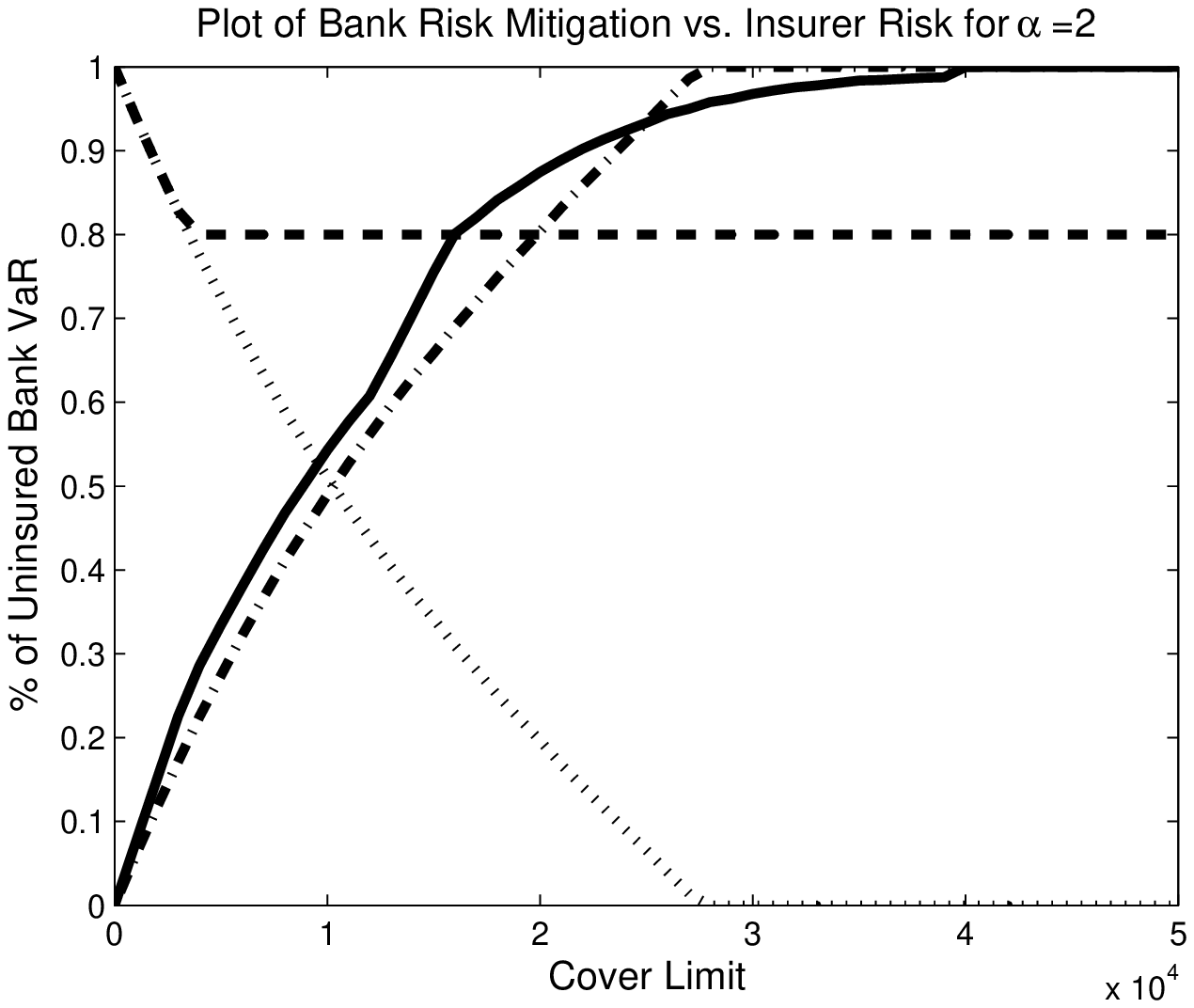}
\includegraphics[scale=0.35]{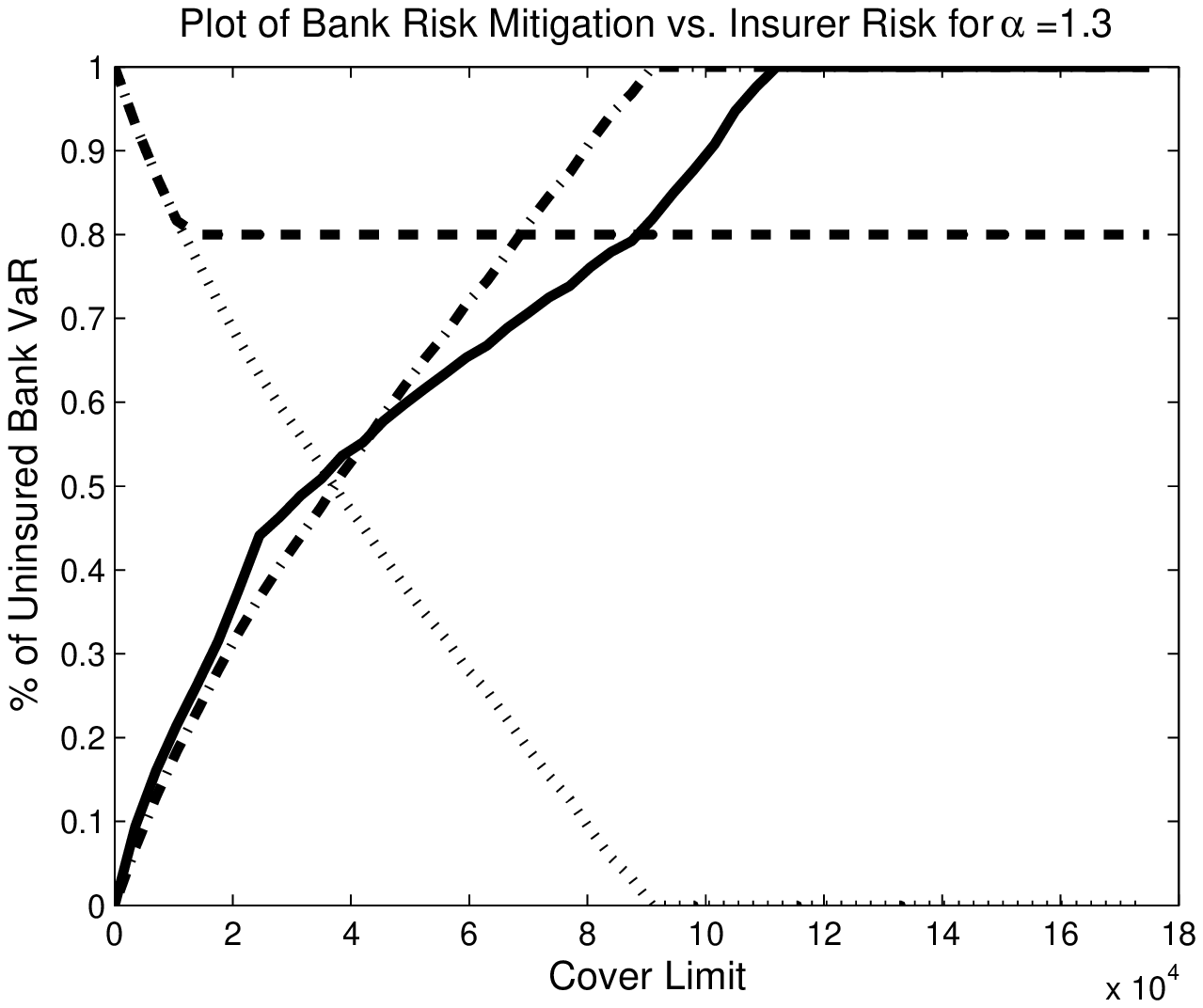}
\includegraphics[scale=0.35]{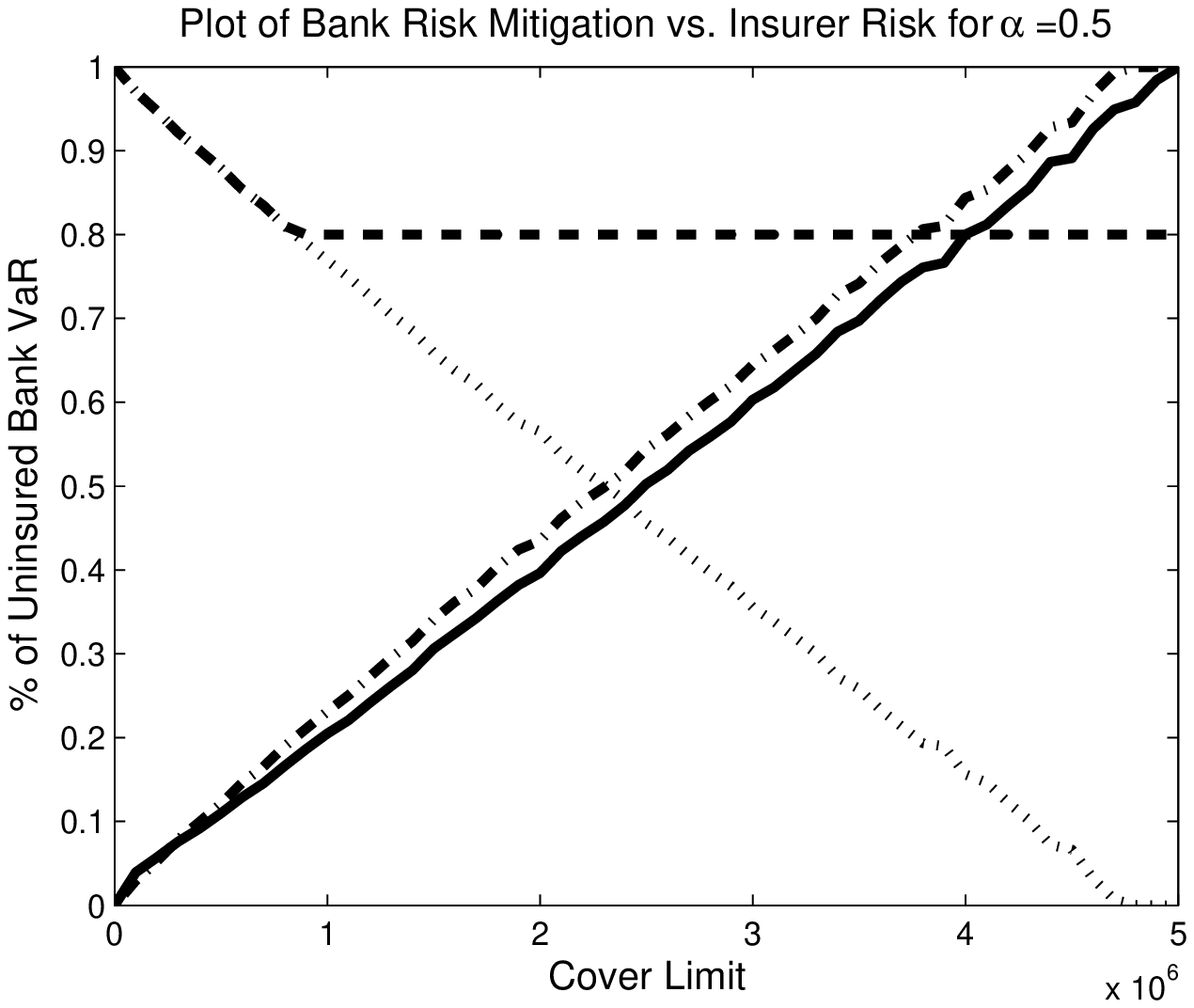}\\
\includegraphics[scale=0.35]{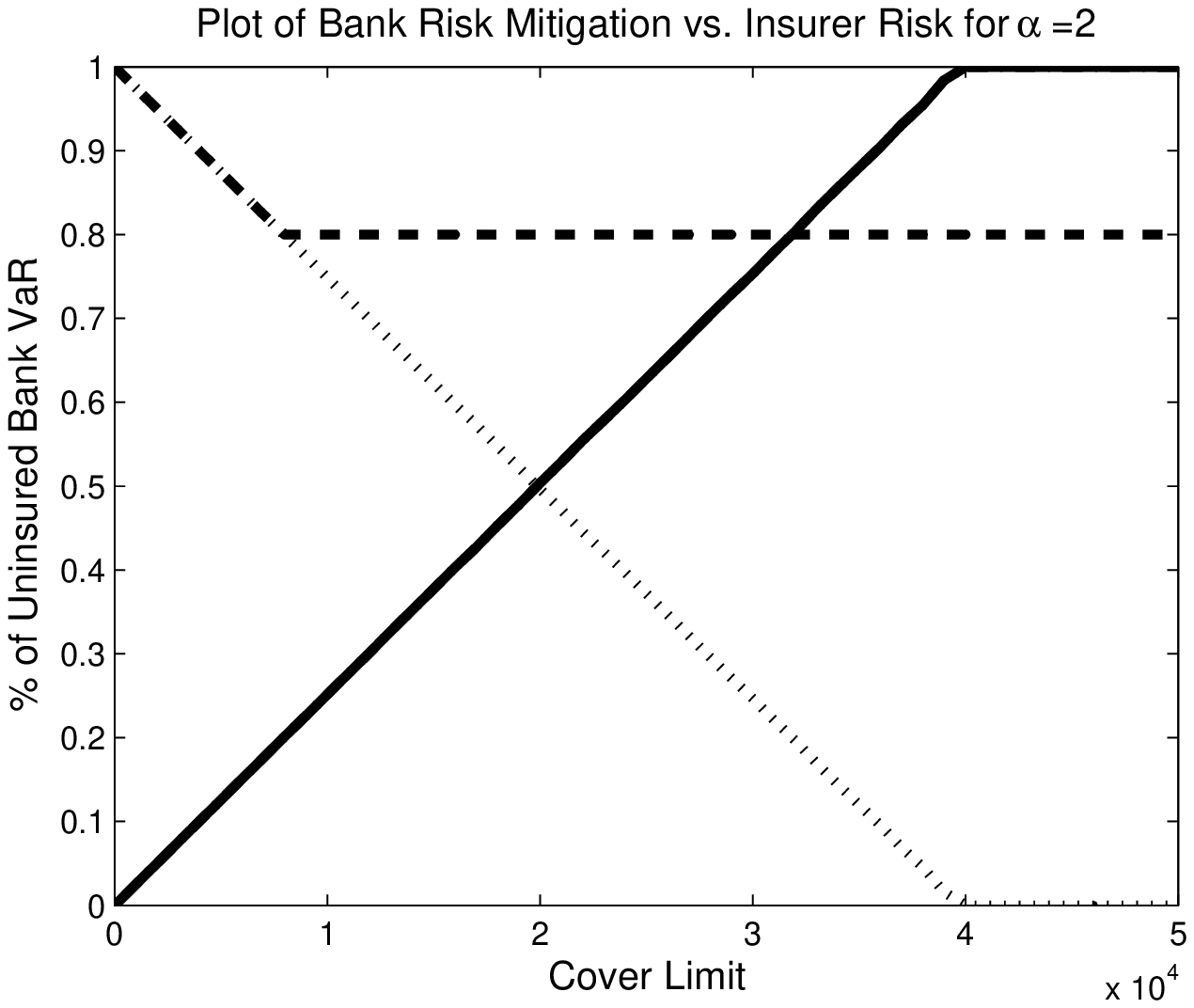}
\includegraphics[scale=0.35]{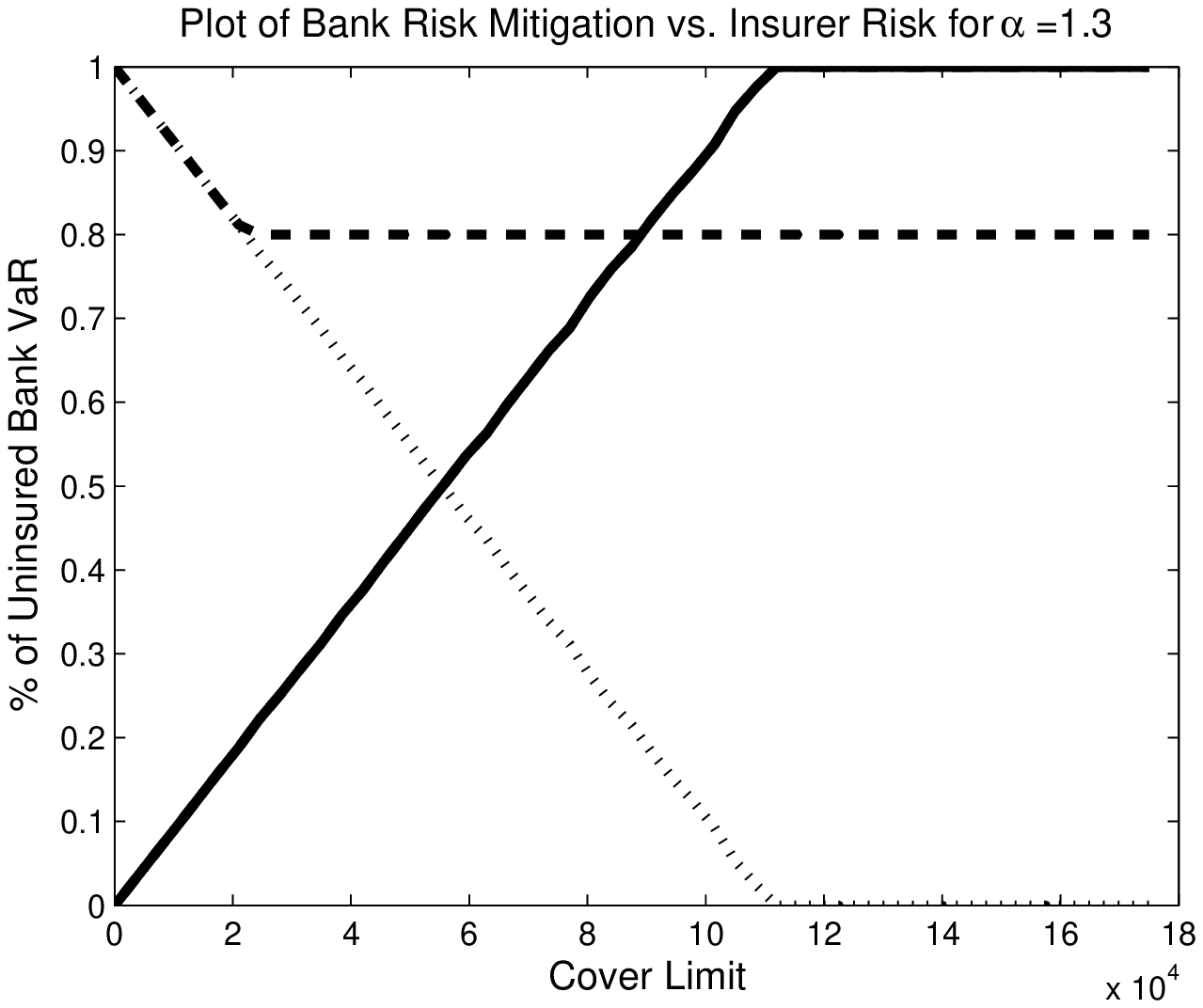}
\includegraphics[scale=0.35]{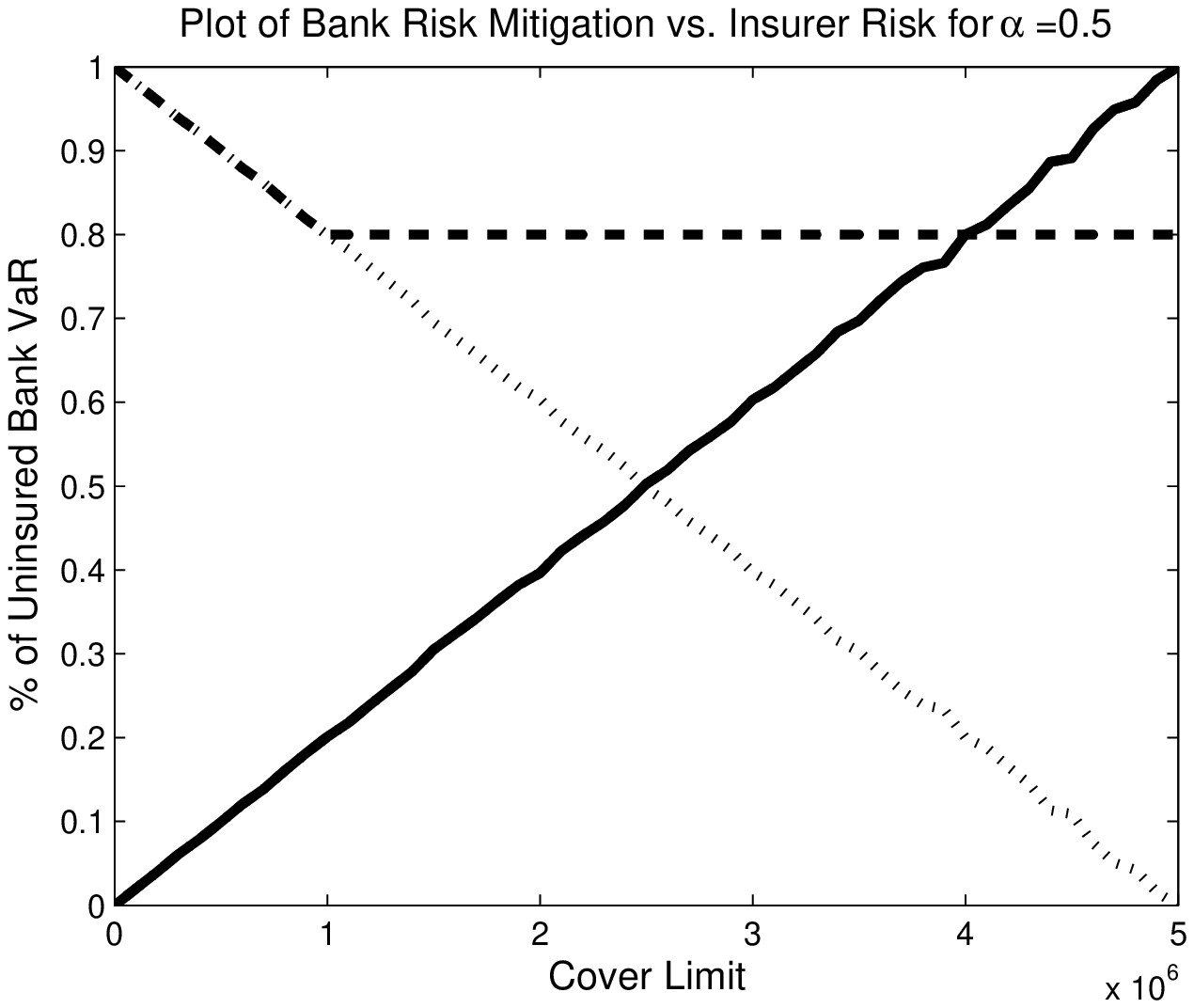}\\
\includegraphics[scale=0.35]{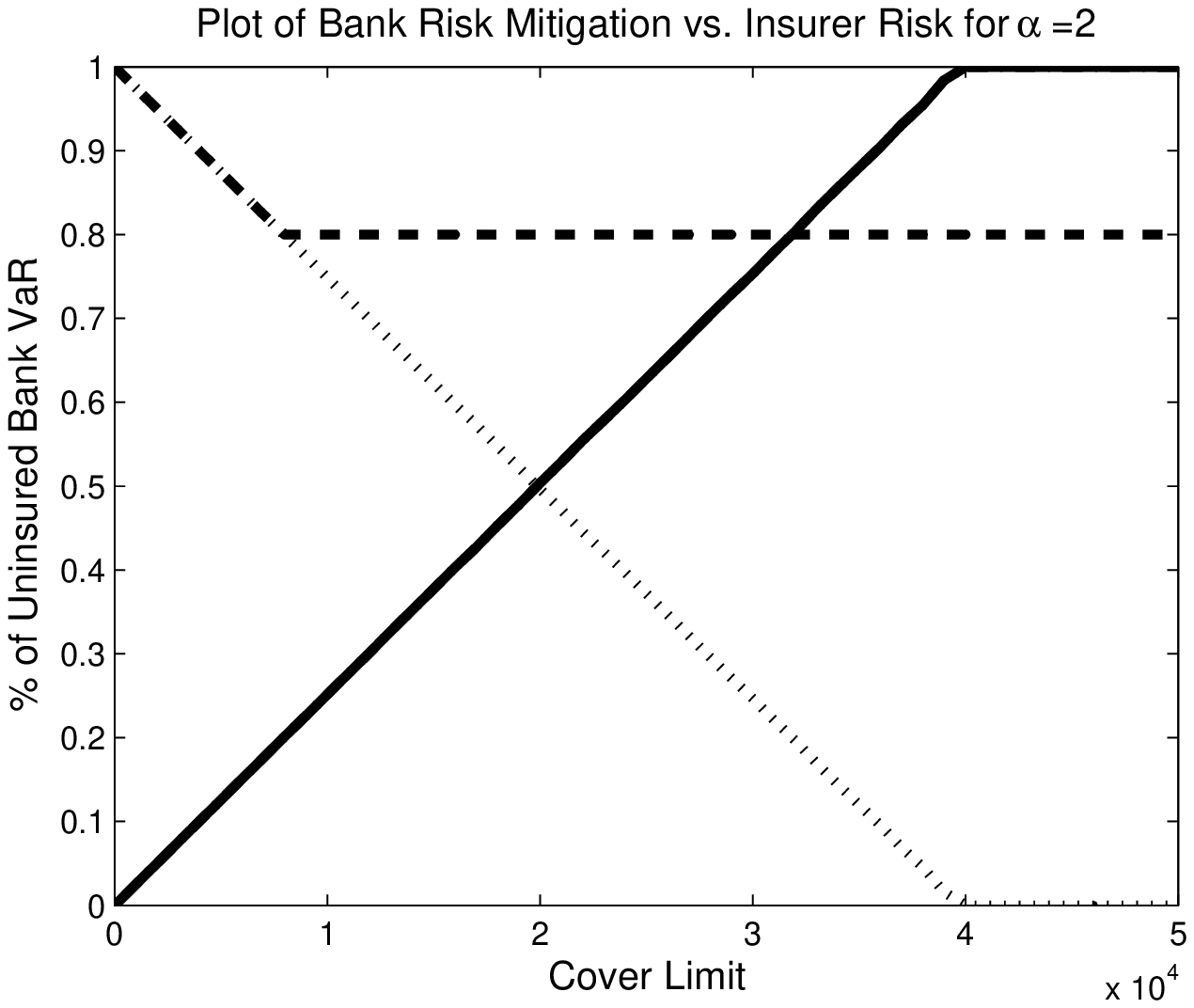}
\includegraphics[scale=0.35]{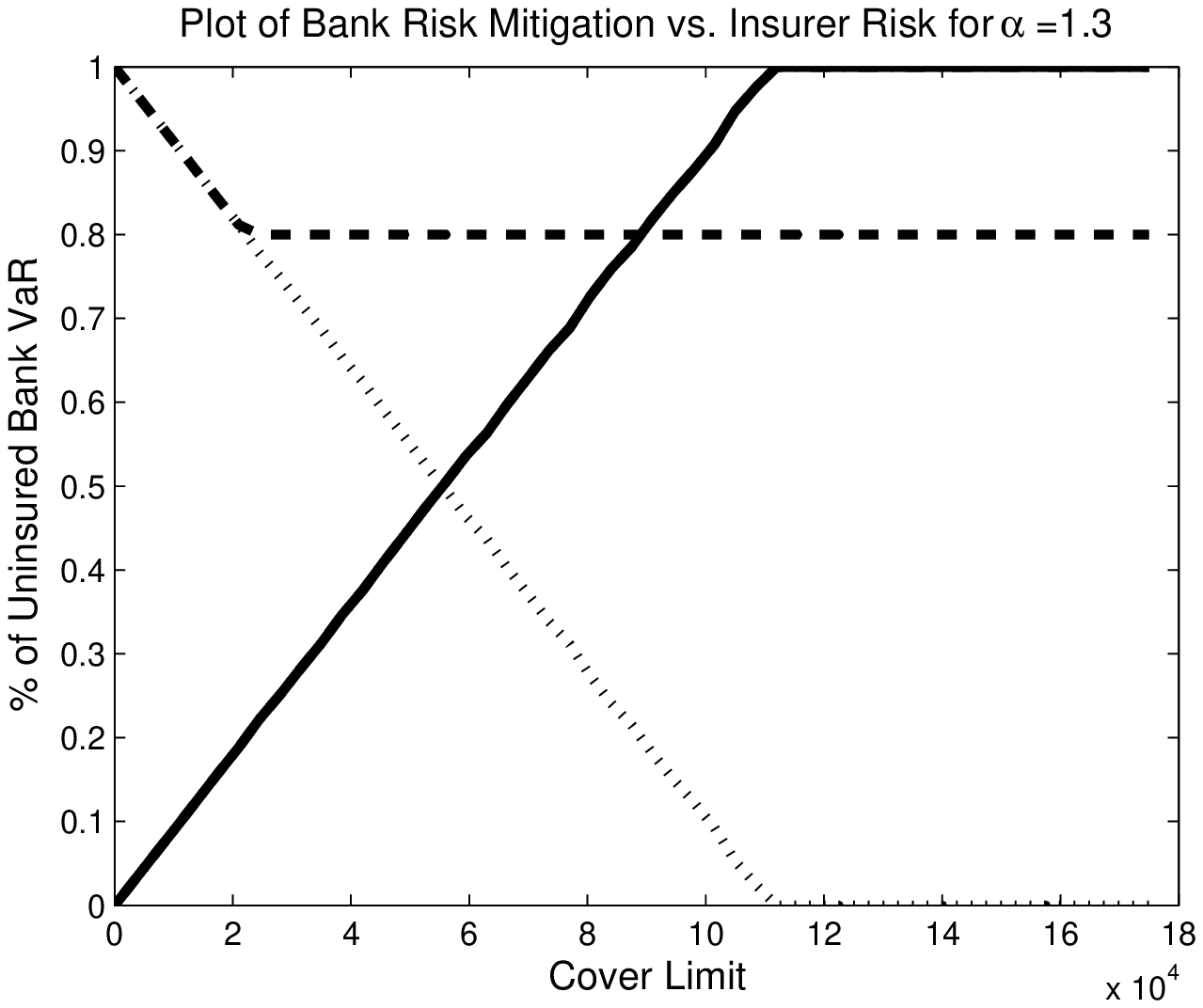}
\includegraphics[scale=0.35]{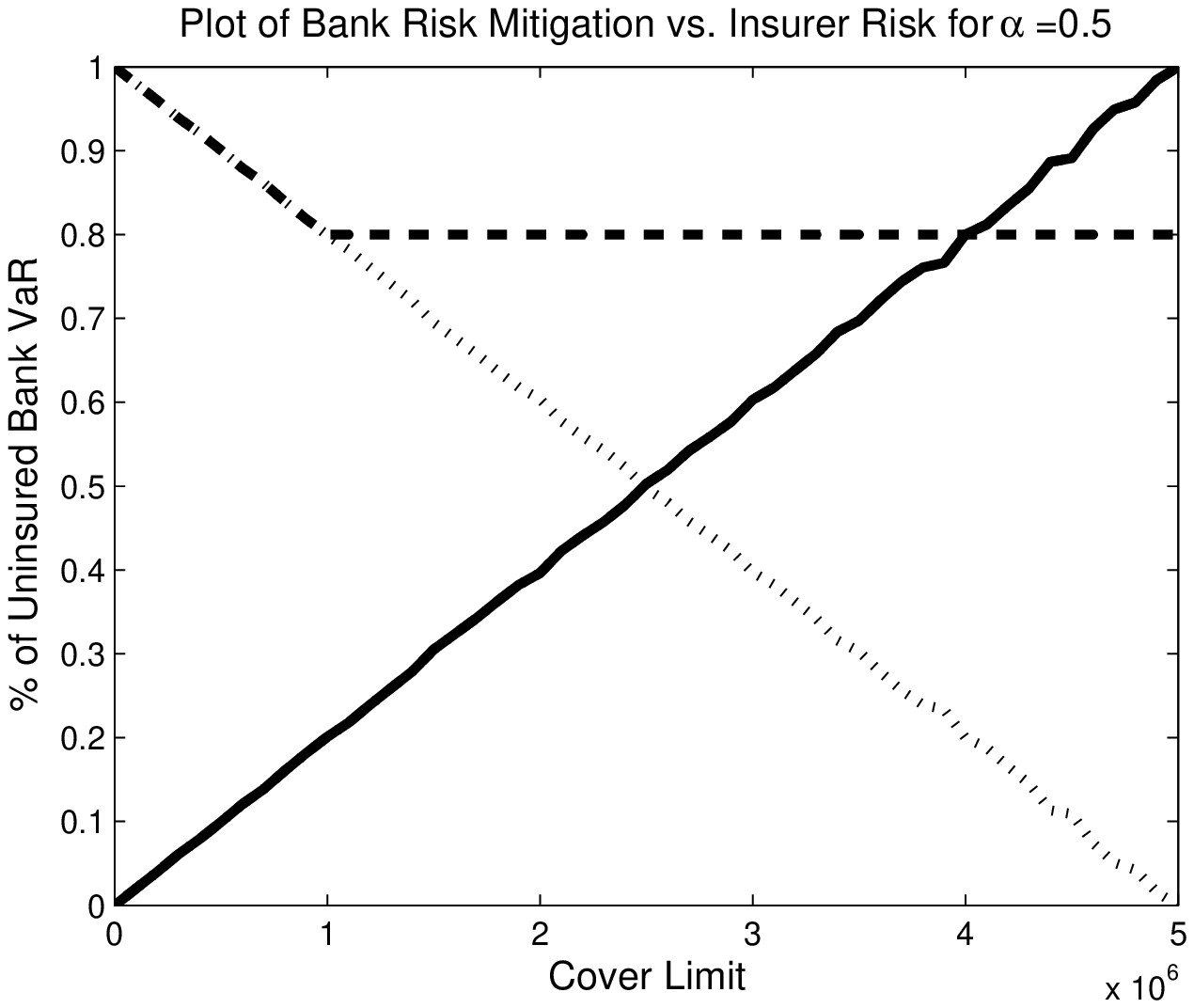}
\caption{\textbf{Top Row} - ILP, \textbf{Middle Row} - ALP, \textbf{Bottom Row} - CLP; \textbf{Column 1} $\alpha = 2$, \textbf{Column 2} $\alpha = 1.3$, \textbf{Column 3} $\alpha = 0.5$. All figures display unregulated Bank risk exposure $\% VaR_{0.95}\left[Z^{(j,i)}\right]$ (dotted line), regulated Bank risk exposure (dashed line),  risk transferred to the Insurer $\% MCR_{0.95}\left[C^{(j,i)}\right]$ (solid line), risk mitigation received by the Bank $\% VaR_{0.95,Mit}\left[Z^{(j,i)}\right]$ (dashed-dot line).}
\label{FigILP_LowAlphaStabVaR}
\end{figure}

To better understand the relationship between policy, cover limit and risk mitigation for the risk exposures, we will provide further investigations of specific models focusing on rare event settings. The results of this additional low frequency analysis are provided in Figure \ref{FigILP_LowAlphaStabVaR}.

Figure \ref{FigILP_LowAlphaStabVaR} demonstrates that the Bank's risk mitigation (measured as $\% VaR_{0.95,Mit}\left[Z^{(j,i)}\right]$, the dashed-dot line) and the risk transferred to the Insurer (measured as $\% MCR_{0.95}\left[C^{(j,i)}\right]$, the solid line) coincide for the ALP and CLP (rows 2 and 3) under both light and heavy tails. This can be attributed to the fact that the ACL on these policies is completely exhausted at each level of TCL. Therefore a \textit{complete} transfer of risk from the bank to the insurer occurs, that is the risk taken on by the insurer through the policy is equal to risk mitigated for the bank. Due to the the complete transfer of risk, we observe a linear relationship between the level of insurance coverage and the risk transferred to the insurer. In other words, an increase in one unit of insurance coverage corresponds to a complete transfer of a comparative unit of risk from the Bank to the Insurer.

Conversely, in Figure \ref{FigILP_LowAlphaStabVaR} we see a significant divergence between the Bank's risk mitigation (dashed-dot line) and the risk transferred to the insurer (solid line) for the ILP (row 1). This can be attributed to the structure of the ILP policy. Under the ILP, a loss is completely mitigated if $X_{s}^{(j)}\leq TCL$ and the loss is reduced to $X_{s}^{(j)}-TCL$ for $X_{s}^{(j)}> TCL$. Therefore we can view the loss process for the Bank as a left-truncated distribution and similarly for the Insurer a right-truncated distribution. However, unlike the ALP/CLP case where the ACL is completely exhausted, there is no upper cap on the total annual coverage for the insurer. This creates an \textit{incomplete} transfer of risk and therefore poses interesting questions around optimal conditions for the application of insurance. 

If we consider for each $\alpha$ level, the region where the risk transferred to the Insurer (solid line) exceeds the Bank's risk mitigation, we can infer that the Insurer will overcharge (in the form of premiums) for the risk mitigation received by the bank. To understand this, we must first consider the motivation behind the pricing structure of the premiums for the Insurer. The Insurer will charge the Bank a premium equivalent to the proportion of risk transferred to the Insurer ($\% MCR_{0.95}\left[C^{(j,i)}\right]$) and \textit{not} the proportion of risk mitigated for the bank ($\% VaR_{0.95,Mit}\left[Z^{(j,i)}\right]$). The region prior to cross over, we see that $\% MCR_{0.95}\left[C^{(j,i)}\right]>\% VaR_{0.95,Mit}\left[Z^{(j,i)}\right]$ so the Bank will be paying for risk mitigation on the value of $\% MCR_{0.95}\left[C^{(j,i)}\right]$ but will only be receiving mitigation for the value $\% VaR_{0.95,Mit}\left[Z^{(j,i)}\right]$. Consequently, this region is not an optimal insurance region for the bank.

Let us assume that the insurer will price the risk \textit{mitigation} according to the level of risk \textit{transferred} to the insurer. As such, if there is a deviation between the level of risk transferred and that of the risk mitigated, that is the transfer of risk is not complete, we will observe a mispricing of the insurance policy. We can assume that this mispricing is realised only by the bank as the market for OpRisk insurance is small and hence the price is set by the supplier (the insurer) not the buyer (the bank). As such, the insurer will always receive the correct premium for the level of risk they retain, while the bank will have to accept whatever premium is set. As such, this presents an arbitrage opportunity for the bank because, if they obtain insurance for the correct portion of risk mitigation, they can essentially profit from the transaction as they can offload their risk at a discount to its true value.

If we consider for each $\alpha$ level, the region where the risk transferred to the insurer (solid line) exceeds the bank's risk mitigation (dashed-dot line), we can infer that the insurer will overcharge for the risk mitigation received by the bank, as $\% MCR_{0.95}\left[C^{(j,i)}\right]>\% VaR_{0.95,Mit}\left[Z^{(j,i)}\right]$. Consequently, this region is not an optimal insurance region for the bank.

\begin{figure}[!ht]
\includegraphics[width=0.5\textwidth, height = 4cm]{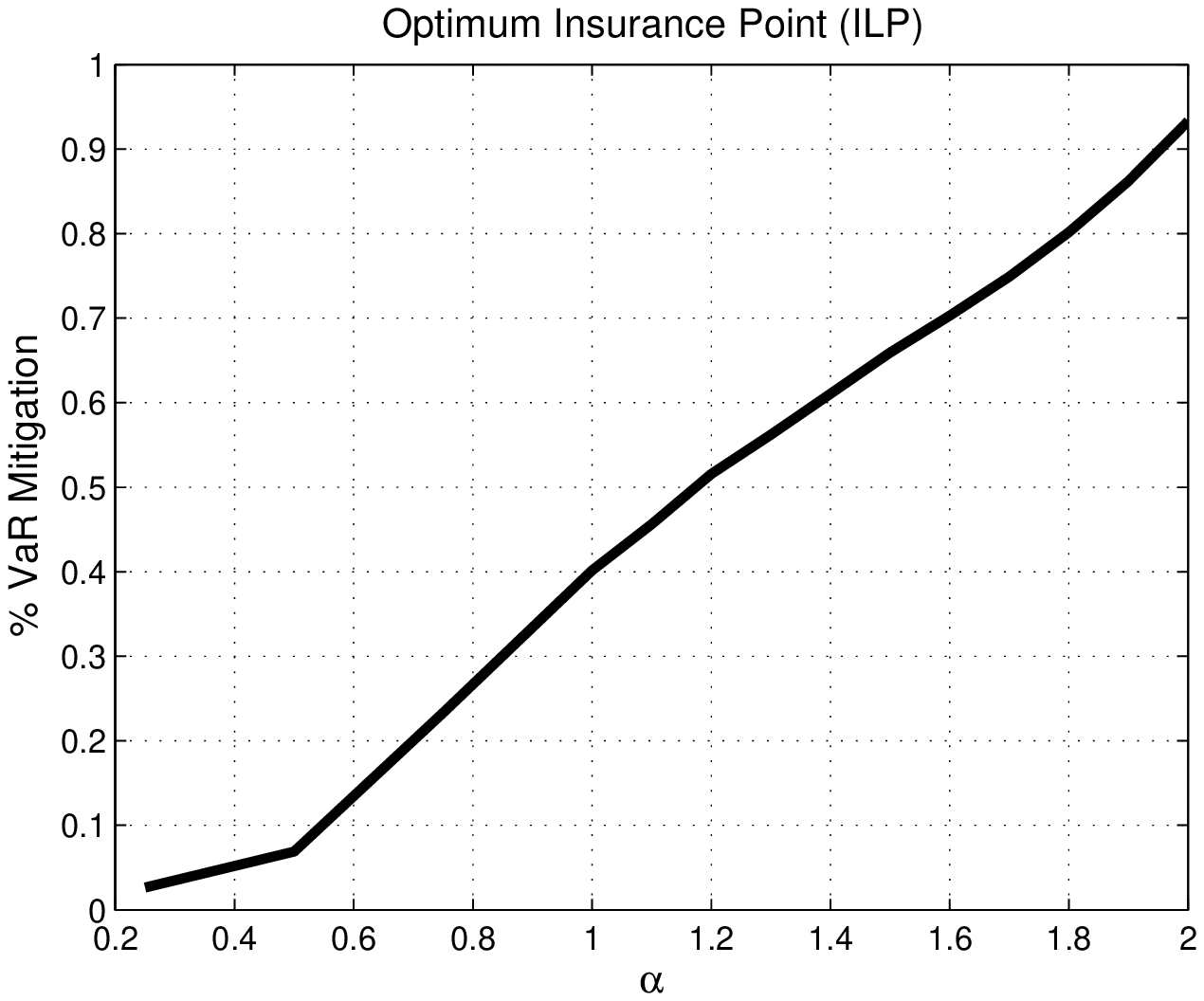}
\includegraphics[width=0.5\textwidth, height = 4cm]{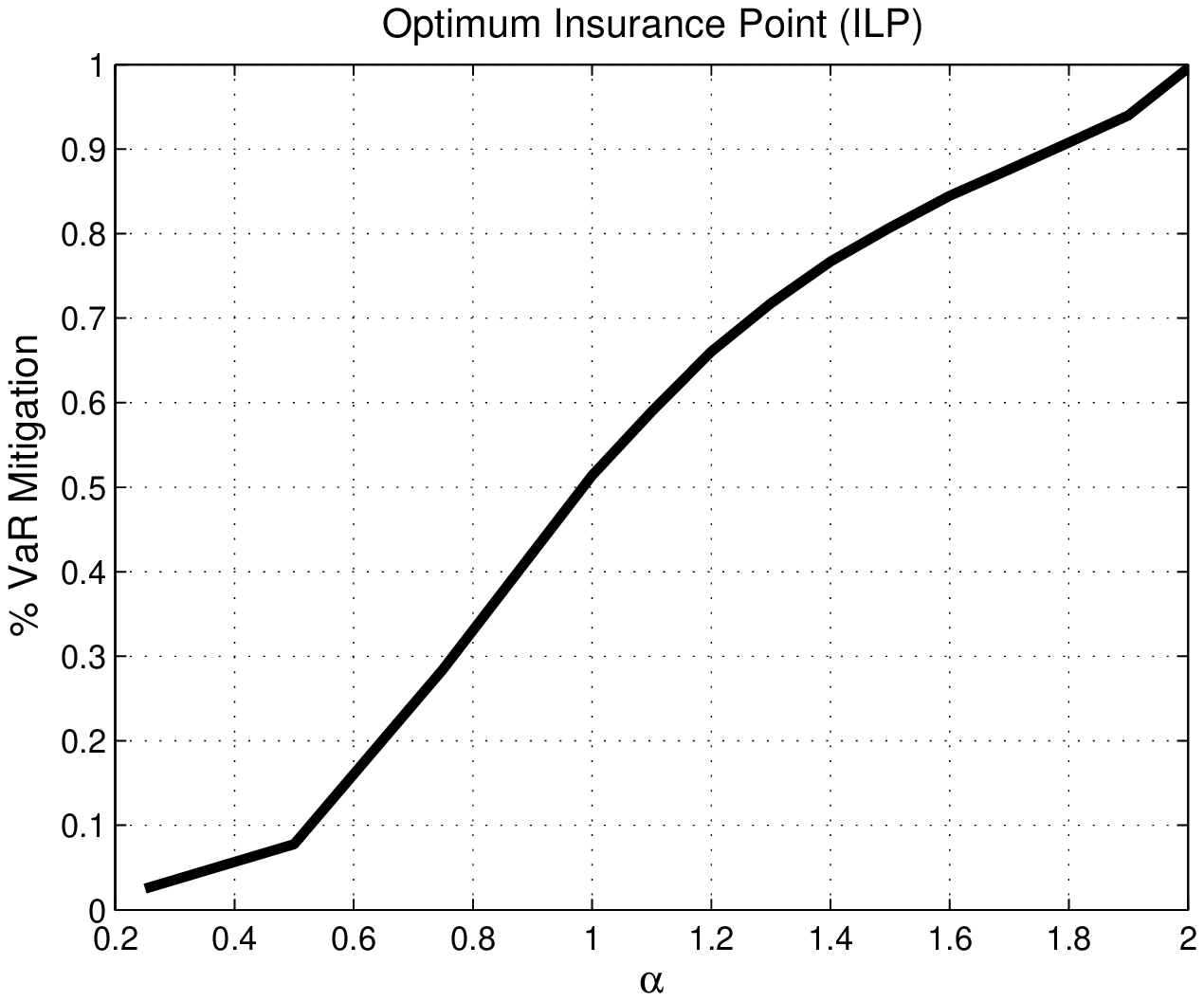}
\caption{Optimum Insurance Point between Bank Risk Mitigation and Insurer Risk $\lambda = 1$ (Left) and $\lambda = 10$ (Right) as a function of \% Var Mitigation.}
\label{FigOPT_LowAlphaStabVaR}
\end{figure}

Conversely, if we consider the region where $\% MCR_{0.95}\left[C^{(j,i)}\right]<\% VaR_{0.95,Mit}\left[Z^{(j,i)}\right]$ (that is where the dashed-dot line exceeds the solid line), the Bank will be paying for less risk than it is offloading. To better illustrate this point, we will consider the relationship between the heaviness of the tails, as measured by $\alpha$, and this so called \textquotedblleft optimum insurance point\textquotedblright, as measured by the risk mitigation received by the bank $\% VaR_{0.95,Mit}\left[Z^{(j,i)}\right]$, which defines the region where it is advantageous for the Bank to undertake insurance on its risk exposure. These results (for low and high frequency settings) are summarised in Figures \ref{FigOPT_LowAlphaStabVaR}.

\begin{figure}[!ht]
\includegraphics[width=0.5\textwidth, height = 4cm]{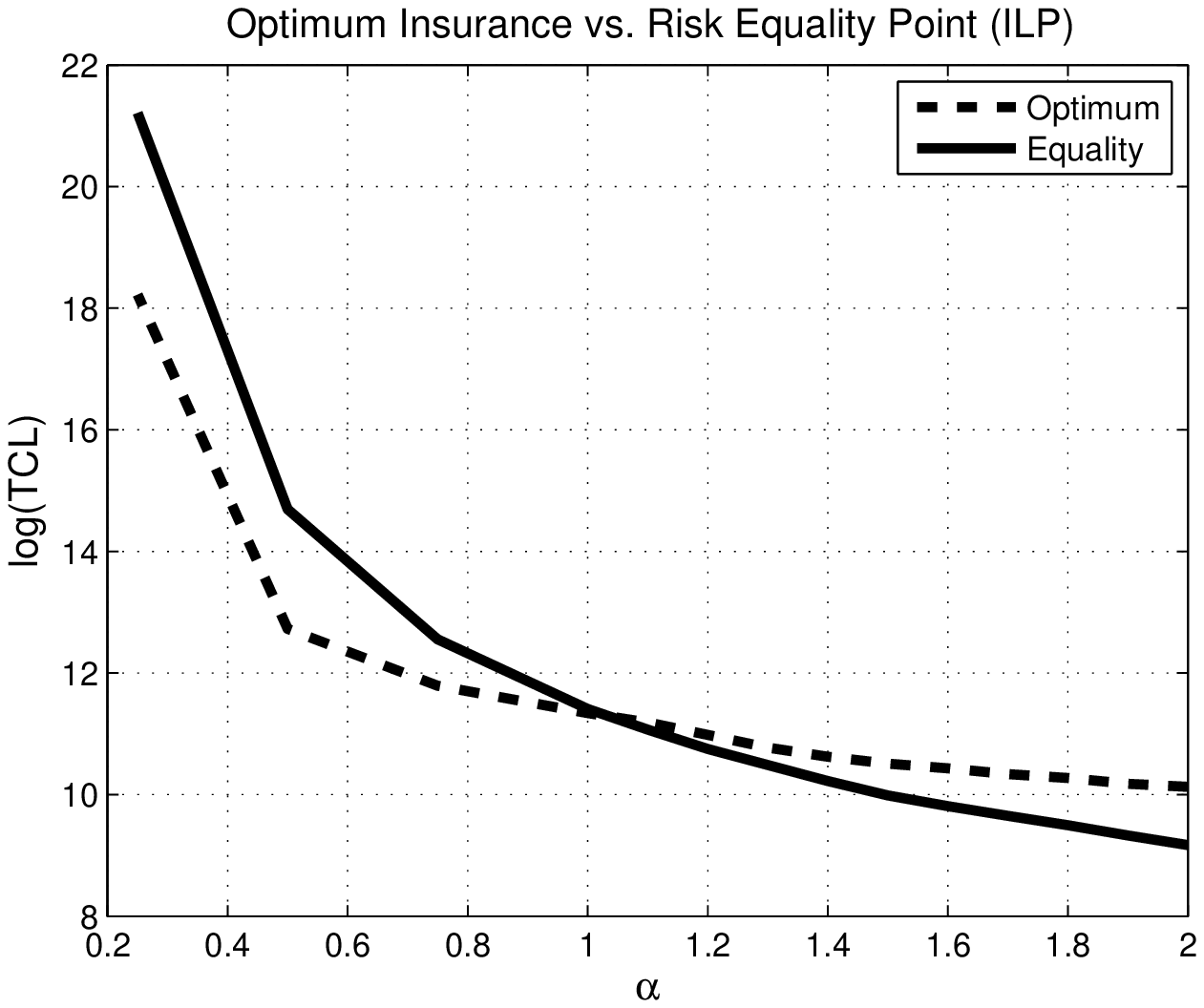}
\includegraphics[width=0.5\textwidth, height = 4cm]{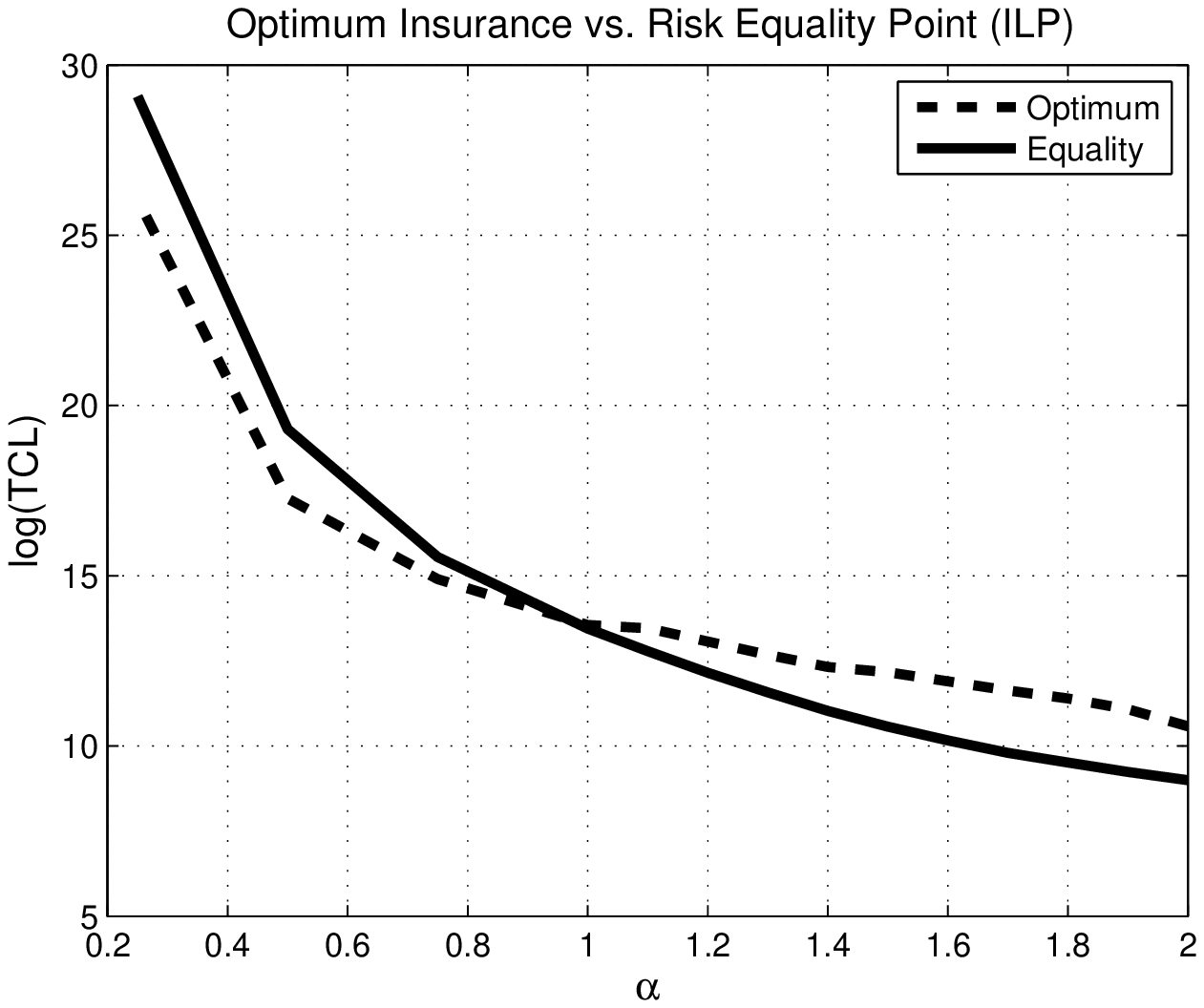}
\caption{Optimum Insurance Point between Bank Risk Mitigation and Insurer Risk $\lambda = 1$ (Left) and $\lambda = 10$ (Right) as a function of TCL.}
\label{FigOPTEQ_LowAlphaStabVaR}
\end{figure}

It can be seen from Figure \ref{FigOPT_LowAlphaStabVaR} that there is a clear relationship between the heaviness of the tails of the severity distribution ($\alpha$) and the optimum point at which to undertake insurance (as measured by the risk mitigation $\% VaR_{0.95,Mit}\left[Z^{(j,i)}\right]$). It can be seen that as the value of $\alpha$ decreases, the optimum value for the bank as a portion of its entire risk exposure decreases. This is a convenient result, as the bank would most likely prefer to reduce its threshold criteria for undertaking insurance as the probability of incurring an extreme loss increases. Similarly, we can evaluate the relationship between $\alpha$ and the optimum insurance point, as measure by TCL, summarised in Figure \ref{FigOPTEQ_LowAlphaStabVaR}.

This relationship provides us with a threshold value of TCL for a Bank to undertake insurance. That is, given a severity model with tail index $\alpha$, a Bank can evaluate a minimum TCL level which they require in order for their insurance mitigation to become \textquotedblleft profitable\textquotedblright.

\subsubsection{Advanced Insurance Policies}
Here we consider the ILP and the CLP insurance policies combined into the advanced insurance structures involving either a haircut or stochastic banding discounting. Again we focus on three detailed studies involving $\alpha$-stable severity tail indexes of $\alpha \in \left\{2,1.3,0.5\right\}$ in the low frequency setting.

\begin{figure}[!ht]
\includegraphics[scale=0.35]{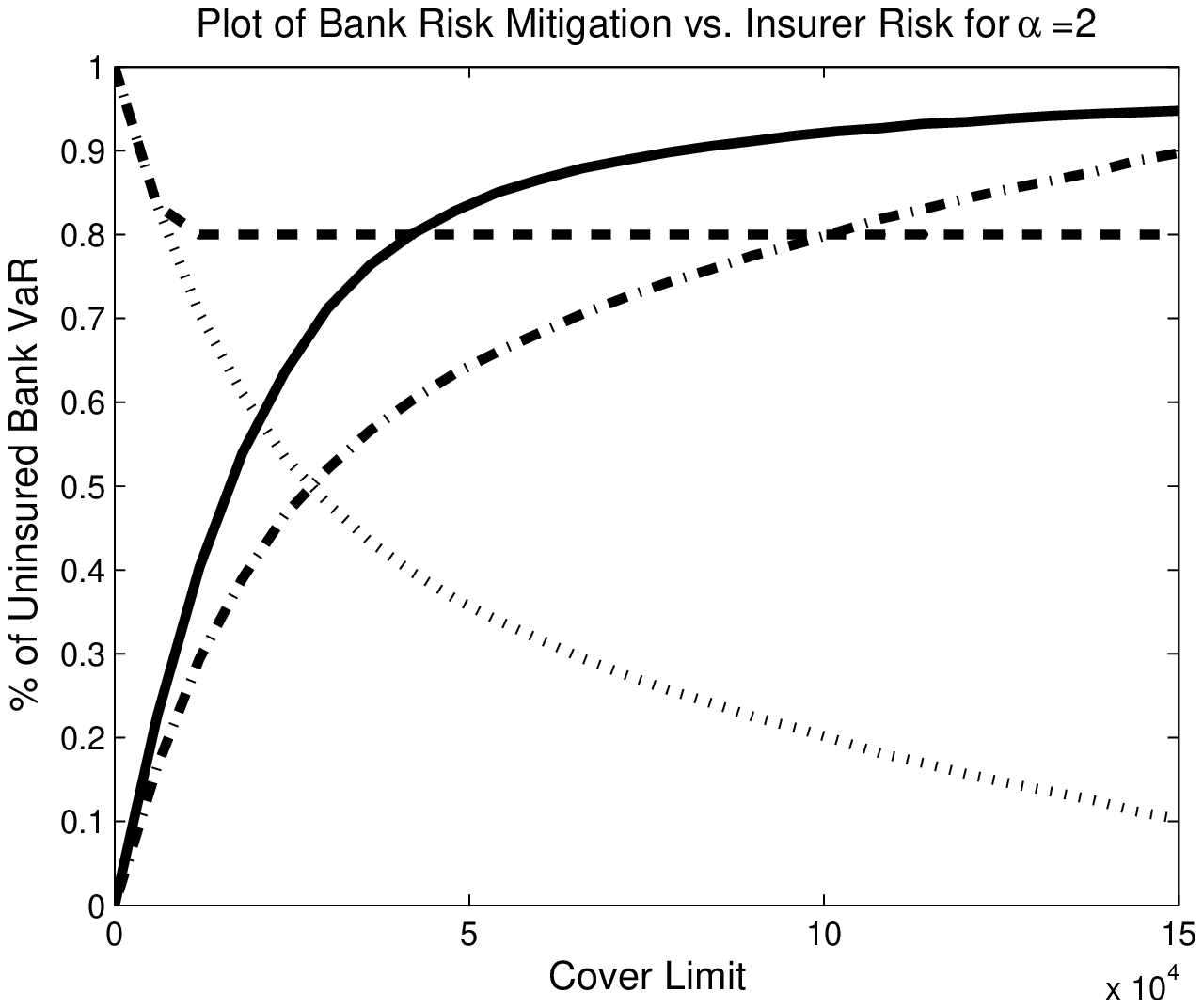}
\includegraphics[scale=0.35]{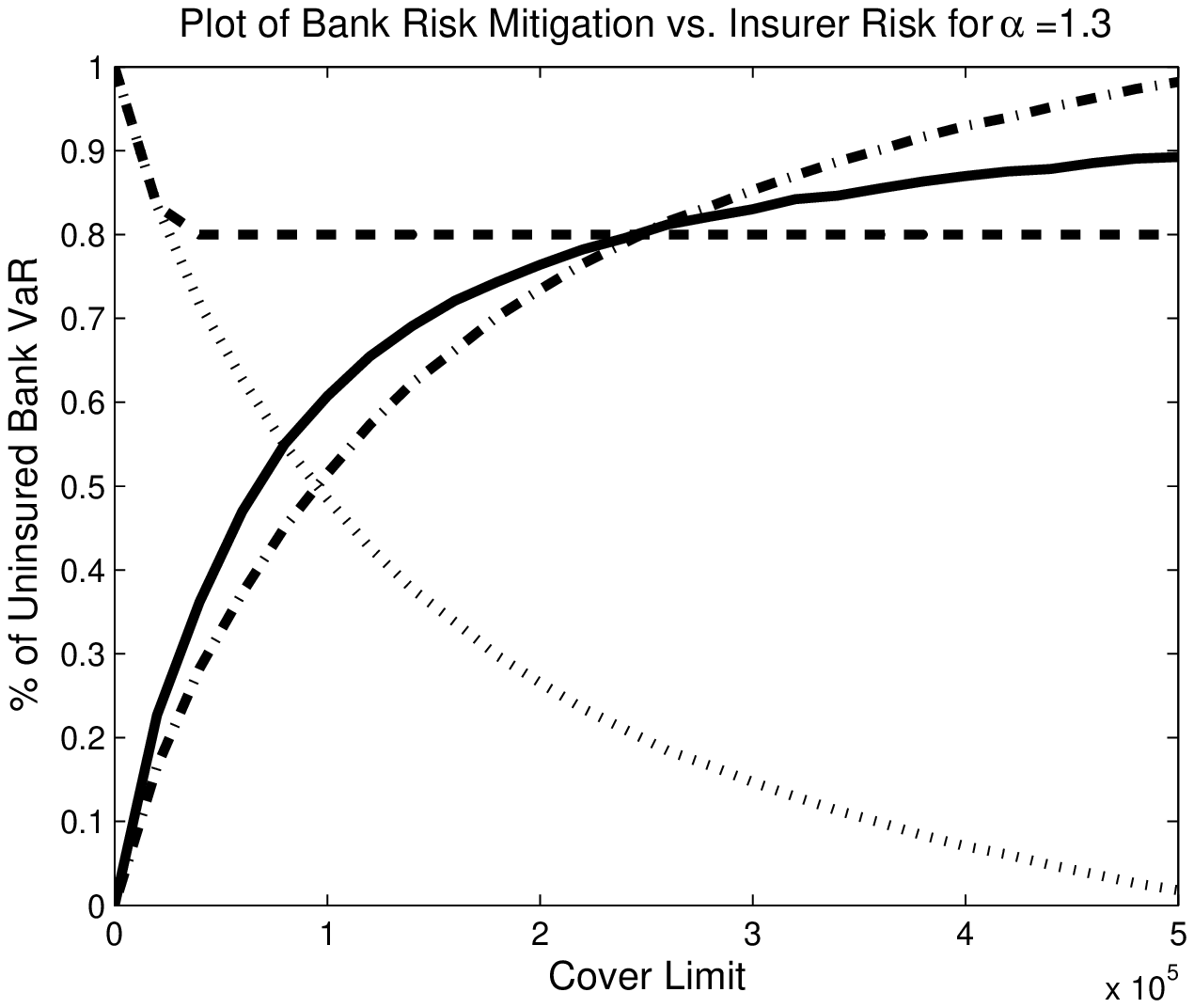}
\includegraphics[scale=0.35]{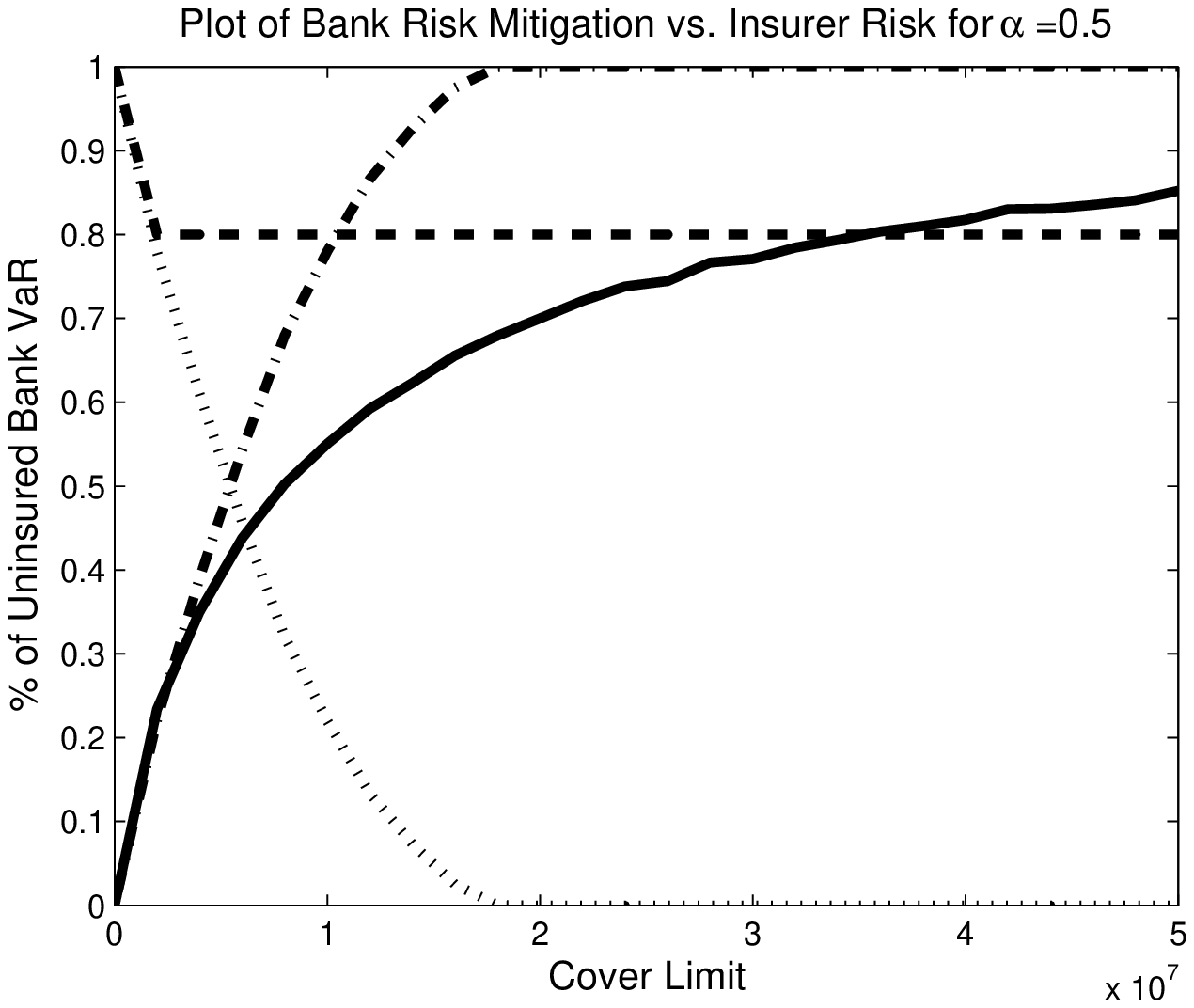}\\
\includegraphics[scale=0.35]{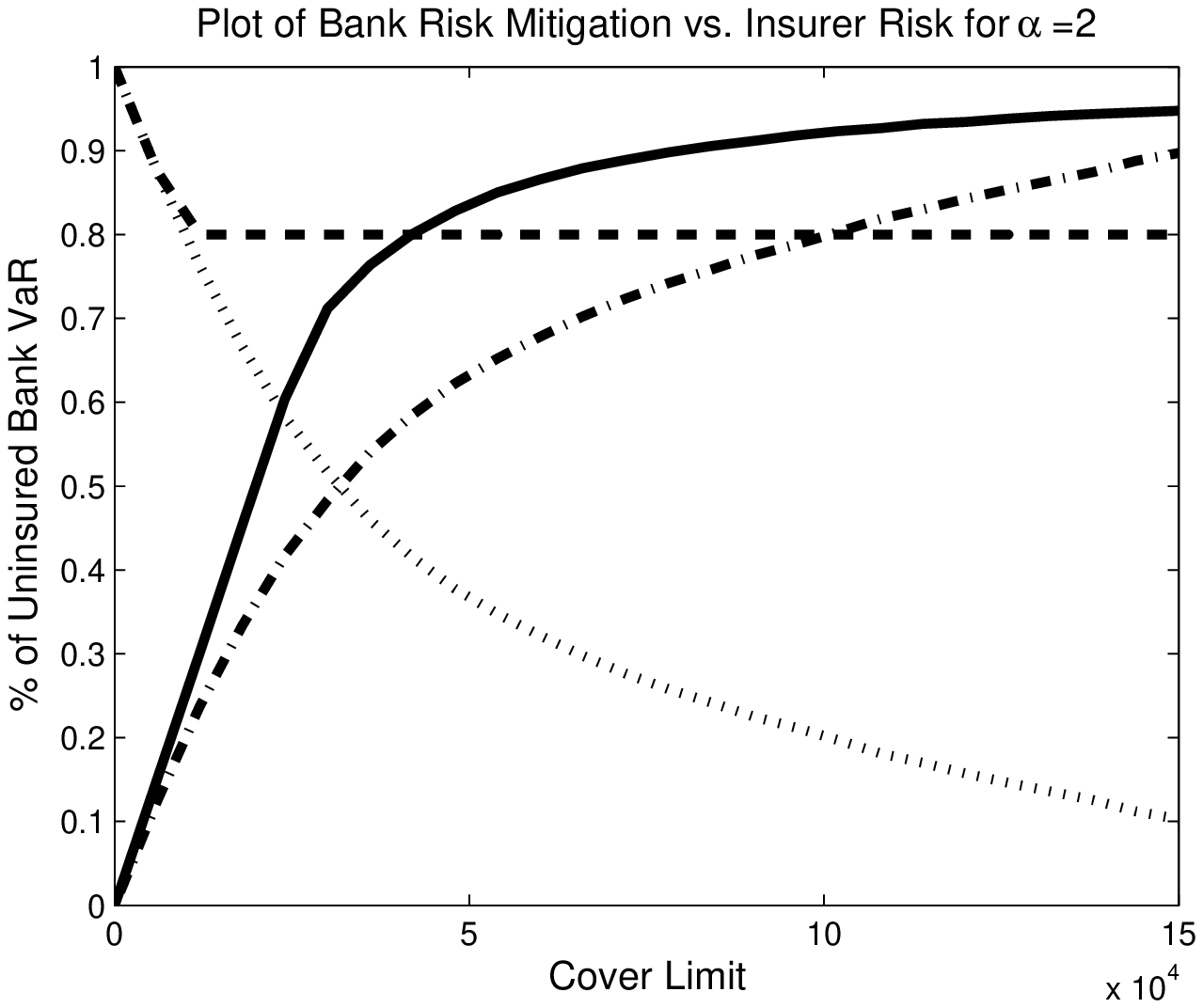}
\includegraphics[scale=0.35]{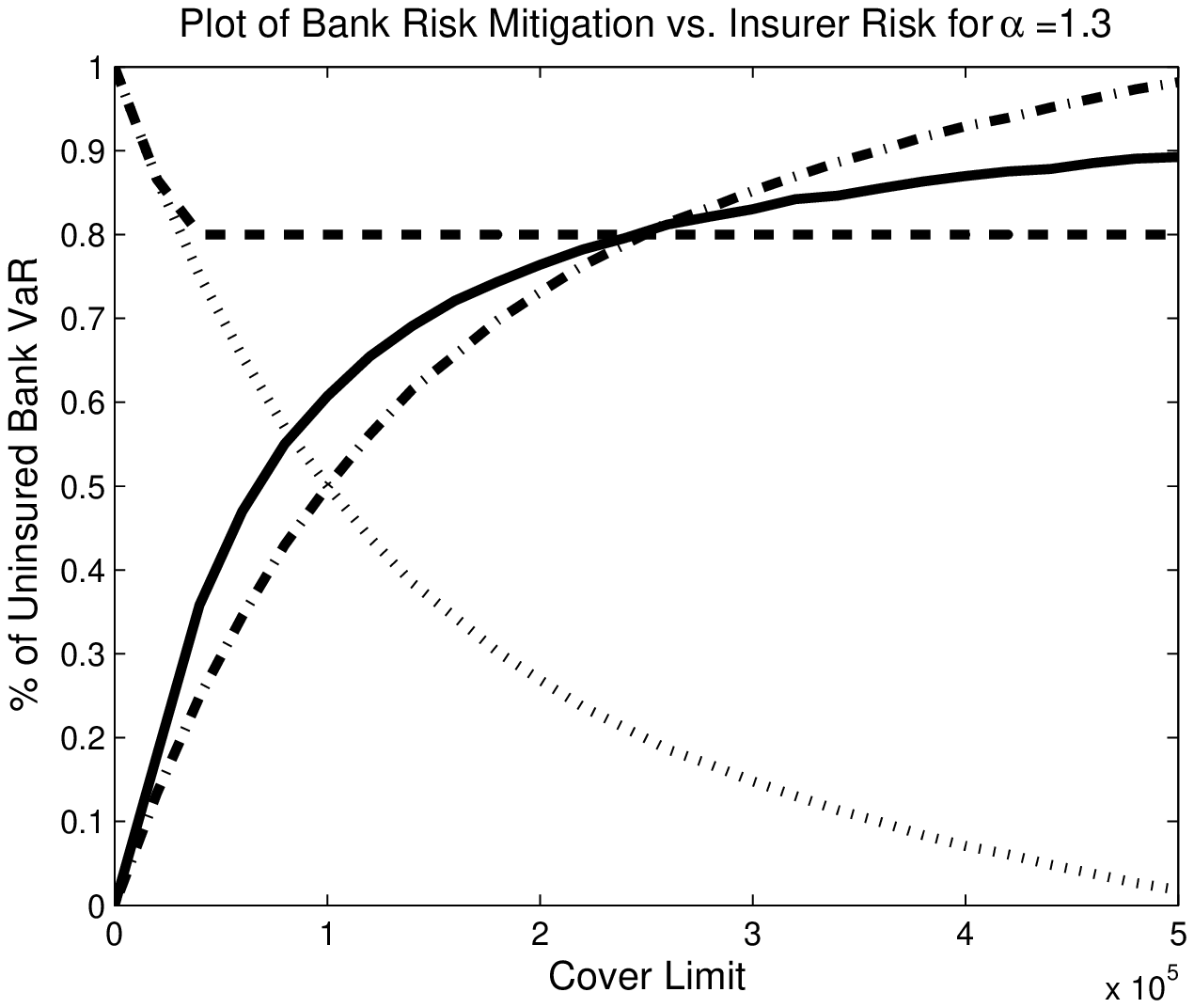}
\includegraphics[scale=0.35]{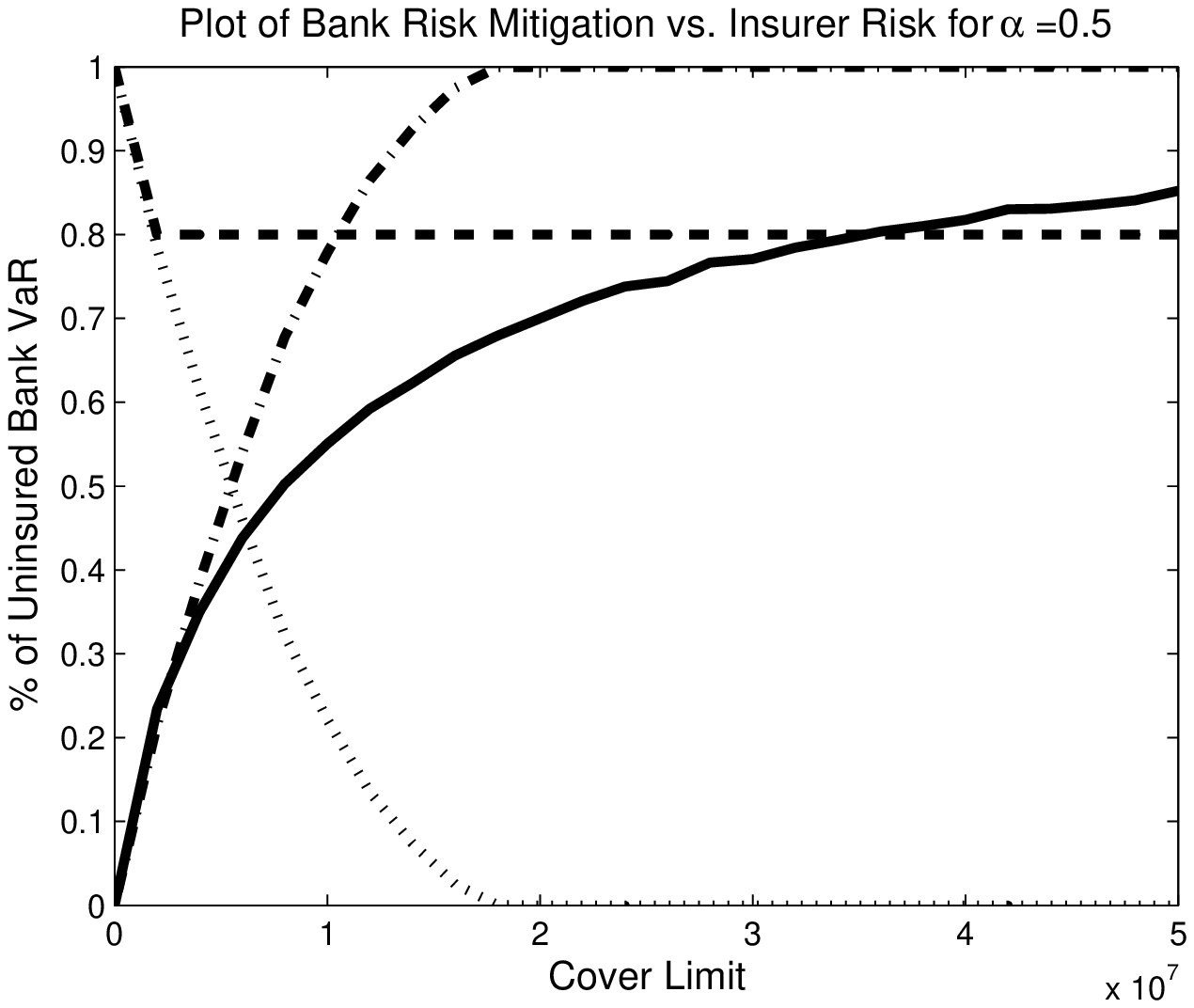}
\caption{\textbf{Top Row} - ILP with Haircut, \textbf{Bottom Row} - CLP with Haircut; \textbf{Column 1} $\alpha = 2$, \textbf{Column 2} $\alpha = 1.3$, \textbf{Column 3} $\alpha = 0.5$. All figures display unregulated Bank risk exposure (dotted line), regulated Bank risk exposure (dashed line),  risk transferred to the Insurer (solid line), risk mitigation received by the Bank (dashed-dot line).}
\label{FigHILP_LowAlphaStabVaR}
\end{figure}

\textbf{Haircut Loss Policies}\\
In Figure \ref{FigHILP_LowAlphaStabVaR}, we demonstrate the impact of a haircut applied to the cover limit.

Firstly, comparing Figure \ref{FigHILP_LowAlphaStabVaR} with that of Figure \ref{FigILP_LowAlphaStabVaR}, we can see that the application of a haircut discounts the level of risk mitigation per unit of TCL applied. However, unlike our previous observations of the low frequency setting, when a haircut is applied to the ILP and CLP we see a saturation in the risk mitigation offered per unit of TCL. This can be attributed to the fact that as we are discounting the TCL to $TCL'$ (and not the ACL) we can infer that $TCL'<ACL$ and hence the CLP policy is now dominated by the value of $TCL'$ and hence coincides with the ILP. Interestingly, in the case of CLP for small values of TCL we can observe the linear relationship attributed to the ALP. This is similar to the caveat previously remarked on in the low frequency setting, that for small TCL the CLP will coincide with the ALP but as the value of TCL increases (due to the discount applied) the CLP will begin to converge to the ILP.

Additionally, we can also observe an increasing divergence betweeen the risk transfered to the Insurer (solid line) and the risk mitigation received by the Bank (dashed-dot line). We observe a growing disparity in the optimum insurance point for the haircut ILP policy as $\alpha$ decreases. Consequently, we can infer that once a haircut discount is applied to the insurance mitigation, the \textquotedblleft profitability\textquotedblright for a bank to undertake insurance for light tailed (small $\alpha$) risk exposures is reduced, while their \textquotedblleft profitability\textquotedblright will be increased for heavy-tailed distributions.

\begin{figure}[!ht]
\includegraphics[scale=0.35]{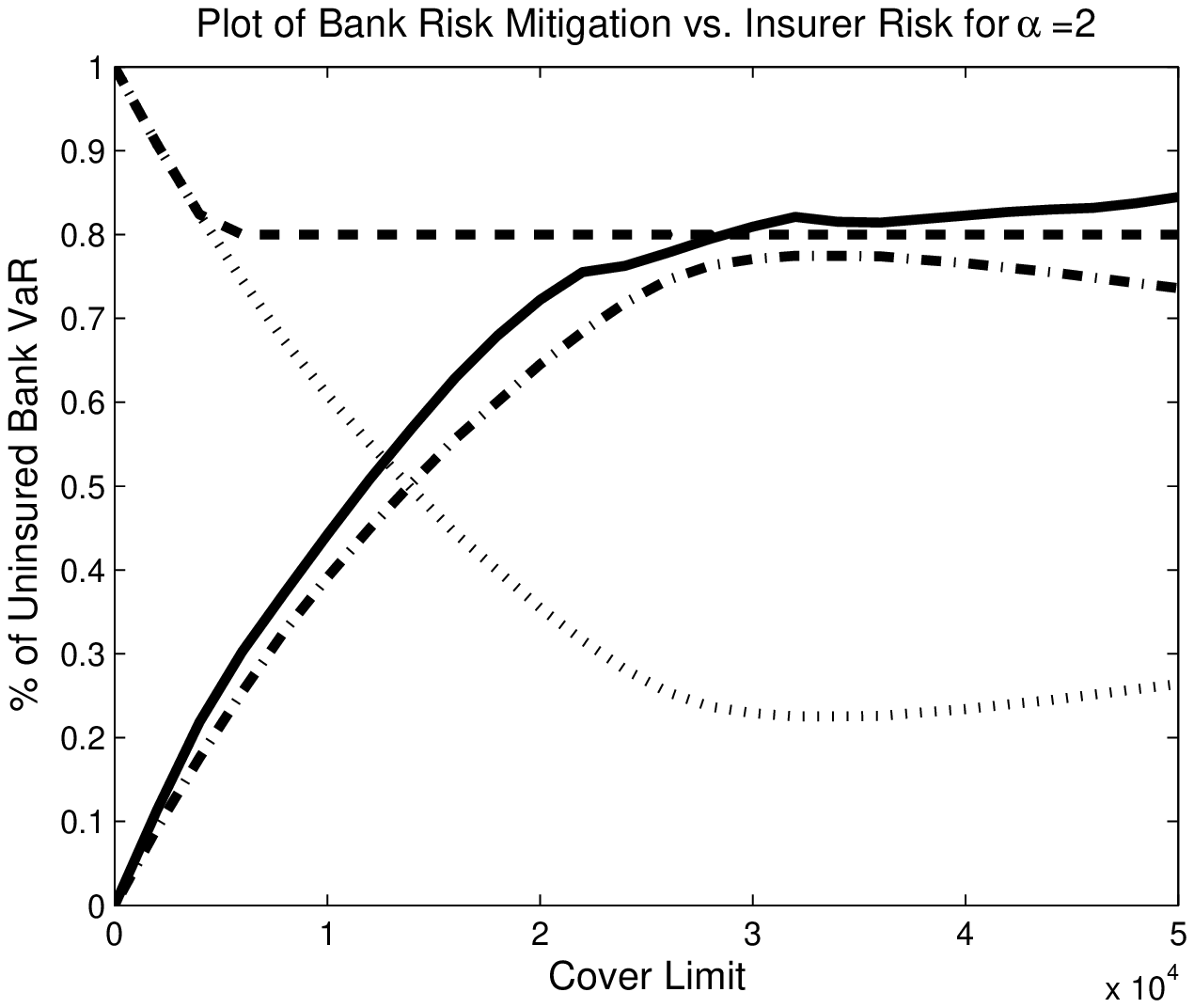}
\includegraphics[scale=0.35]{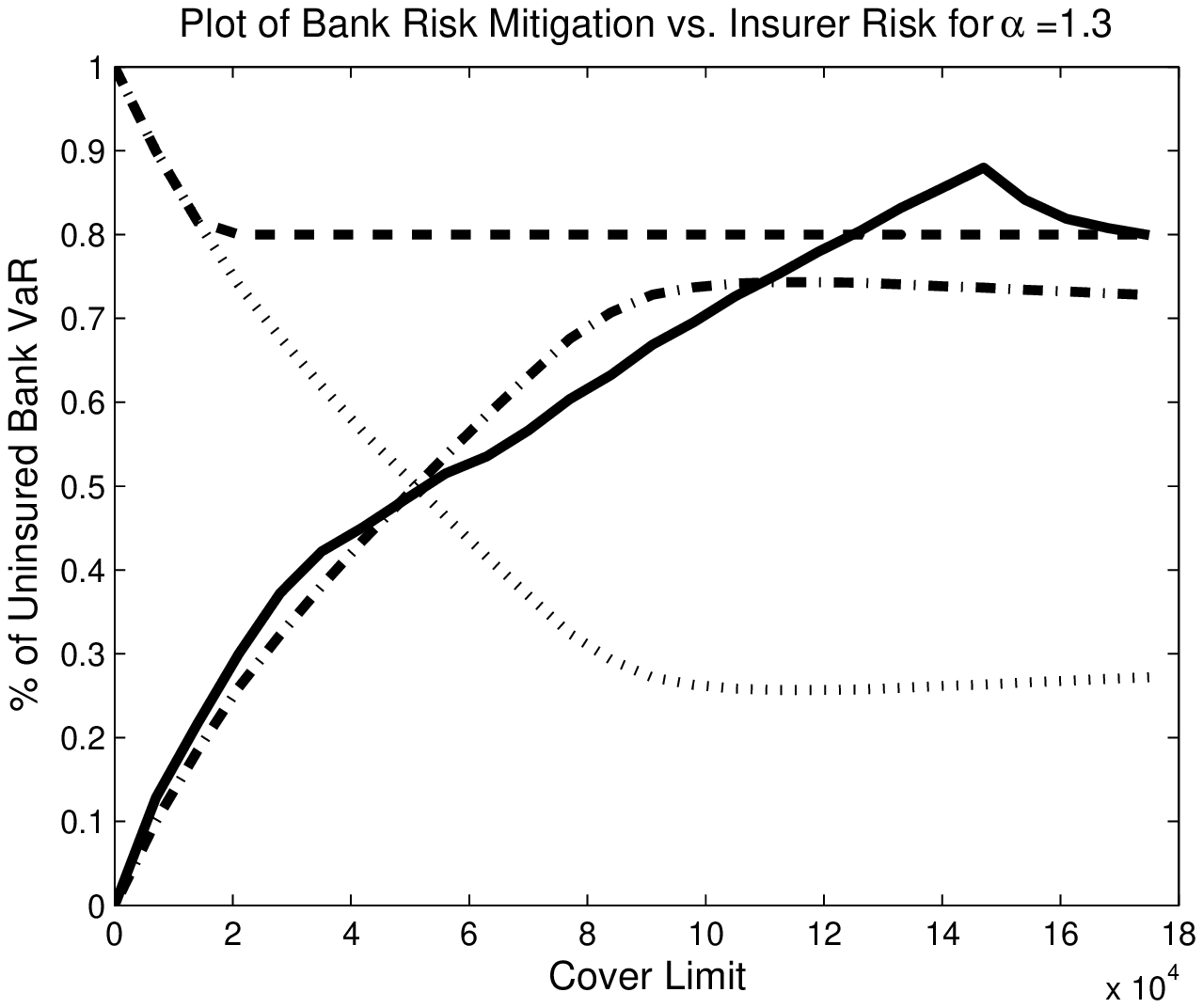}
\includegraphics[scale=0.35]{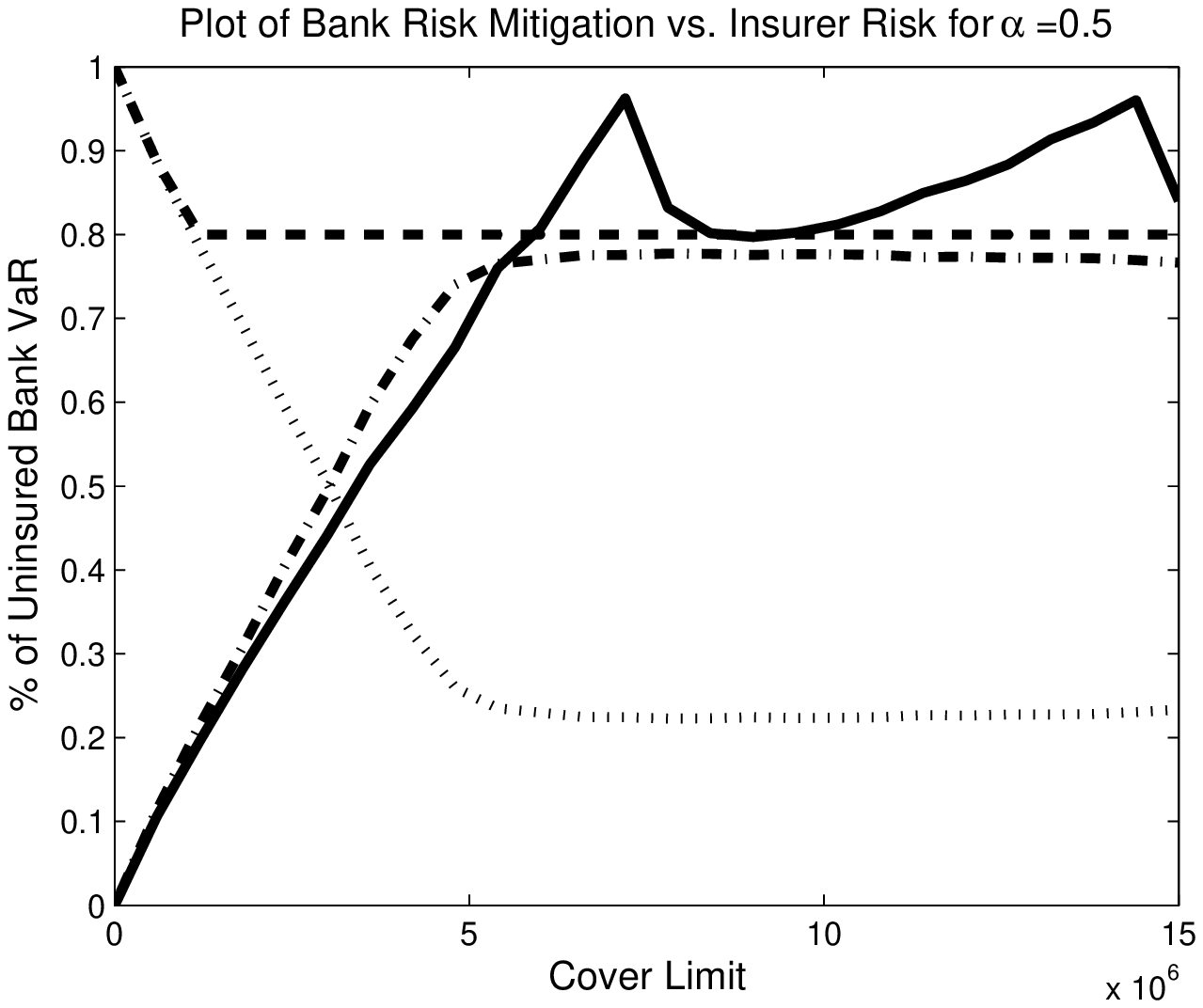}\\
\includegraphics[scale=0.35]{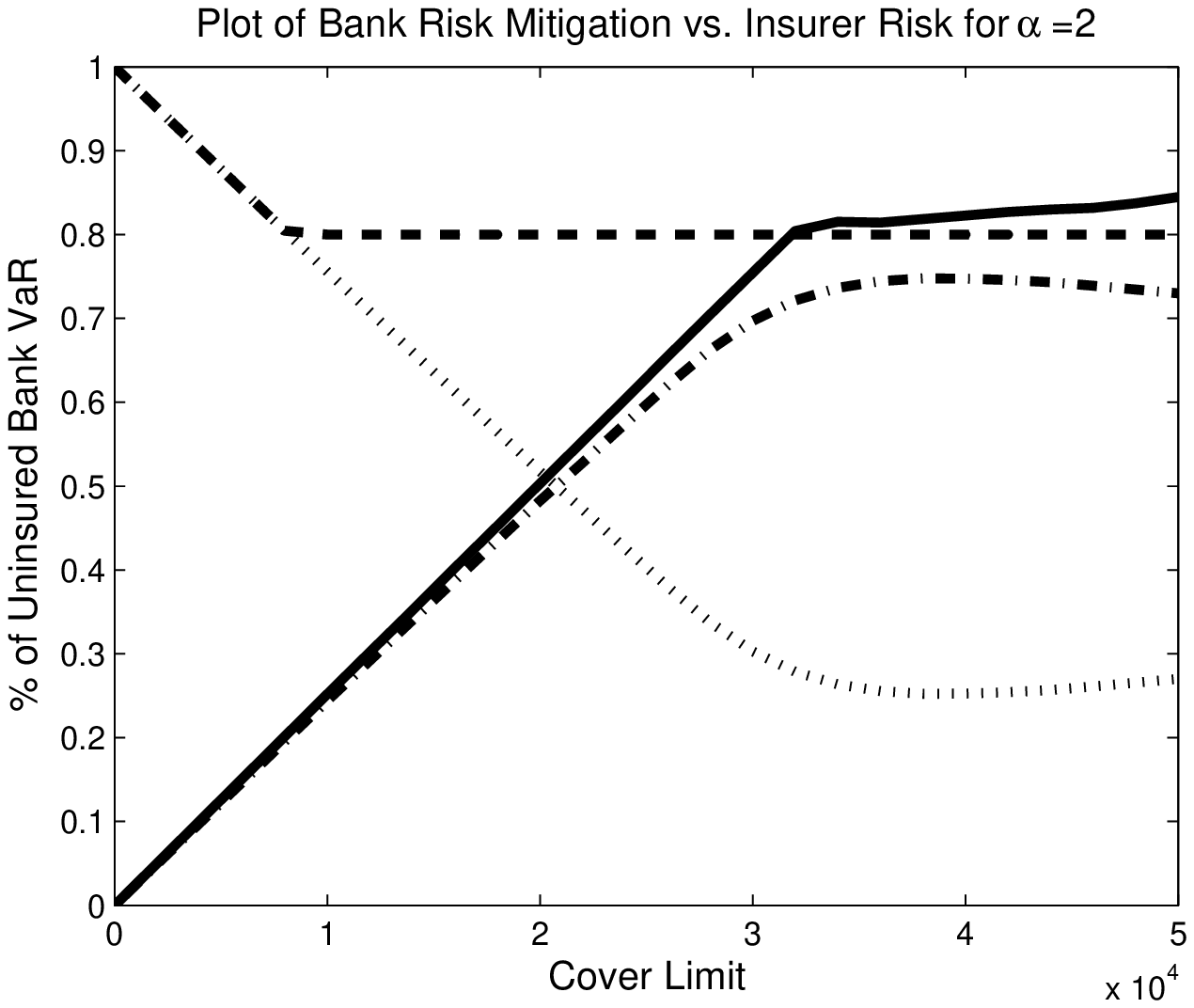}
\includegraphics[scale=0.35]{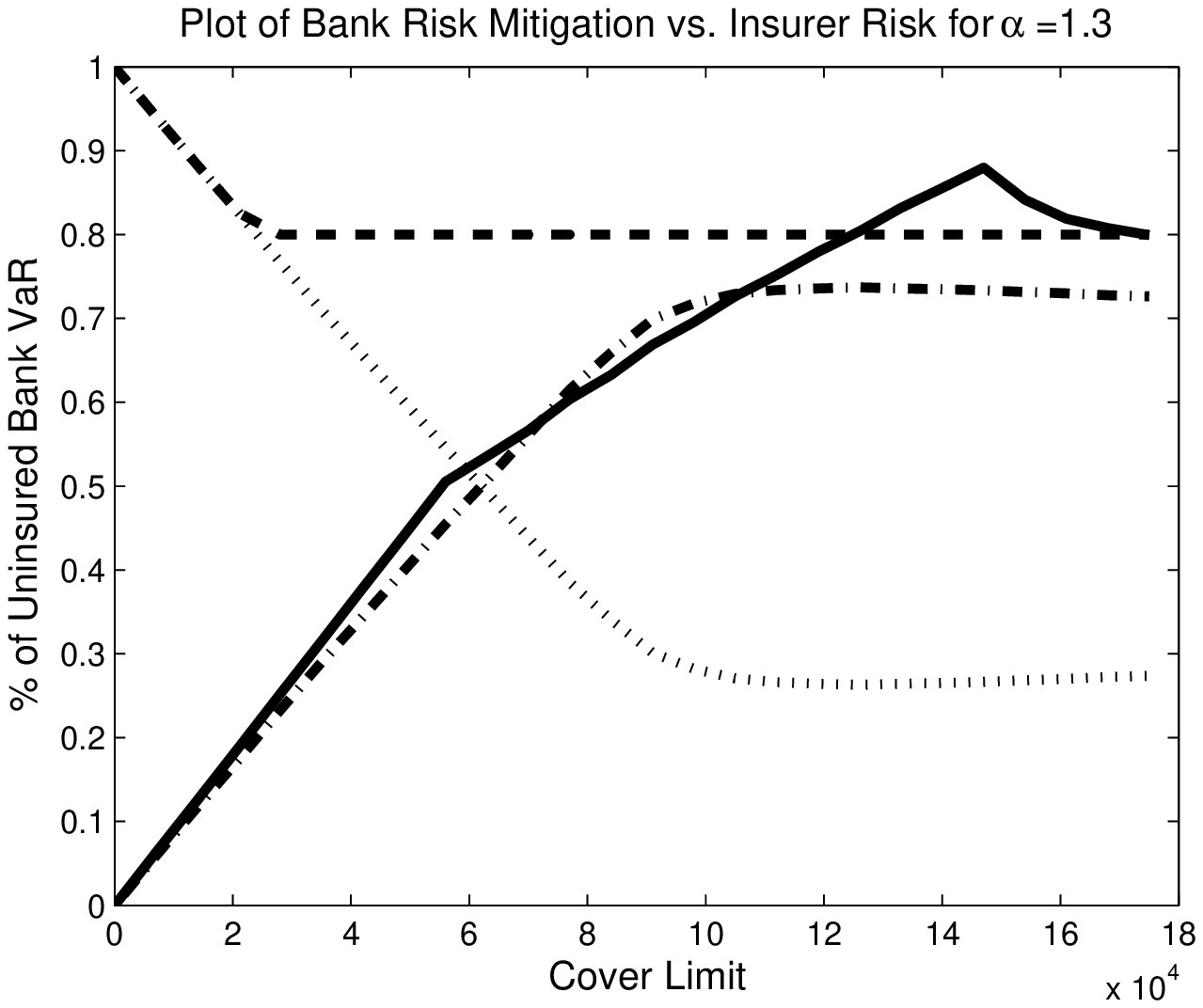}
\includegraphics[scale=0.35]{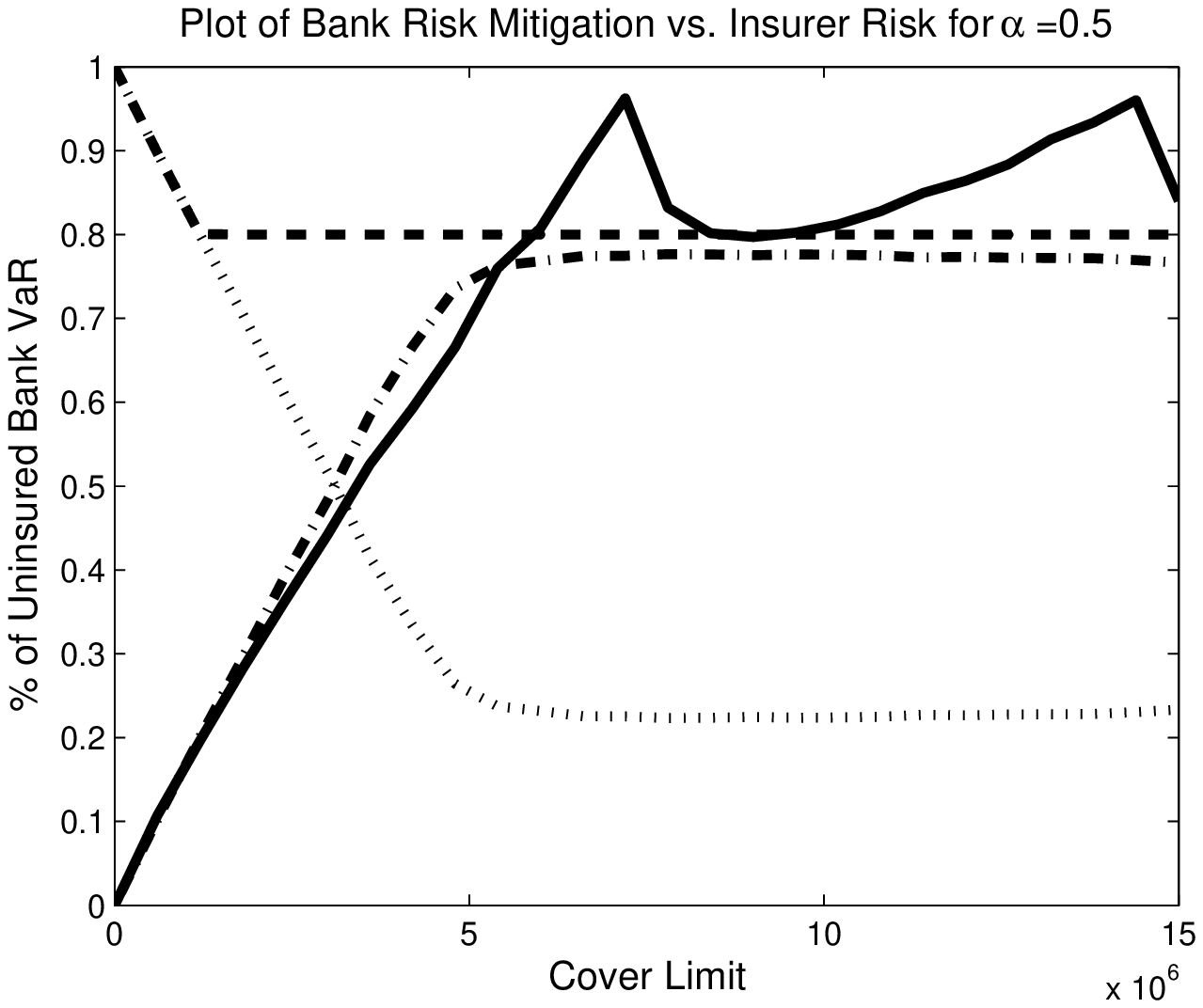}
\caption{\textbf{Top Row} - ILP with Linear Banding, \textbf{Bottom Row} - CLP with Linear Banding; \textbf{Column 1} $\alpha = 2$, \textbf{Column 2} $\alpha = 1.3$. All figures display unregulated Bank risk exposure (dotted line), regulated Bank risk exposure (dashed line),  risk transferred to the Insurer (solid line), risk mitigation received by the Bank (dashed-dot line).}
\label{FigLinB_LowAlphaStabVaR}
\end{figure}

\textbf{Banded Loss Policies}\\
In Figure \ref{FigLinB_LowAlphaStabVaR} we demonstrate a linear banding of the cover limit. 

From the perspective of the bank, in Figure \ref{FigLinB_LowAlphaStabVaR} we observe that there appears to be a natural limit to the amount of risk mitigation available from the insurer (dash-dot line) under the banding discount. Intuitively, this is due to the fact that the insurer will not provide complete compensation for every loss event and as such the bank is never capable of fully mitigating its risk exposure, thus demonstrating the effect of payment uncertainty on risk transfer efficiency.

From the perspective of the insurer, this banding structure also creates a limit for the potential risk exposure transferred (solid line). In addition, we observe the banding structure present in the model as illustrated by the peaks in the insurer's $\% MCR_{0.95}\left[C^{(j,i)}\right]$. Obviously, this illustrates the transfer of risk from one band to another and the associated discounting change that results. For the infinite moments model ($\alpha=0.5$) we can see that the banding shift is not realised until the bank risk has approached its limit, while for less heavy-tailed models the banding shifts can be seen occuring prior to the maximisation of risk mitigation received by the bank. Finally, this banding shift does not appear to affect the risk mitigation (dashed-dot line) received by the bank. Note that as the number of bands $D \rightarrow \infty$ we will see a convergence of the BLP to that of the basic policy with no discounts.

Interstingly, unlike the observations in the case of haircuts, we do not observe an increasing divergence between risk transferred (solid line) and the risk mitigation (dashed-dot), and consequently, our analysis now yields an optimum insurance region. In this region, the Bank will only be charged a fair (or profitable) premium for TCL values ranging between some minimum and maximum. Additionally, we can observe that due to payment uncertainty there is no situation in which a bank can expect to be charged a fair premium for their risk transfer if the exposure is light tailed ($\alpha$ in a region of $2$).

\section{Conclusions}
This paper studied the behaviour of different insurance policies in the context of capital reduction for a range of possible extreme loss models and insurance policy scenarios in a multi-period, multiple risk settings. We have developed novel stochastic insurance models for Basel II LDA modelling and investigated their ability to provide insurance mitigation. Also, we have examined Solvency II and quanitified important insurer capital quantities related to the MCR and SCR. In addition we have introduced and extensively studied for OpRisk models the family of $\alpha$-stable severity models in the LDA structure of Basel II, demonstrating that they provide a flexible family of models capable of capturing many important aspects of OpRisk loss processes, such as extreme loss scenarios. In the process we provide closed-form solutions for the distribution of loss processes and claims processes in an LDA structure as well as closed-form analytic solutions for the Expected Shortfall, SCR and MCR under Basel II and Solvency II.

\end{document}